\UseRawInputEncoding
\documentclass[11pt,reqno]{amsart}
\usepackage{amsthm,amsfonts,amssymb,euscript,fancyhdr,datetime,xcolor,graphicx,float}
\usepackage[colorlinks,linkcolor=blue,citecolor=blue]{hyperref}

\newcommand\smallO{
  \mathchoice
    {{\scriptstyle\mathcal{O}}}
    {{\scriptstyle\mathcal{O}}}
    {{\scriptscriptstyle\mathcal{O}}}
    {\scalebox{.6}{$\scriptscriptstyle\mathcal{O}$}}
  }

\usepackage[font=small,labelfont=bf]{caption} 

\newcommand{\gd}{{g \mkern-8mu /\ \mkern-5mu }}
\newcommand{\di}{\mbox{$d \mkern-9.2mu /$\,}}

\def\Kerr{{\mathrm{Kerr}}}

\def\a{{\alpha}}

\def\be{{\beta}}

\def\ga{\gamma}

\def\de{\delta}

\def\si{\sigma}
\def\Si{\Sigma}
\def\om{\omega}
\def\Om{\Omega}
\def\th{\theta}

\def\varep{\varepsilon}
\def\pr{{\partial}}

\def\les{\lesssim}
\def\rh{{\rho}}
\def\ind{{\in \mkern-16mu /\ \mkern-4mu}}
\def\XX{{\mathcal{X}}}
\def\YY{{\mathcal{Y}}}

\def\gdcd{{\dot{\gd_c}}}
\def\Omd{{\dot{\Om}}}
\providecommand{\lrpar}[1]{\left( #1\right)}

\def\etabd{{\dot{\etab}}}
\def\etad{{\dot{\eta}}}

\def\Kd{{\dot{K}}}
\def\Ldo{{\overset{\circ}{\Ld}}}
\def\phid{{\dot{\phi}}}
\def\omd{{\dot{\om}}}
\def\abd{{\dot{\ab}}}

\def\ombd{{\dot{\omb}}}
\def\ad{{\dot{\alpha}}}

\def\CC{{\mathcal C}}

\def\MM{{\mathcal M}}

\def\HH{{\mathcal H}}

\def\OO{{\mathcal O}}
\def\SS{{\mathcal S}}

\def\KK{{\mathcal K}}

\def\DD{{\mathcal D}}

\def\RR{{\mathcal R}}
\def\QQ{{\mathcal Q}}

\def\HHb{\underline{\mathcal H}}

\def\D{{\bf D}}

\def\H{{\mathcal H}}

\def\g{{\bf g}}

\def\SSS{{\mathbb{S}}}
\def\RRR{{\mathbb R}}

\DeclareMathOperator{\Div}{\mathrm{div}}
\DeclareMathOperator*{\Curl}{\mathrm{curl}}
\def\half{\frac{1}{2}}
\newcommand{\pd}{\pd \mkern-9mu/\ \mkern-7mu}
\newcommand{\Lied}{\mathcal{L} \mkern-9mu/\ \mkern-7mu}

\newcommand{\DDd}{\DD \mkern-10mu /\ \mkern-5mu}

\newcommand{\Du}{\underline{D}}

\newcommand{\Divd}{\Div \mkern-17mu /\ }
\newcommand{\Divdo}{{\overset{\circ}{\Div \mkern-17mu /\ }}}
\newcommand{\Curld}{\Curl \mkern-17mu /\ }

\newcommand{\Nd}{\nabla \mkern-13mu /\ }

\newcommand{\Ld}{\triangle \mkern-12mu /\ }

\newcommand{\trchi}{{\tr \chi}}
\newcommand{\trchib}{{\tr \chib}}

\newcommand{\chihd}{{\dot{\chih}}}
\newcommand{\chibhd}{{\dot{\chibh}}}
\newcommand{\omtrchid}{\dot{(\Om \tr \chi)}}
\newcommand{\omtrchibd}{\dot{(\Om \tr \chib)}}
\def\ni{\noindent}

\def\Lb{{\,\underline{L}}}

\def\tr{\mathrm{tr}}

\def\chih{{\widehat \chi}}
\def\chib{{\underline \chi}}
\def\chibh{{\underline{\chih}}}

\def\etab{{\underline \eta}}
\def\omb{{\underline{\om}}}

\def\aa{{\underline{\a}}}

\def\th{\theta}

\def\Rbf{{\mathbf{R}}}

\newcommand{\gac}{{\overset{\circ}{\ga}}}

\newcommand{\ab}{{\underline{\alpha}}}
\newcommand{\beb}{{\underline{\beta}}}

\newtheorem{theorem}{Theorem}[section]
\newtheorem{lemma}[theorem]{Lemma}
\newtheorem{proposition}[theorem]{Proposition}
\newtheorem{corollary}[theorem]{Corollary}

\newtheorem{remark}[theorem]{Remark}

\numberwithin{equation}{section}

\makeatletter
\def\@setthanks{\vspace{-\baselineskip}\def\thanks##1{\@par##1\@addpunct.}\thankses}
\makeatother
\begin{document}

\title[Characteristic gluing problem for Einstein equations]{The characteristic gluing problem \\ for the Einstein equations and applications}
\author[S. Aretakis, S. Czimek and I. Rodnianski]{Stefanos Aretakis $^{(1)}$, Stefan Czimek $^{(2)}$, and Igor Rodnianski $^{(3)}$} 

\thanks{\noindent$^{(1)}$ Department of Mathematics, University of Toronto, 40 St George Street, Toronto, ON, Canada, \texttt{aretakis@math.toronto.edu}. \\
$^{(2)}$ Institute for Computational and Experimental Research in Mathematics, Brown University, 121 South Main Street, Providence, RI 02903, USA,  \texttt{stefan\_czimek@brown.edu}. \\
$^{(3)}$ Department of Mathematics, Princeton University, Fine Hall, Washington Road, Princeton, NJ 08544, USA, \texttt{irod@math.princeton.edu}. }

\begin{abstract} In this paper we introduce the characteristic gluing problem for the Einstein vacuum equations. We present a codimension-$10$ gluing construction for characteristic initial data which are close to the Minkowski data and we show that the $10$-dimensional obstruction space consists of gauge-invariant charges which are conserved by the linearized null constraint equations. By relating these $10$ charges to the ADM energy, linear momentum, angular momentum and the center-of-mass we prove that asymptotically flat data can be characteristically glued (including the $10$ charges) to the data of a suitably chosen Kerr spacetime, obtaining as a corollary an alternative proof of the Corvino--Schoen spacelike gluing construction. Moreover, we derive a localized version of our construction where the given data restricted on an angular sector is characteristically glued to the Minkowski data restricted on another angular sector. As a corollary we obtain an alternative proof of the Carlotto--Schoen localized spacelike gluing construction. Our method yields no loss of decay in the transition region, resolving an open problem. We also discuss a number of other applications.
\end{abstract}

\maketitle
\setcounter{tocdepth}{3}
\tableofcontents

\section{Introduction and statement of main results}

\subsection{Introduction to the characteristic gluing problem} \ni The \emph{Cauchy problem} in general relativity is concerned with a construction of solutions of the Einstein (vacuum) equations 
$$ \mathbf{Ric}_{\alpha\beta}(\mathbf{g}) - \frac 12 \mathbf{g}_{\alpha\beta} \mathbf{R}_{\mathrm{scal}}(\mathbf{g}) = 0, $$
a metric $\mathbf{g}$ and a 4-dimensional Lorentzian manifold ${\mathcal M}$ with prescribed initial data consisting of the 
Riemannian metric $g$, 3-dimensional {\it spacelike} hypersurface $\Sigma$ and a second fundamental form $k$. As is well known, the 
data can not be prescribed arbitrarily and the Einstein equations force it to satisfy the {\it constraint equations}:
\begin{align}
&R_{\mathrm{scal}}(g)-|k|^2+({\mbox {tr}} k)^2=0,\notag\\
&{\mbox{div}} k-\nabla {\mbox {tr}} k=0.\label{eq:constr}
\end{align}
The study of the Cauchy problem thus naturally starts with the corresponding construction and the study of the space and properties of solutions 
of the constraint equations. The first, and still widely used, systematic attempt at constructing solutions was the Lichnerowitz-York 
{\it conformal method} \cite{Lich,York}
 which casts \eqref{eq:constr} as an elliptic system for a conformal factor $\phi$ and a traceless, divergence 
free 2-tensor $\sigma$ and {\it given} conformal class $g_0$ and a scalar function $\tau$, so that 
$$
g=\phi^4 g_0,\quad k=\phi^{-2}\sigma+\frac \tau 3\phi^4g_0,
$$
and $\phi$ and $\sigma$ satisfy
$$
\frac 18\Delta_{g_0}\phi - R_{\mathrm{scal}}(g_0) \phi+|\sigma|^2\phi^{-7}-\frac 23 \tau^2\phi^5=0,\qquad {\text div}_{g_0}\sigma=0.
$$
 The elliptic character of the resulting system suggests that the constraint equations possess certain rigidity properties. For instance, one might expect a {\it unique continuation} property: for a given $\Sigma$
(or even just merely fixing the topology of $\Sigma$,) any solution is uniquely determined as soon as it is defined on an open subset of $\Sigma$. The positive mass theorem \cite{SchoenYau1,SchoenYau2,Witten} could be viewed as a manifestation of such rigidity. Yet, the constraint equations are not fully rigid.
In a work that ran somewhat contrary to the previous experience with the constraint equations, Corvino and Corvino-Schoen \cite{Corvino,CorvinoSchoen} showed that any asymptotically flat solution of the constraint equation on $\Sigma=\Bbb R^3$ can be {\it glued}, far out enough, to a particular member 
of the Kerr family. This statement, in addition to contradicting rigidity, should also be surprising in view of its locality: a solution is changed 
to another solution without any change occurring on a given compact set. This is not a property of an elliptic equation and can only be explained by the additional freedom contained in the choice of the conformal class $g_0$ or, alternatively, the underdeterminancy 
of the system \eqref{eq:constr}. 

While spacelike hypersurfaces are instrumental in the formulation of the Cauchy problem, {\it null} hypersurfaces have a privileged role in
general relativity, in particular, but not restricted to, in {\it characteristic problems}. In that context, a null hypersurface $\H$, foliated by 
2-dimensional compact sections, say spheres, $S_v$, possesses a degenerate metric $\gd$, defining Riemannian metrics on each $S_v$,
and two null second fundamental forms $\chi$ and $\chibh$ which can be decomposed as 
$$
\chi = \chih +\half \trchi \gd, \,\, \chib= \chibh + \half \trchib \gd.
$$
We may assume (slightly oversimplifying) that a space-time metric ${\bf g}$ in the neighborhood of $\H$ is given by 
$$
{\bf g}=-dudv+\gd_{AB} d\theta^A d\theta^B,
$$
with $\theta^A$ some coordinate system on $S_v$. Then, with $D$ denoting the Lie derivative with respect to the null generator
of $\H$, $\frac{\partial}{\partial v}$, the basic content of {\it null constraint} equations is contained in 
\begin{align}
&D\gd=\chi,\notag\\
&D\trchi+\frac 12(\trchi)^2=-|\chih|^2,\label{eq:sys}\\
&D\chib=\chib\otimes\chi-\lrpar{K+\frac 12 \trchi\trchib-\frac 12 \chi\cdot\chib}\gd \notag,
\end{align} 
with $K$ being the Gauss curvature of $\gd$. These equations reflect the fact that $\H$ is embedded in an ambient spacetime 
$({\mathcal M}, \bf{g})$ satisfying the Einstein vacuum equations. The system above corresponds to the $C^1$ embedding and 
needs to to be supplemented by additional (similar) equations for higher order transversal quantities if one is interested in higher 
order regularity embeddings.

It turns out that the set of all solutions of the above system can be parametrized
by prescribing the conformal class $[\gd]$ of $\gd$ on $S_v$ for all $v$ as well as the {\it initial data} on, say, $S_0$:
the conformal factor $\phi$, $\trchi$ and $\chib$, \cite{ChrFormationBlackHoles}. What is more is that with these data, 
the equations become a system of ODE's in the variable $v$ for each $\theta$. This indicates that, unlike the spacelike 
constraint problem, the solutions of the null constraint equations can be easily constructed. 

However, existence is not the only property of interest. Of particular importance is the question of {\it flexibility} of solutions which asks if one can deform one 
solution to another. In this paper, we will view this question from a perspective of a {\it characteristic gluing problem}
in which we will attempt to connect 
two codimension-$2$ surfaces, say spheres $S_1$ and $S_2$, with the corresponding induced data  (by the ambient space-time, to any 
prescribed order) by a solution of the null constraint equations. As we shall see, at the linear level, solutions of the null constraint 
equations can be constructed explicitly and be parametrized in terms of weighted integrals of the linearized conformal class $[\gd]$,
which can be viewed as a linear control for the system of null constraints. The resulting resulting representation formulas either come 
with different weights, which effectively means that they be can be glued independently, or, if they come with the same weights, 
correspond to linearly conserved quantities which, while they can not be freely glued, turn out to be either gauge dependent or 
give rise to the global linearly conserved charges -- 10-dimensional space of obstructions to gluing. 

One of the key aspects of the characteristic gluing problem, which comes as a consequence of the ODE nature of the null constraint equations, is that it is {\it local}. That is, a solution connecting $S_1$ and 
$S_2$, if it exists, can be constructed {\it only} between $S_1$ and $S_2$, without any regard to what happens to the past of $S_1$ and to the future of $S_2$. This locality, coupled with the domain of dependence and domain of influence arguments, makes the characteristic gluing problem an intriguing tool for applications to spacelike gluing problems.  The above relation between the characteristic and spacelike 
data is captured in the following simple picture. Solutions of the null constraint equations on an outgoing null hypersurface 
${\mathcal {H}_{[1,2]}}$, connecting $S_1$ and $S_2$ and on incoming null hypersurface ${\mathcal {\underline{H}}_{[2,3]}}$,
connecting $S_2$ and $S_3$, uniquely determine the solution of the Einstein vacuum equations in their past and, in particular, on 
a piece of spacelike hypersurface containing $S_1$ and $S_3$ (as long as they lie in the range guaranteed by the local existence and 
uniqueness results.)

\begin{figure}[H]
\begin{center}
\includegraphics[width=7.5cm]{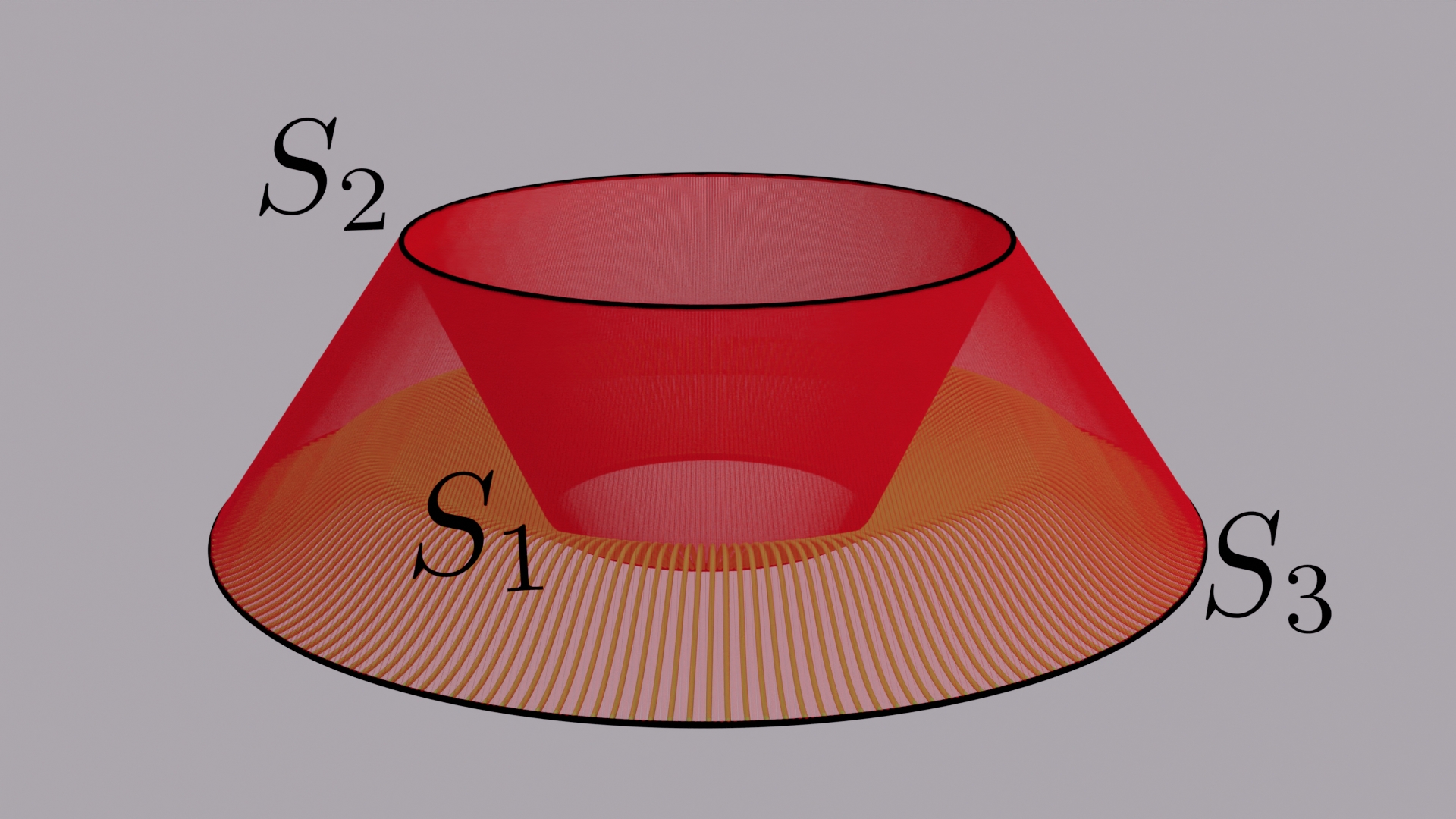} 
\vspace{0.4cm} 
\caption{The solution of the null structure equations on ${\mathcal {H}_{[1,2]}}$ and ${\mathcal {\underline{H}}_{[2,3]}}$ determine the data on a spacelike hypersurface containing $S_1$ and $S_3$.}\label{FIGintrospacelike}
\end{center}
\end{figure}

\subsection{Brief discussion of the results}  We start by introducing the characteristic gluing problem, which we view as the problem of 
constructing a null hypersurface $\mathcal{H}$ together with a solution of the null constraint equations, 
connecting spheres $S_1$ and $S_2$ with given data on them. The $C^k$-gluing corresponds to the existence of the $C^k$-germ 
of the (infinitesimal) embedding of $\mathcal{H}$  into an ambient Einstein vacuum spacetime. 

We then identify all obstructions to $C^2$-gluing of the data of a sphere $S_1$ to the data of a sphere $S_2$ along a null hypersurface $\mathcal{H}$, under the assumption that the provided data on $S_1$ and $S_2$ is small, that is, close to the respective Minkowski data. Moreover, we analyze the $C^{m+2}$-gluing of data of two sphere $S_1$ and $S_2$ (close to the respective Minkowski data) along two null hypersurfaces bifurcating from an auxiliary sphere, for any integer $m\geq0$. This analysis reveals a {\it threshold}: modulo the matching 
of global charges and gauge transformations of $S_2$, $C^2$ gluing between $S_1$ and $S_2$ is possible; the higher regularity gluing 
possesses additional obstructions and requires first gluing $S_1$ to the above mentioned auxiliary sphere along the transversal null 
hypersurface ${\mathcal{\underline H}}$.

As mentioned above, the obstructions stem from \emph{conservation laws} of the linearized null constraint equations at Minkowski spacetime. We prove that these conservation laws determine a 10-dimensional space
of \emph{gauge-invariant charges} and an infinite-dimensional space of \emph{gauge-dependent} charges. We demonstrate
that the the gauge-dependent charges can be matched by applying sphere perturbations and sphere diffeomorphisms to the sphere data.  
We provide a geometric interpretation the 10-dimensional space of gauge-invariant charges by connecting them to the ADM energy,
linear momentum, angular momentum and center-of-mass, and we use this identification to characteristically glue asymptotically flat spacetimes to a suitably chosen member of the Kerr family. As a corollary, we get an alternative proof of the Corvino--Schoen spacelike gluing construction.

 Moreover, we show how to localize small characteristic data along an angular sector of a null hypersurface $\mathcal{H}$ by constructing a solution to the null constraint equations such that 1) it agrees  with the original solution on an initial section $S_1$ of $\mathcal{H}$ and on an angular sector $K'$ of the null hypersurface of $\mathcal{H}$, and 2) it agrees with the trivial Minkowski data on the complement of a slightly larger angular sector $K'\subset\subset K$ on a later section $S_2$ of $\mathcal{H}.$  Our construction yields sharp estimates in the transition region $K\setminus K'$ between the two angular sectors along $\mathcal{H}$ and can be used to retrieve the Carlotto--Schoen \cite{CarlottoSchoen} construction without any loss of decay in the transition region, hence resolving an open problem in this direction. We note that from the point of view of construction and gluing of solution of the null constraints on a null hypersurface, the 
 phenomenon of {\it sector  localization} is completely natural: construction proceeds by solving transport equations along {\it each} of the null 
 generators of $\mathcal{H},$ {\it ``almost"} independently of each other.
 
 The solution of the localized characteristic gluing problem described above is based on a more general observation: the 10-dimensional 
 space of obstructions preventing us from gluing the sphere $S_1$ to $S_2$ can be eliminated if we do not insist on fixing {\it all} the 
 data on $S_2$. The latter can take many different forms but one particular case is when we fix the data everywhere but on an 
 open subset of $S_2$. This flexibility is sufficient to eliminate the obstructions, a solution can be constructed and controlled, 
 including on that open subset of $S_2$, in terms of the given data.

As was explained 
 earlier, the Einstein equations provide an intimate link between characteristic data, given on 2 transversal null hypersurfaces, and 
 and spacelike data living on a (piece of) any 3-dimensional spacelike hypersurface contained in their past. The two characteristic gluing results described above provide applications to the corresponding spacelike gluing problems exploiting this link (in two different ways.)
 We also give additional applications to the extension/fill-in problems and construction of ``exotic" spacelike data.

\subsection{Previous gluing constructions} \label{SEChistory9999} 

\ni Gluing constructions in general relativity are, up to now, chiefly focused on the gluing of \emph{spacelike} initial data subject to the elliptic constraint equations. 

On the one hand, constructions based on the gluing of connected sums (see also the works \cite{SchoenYauPSCM,GromovL} on codimension-$3$ surgery for manifolds of positive scalar curvature) were studied by Chru\'sciel--Isenberg--Pollack \cite{CIP1,CIP2}, Chru\'sciel--Mazzeo \cite{ChruscielMazzeo}, Isenberg--Maxwell--Pollack \cite{IMP3}, Isenberg--Mazzeo--Pollack \cite{IMP1,IMP2}.

On the other hand, in their ground breaking work, Corvino \cite{Corvino} and Corvino--Schoen \cite{CorvinoSchoen} used the \emph{geometric under-determinedness} of the spacelike constraint equations to study the (codimension-$1$) gluing problem. 
In particular, they showed that asymptotically flat spacelike initial data can be glued across a compact region to exactly Kerr spacelike initial data; see Corollary \ref{THMspacelikeGLUINGtoKERRv2INTRO} below. Further constructions and refinements based on this approach were established by Chru\'sciel--Delay \cite{ChruscielDelay1,ChruscielDelay}, Chru\'sciel--Pollack \cite{ChruscielPollack}, Cortier \cite{Cortier}, Hintz \cite{Hintz}. Another milestone was the result \cite{CarlottoSchoen} by Carlotto--Schoen which showed that spacelike initial data can be glued -- along a non-compact cone -- to spacelike initial data for Minkowski.

The \emph{characteristic} gluing problem was previously studied by the first author \cite{CiteGluing, CiteElliptic} in the much simpler setting of the linear homogeneous wave equation on general (but fixed) Lorentzian manifolds. Similarly to the present paper, \cite{CiteGluing} determined that the only obstruction to this gluing are conservation laws along null hypersurfaces. It was subsequently shown that these conservation laws have important applications in the study of the evolution of scalar perturbations on both sub-extremal \cite{CiteSS,CitePriceLaw, CiteKerr, Ma} and extremal \cite{CiteExtremal1,CiteExtremal2,CiteExtremal3,PRL} black hole spacetimes.

\subsection{Results on the gluing of characteristic and spacelike initial data}

Consider two vacuum spacetimes $(\MM_1,\g_1)$ and $(\MM_2,\g_2)$. Let $S_1$ and $S_2$ be two spacelike $2$-spheres in $\MM_1$ and $\MM_2$, respectively, given as intersection spheres of respective local double null coordinate systems. We define \emph{sphere data} $x_1$ on $S_1$ and $x_2$ on $S_2$ to be given by the respective restriction of the metric components, the Ricci coefficients and the components of the Riemann curvature tensor of the spacetimes to the respective spheres (with respect to the local double null coordinates). \emph{Sphere perturbations} of the sphere data $x_2$ on $S_2$ in the vacuum spacetime $\MM_2$ are defined as follows. Consider the ingoing null hypersurface $\HHb_2$ in $\MM_2$ through the sphere $S_2$. Then the induced sphere data $x'_2$ on a section $S_2'$ of $\HHb_2$ is called a sphere perturbation of $x_2$ on $S_2$ in $\MM_2$. We moreover define a \textit{sphere diffeomorphism} of $x_2$ by pulling back $x_2$ under a diffeomorphism of $S_2$ (see Section \ref{SECspherePERT}).   The following is the first main result of this paper, see Section \ref{SECcharGluing} for its proof.

\begin{theorem}[Perturbative codimension-$10$  characteristic gluing] \label{THMmain1}
 Let $\de>0$ be a real number. Consider sphere data $x_1$ on a sphere $S_1$, and characteristic initial data $x_{[-\de,\de],2}$ along an ingoing null hypersurface $\HHb_{2}$, and let $S_2$ be a section of $\HHb_{2}$ with sphere data $x_{0,2}$ (see Sections \ref{SECdoublenull}, \ref{SECspheredataCharges} and \ref{SECnorms}). Assume that both $x_1$ and $x_{[-\de,\de],2}$ are respectively sufficiently close to the sphere data on a sphere of radius $1$ and characteristic initial data on the ingoing null hypersurface $\HHb_{2}$ passing through the sphere of radius $2$ in Minkowski (with respect to the standard double null coordinates, see Section \ref{SECdoublenull}). Then there is a null hypersurface $\HH'_{[1,2]}$ connecting the sphere data $x_1$ on $S_1$ to a perturbation $S_2' \subset \HHb_{2}$ of the sphere $S_2$ with sphere data $x_2'$ (subject also to sphere diffeomorphisms), satisfying the null constraint equations such that -- up to the $10$ gauge-invariant charges explicitly defined at $S_2'$ -- all derivatives tangential to $\HH'_{[1,2]}$ of the sphere data $x_1$ and $x'_2$ are glued.
\end{theorem}

\begin{figure}[H]
\begin{center}
\includegraphics[width=7.5cm]{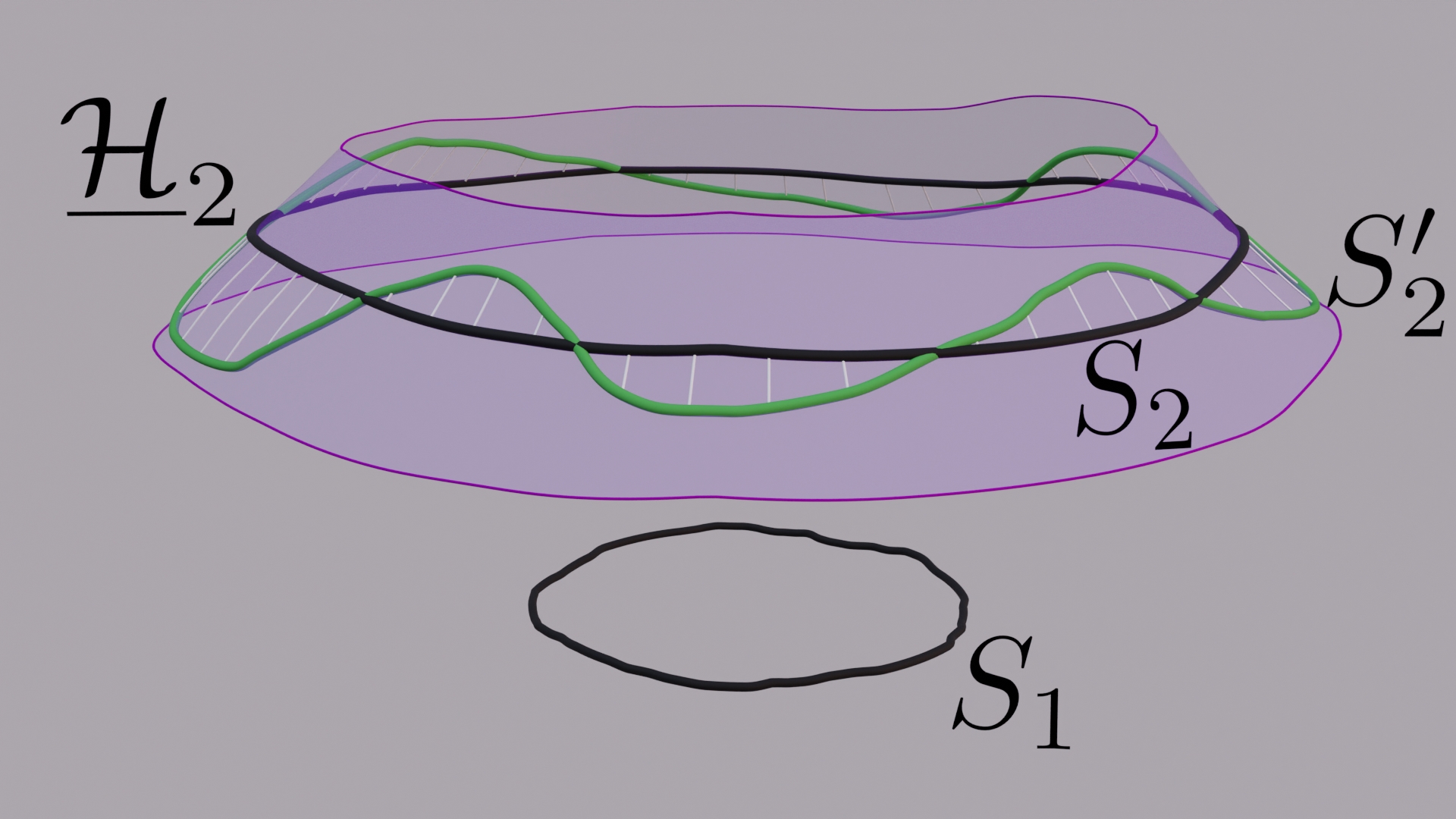} \,\,\, \includegraphics[width=7.5cm]{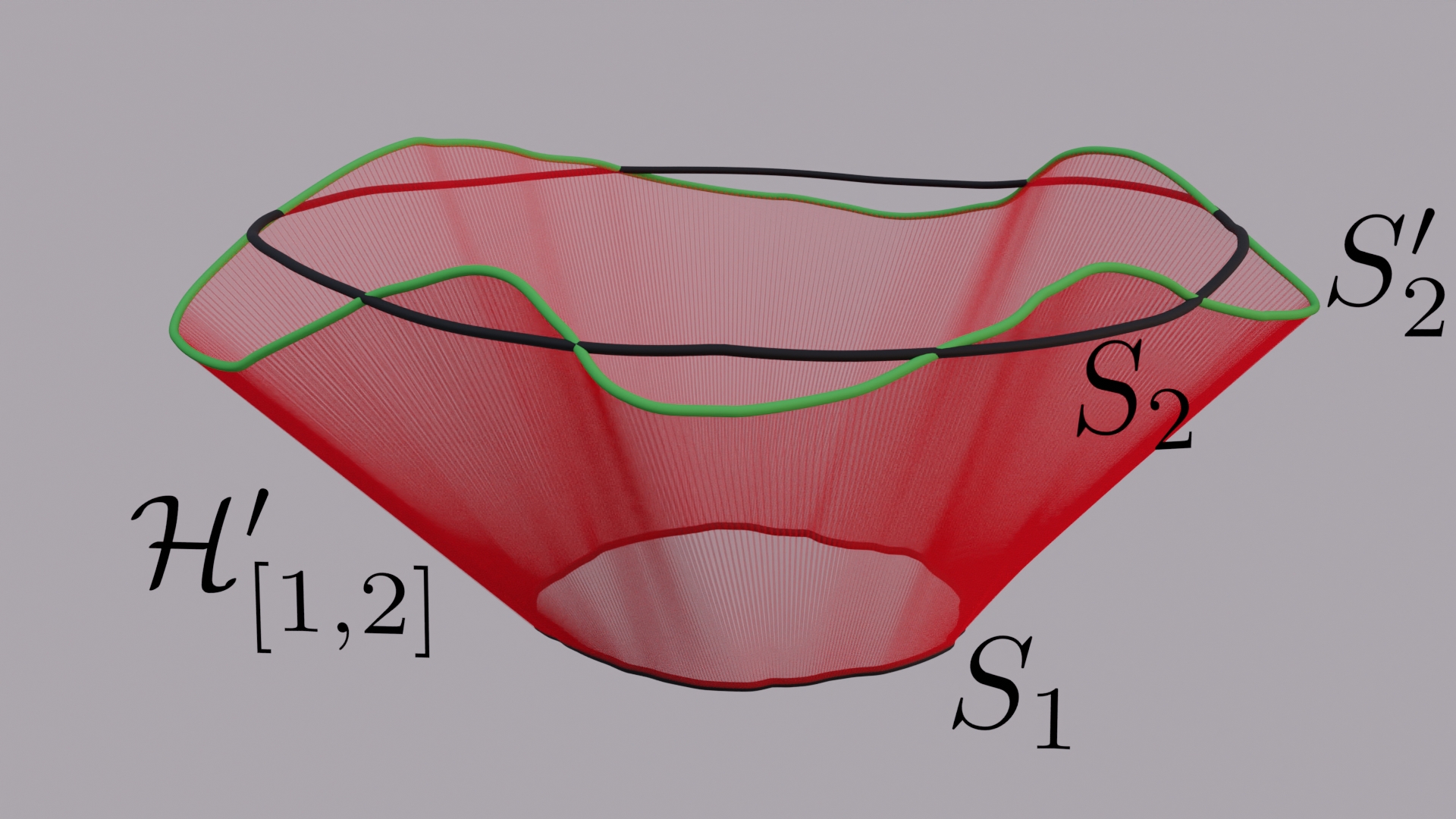}
\vspace{0.4cm} 
\caption{The codimension-$10$ perturbative characteristic gluing of $S_1$ and $S_2$. The left picture depicts a sphere perturbation $S_2'$ of $S_2$. The right picture illustrates the codimension-10 characteristic gluing of $S_1$ and $S_2'$.}\label{FIG1}
\end{center}
\end{figure}

\ni In Theorem \ref{THMmain1} it is equivalently possible to perturb along the ingoing null hypersurface $\HHb_{1}$ passing through the sphere $S_1$ and keep the sphere $S_{2}$ fixed; this formulation is used in Theorem \ref{PROPmain2}. The gluing in Theorem \ref{THMmain1} is -- up to the $10$-dimensional space of charges -- at the level of $C^2$ gluing for the metric components.

\begin{figure}[H]
\begin{center}
\includegraphics[width=7.5cm]{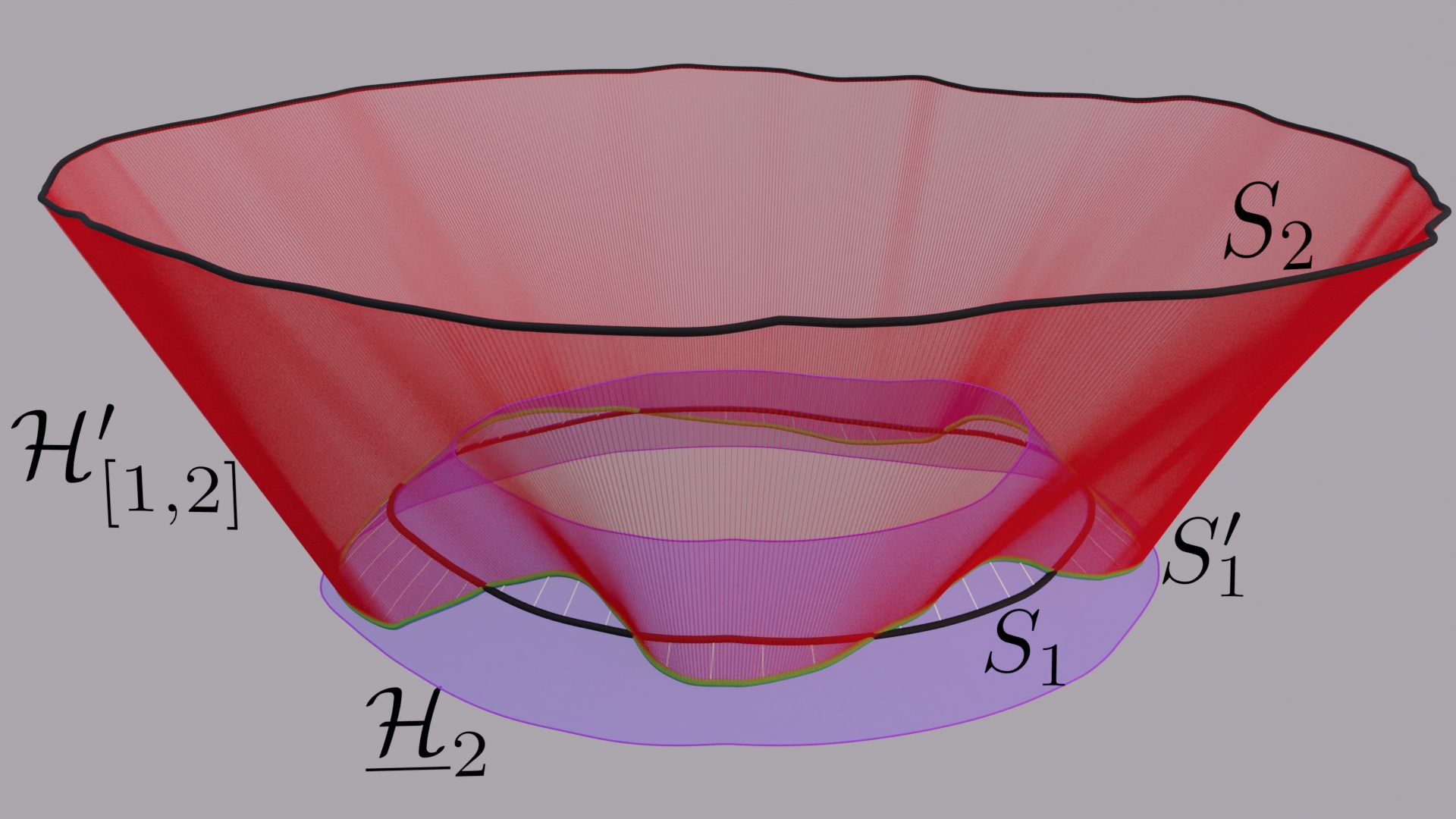} 
\vspace{0.4cm} 
\caption{The codimension-10 characteristic gluing of a sphere perturbation $S_1'$ and $S_2$.}\label{FIG2}
\end{center}
\end{figure}

\ni Higher-order derivatives of the sphere data transversal to the gluing null hypersurface are not glued at $S_2'$. This is due to the existence of additional higher-order conserved charges which involve these transversal derivatives. The next theorem, proved in Section \ref{SECbifurcateGluing}, resolves both of these issues by considering gluing along two null hypersurfaces bifurcating from an auxiliary sphere. This bifurcate gluing is -- up to the $10$-dimensional space of charges -- at the level of $C^{m+2}$-gluing for the metric components, where $m\geq0$ is the order of the higher-order sphere data (see Section \ref{SECspheredataCharges}).

\begin{theorem}[Codimension-$10$ bifurcate characteristic gluing] \label{THMbifurcate}Consider two spheres $S_1$ and $S_2$ equipped with sphere data $x_1$ and $x_2$, respectively, as well as with prescribed higher-order derivatives in all directions (see Sections \ref{SECdoublenull}, \ref{SECspheredataCharges} and \ref{SECnorms}). If the higher-order data on $S_1$ and $S_2$ is sufficiently close to the respective higher-order data on the (round) spheres of radius $1$ and $2$ in Minkowski spacetime, then it is possible to characteristically glue up to a $10$-dimensional space of charges the higher-order data of $S_1$ and $S_2$  along a bifurcate null hypersurface $\HH\cup \HHb$ emanating from an auxiliary sphere $S_{\mathrm{aux}}$. The higher-order sphere data $x_{\mathrm{aux}}$ on $S_{\mathrm{aux}}$ is close to the higher-order sphere data on the round sphere of radius $1.5$ in Minkowski spacetime.
\end{theorem}

\begin{figure}[H]
\begin{center}
\includegraphics[width=7.5cm]{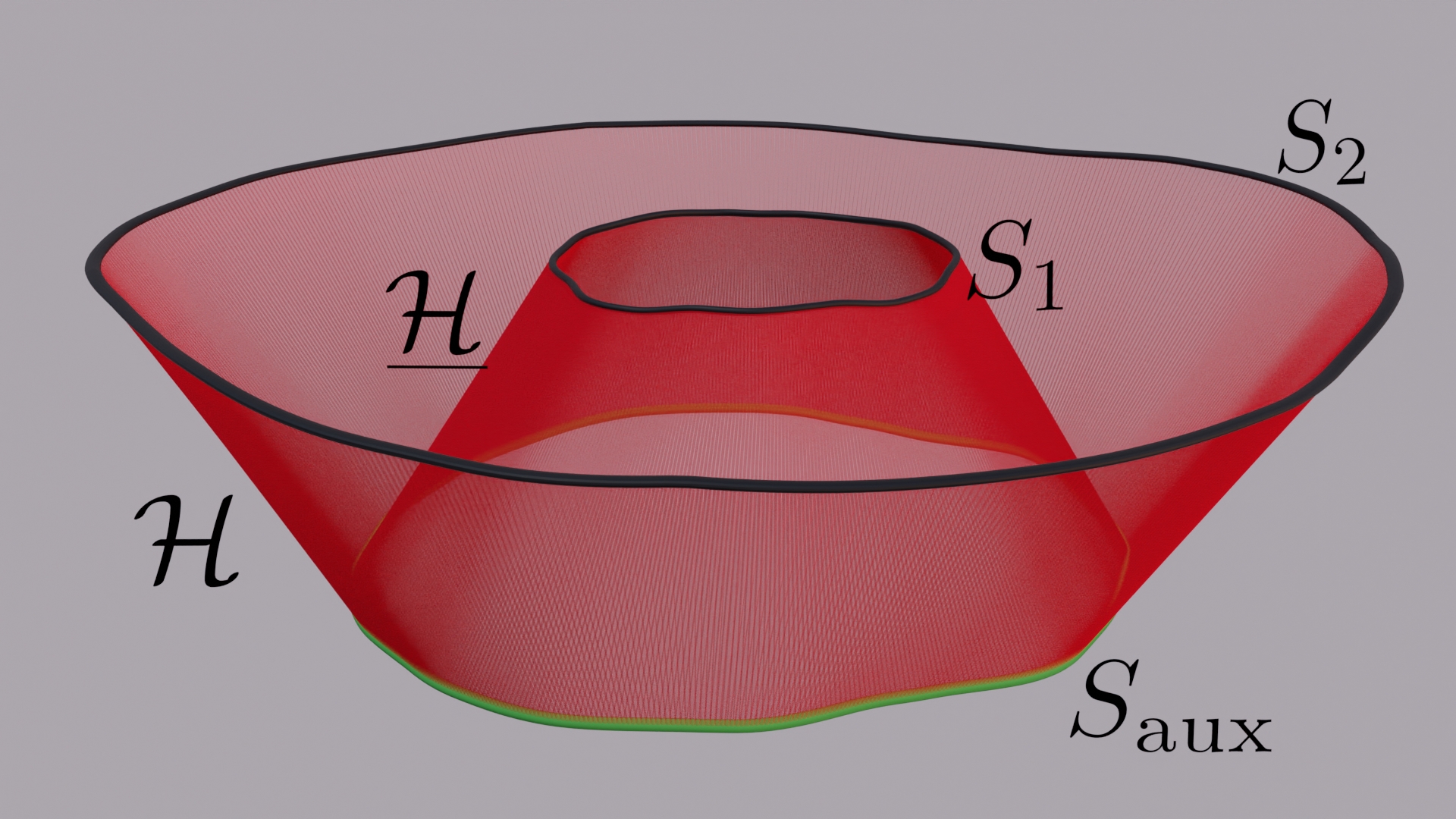} 
\vspace{0.4cm} 
\caption{The codimension-10 bifurcated characteristic gluing of $S_1$ and $S_2$. The bifurcate null hypersurface emanates from an auxiliary sphere $S_{\mathrm{aux}}$.}\label{FIG3}
\end{center}
\end{figure}

\ni In the bifurcate characteristic gluing of Theorem \ref{THMbifurcate}, we are \emph{not} perturbing $S_1$ to $S_{\mathrm{aux}}$ inside an ambient spacetime (as in Theorem \ref{THMmain1}), but we \emph{construct} the solution to the null structure equations along $\HHb$ in our bifurcate gluing.

As mentioned, the above characteristic gluing holds up to a $10$-dimensional space of charges. These charges are calculated as integrals over spacelike $2$-spheres and are denoted by the real number $\mathbf{E}$ and the $3$-dimensional vectors $\mathbf{P}, \mathbf{L}$ and $\mathbf{G}$. At the linear level, the charges $\mathbf{E}$ and $\mathbf{P}$ are proportional to the modes $l=0$ and $l=1$ of $\rh+ r \Divd \beta$, while $\mathbf{L}$ and $\mathbf{G}$ are proportional to the magnetic and electric parts of the mode $l=1$ of $\beta$; see Section \ref{SECspheredataCharges}. Our definitions of the charges are, to leading order, consistent with previous definitions in general relativity of mass, linear and angular momentum in terms of integrals over spheres; see, for example, \cite{Komar,Wald}. 

Based on the above interpretation and Theorem \ref{THMmain1} we prove the following result in Section \ref{SECcharGluingKerr}.
\begin{theorem}[Perturbative characteristic gluing to Kerr] \label{PROPmain2}
Let $\de>0$ be a real number. Consider a strongly asymptotically flat family of characteristic initial data $x_{R\cdot [-\de,\de],R}$ on ingoing null hypersurfaces $\HHb_{R\cdot [-\de,\de],R}$ (see Sections \ref{SECdoublenull}, \ref{SECspheredataCharges} and \ref{SECnorms}). For $R\geq1$ sufficiently large, there exist 1.) a sphere perturbation $S^{\mathrm{pert}}_{-R,R} \subset \HHb_{R\cdot [-\de,\de],R}$ of $S_{-R,R}$ with sphere data subject to sphere diffeomorphisms, 2.) a sphere $S_{-R,2R}^{\Kerr}$ in a Kerr spacetime, and 3.) a null hypersurface $\HH_{-R,[R,2R]}$, solving the constraint equations, and connecting $S^{\mathrm{pert}}_{-R,R}$ and $S_{-R,2R}^{\Kerr}$ and their respective sphere data.
\end{theorem}
\ni The gluing in Theorem \ref{PROPmain2} is at the level of $C^2$ for the metric components. In Theorem \ref{PROPmain2} we glue to a reference sphere in Kerr (see Section \ref{SECKerrspacelike}). We could alternatively also glue to a perturbation thereof to avoid perturbing $S_{-R,R}$ to $S_{-R,R}^{\mathrm{pert}}$, see the remark after Theorem \ref{THMmain1}. We note that in the proof of Theorem \ref{PROPmain2} it is not necessary to have a family of characteristic initial data. Indeed, the proof works for one fixed, sufficiently small characteristic initial datum.

Applying the bifurcate characteristic gluing of Theorem \ref{THMbifurcate}, we have the following result which is at the level of $C^{m+2}$-gluing of metric components, where $m\geq0$ is the order of the family of higher-order sphere data.
\begin{theorem}[Bifurcate characteristic gluing to Kerr] 
\label{THMcharGluingTWOfirstintroversion199901} 
On spheres $S_R$ let $x_R$ be a strongly asymptotically flat family of higher-order sphere data (see Section \ref{SECnorms}). For $R\geq1$ sufficiently large, we can characteristically glue, to the same higher-order, along two null hypersurfaces emanating from an auxiliary sphere, the sphere $S_R$ to a sphere $S_{2R}^{\Kerr}$ in some Kerr spacetime.
\end{theorem}

\ni As for Theorem \ref{PROPmain2}, for Theorem \ref{THMcharGluingTWOfirstintroversion199901} it is not necessary to have a family of higher-order sphere data, but the proof applies also to one single sufficiently small sphere datum.

As a corollary to Theorem \ref{THMcharGluingTWOfirstintroversion199901}, we give an alternative proof of the \emph{spacelike gluing to Kerr} \cite{Corvino,CorvinoSchoen,ChruscielDelay} for strongly asymptotically flat spacelike initial data.
\begin{corollary}[Spacelike gluing to Kerr]
\label{THMspacelikeGLUINGtoKERRv2INTRO}\label{CORmain3} Let $m\geq0$ be an integer. Let $(\Si,g,k)$ be smooth strongly asymptotically flat spacelike initial data with ADM energy $\mathbf{E}_{\mathrm{ADM}}>0$. Then, sufficiently far out, $(g,k)$ can be glued (with $C^{{m+2}}$-regularity) across a compact region to spacelike initial data for some Kerr spacetime with ADM asymptotic invariants close to those of $(\Si,g,k)$.
\end{corollary} 

\begin{figure}[H]
\begin{center}
\includegraphics[width=7.5cm]{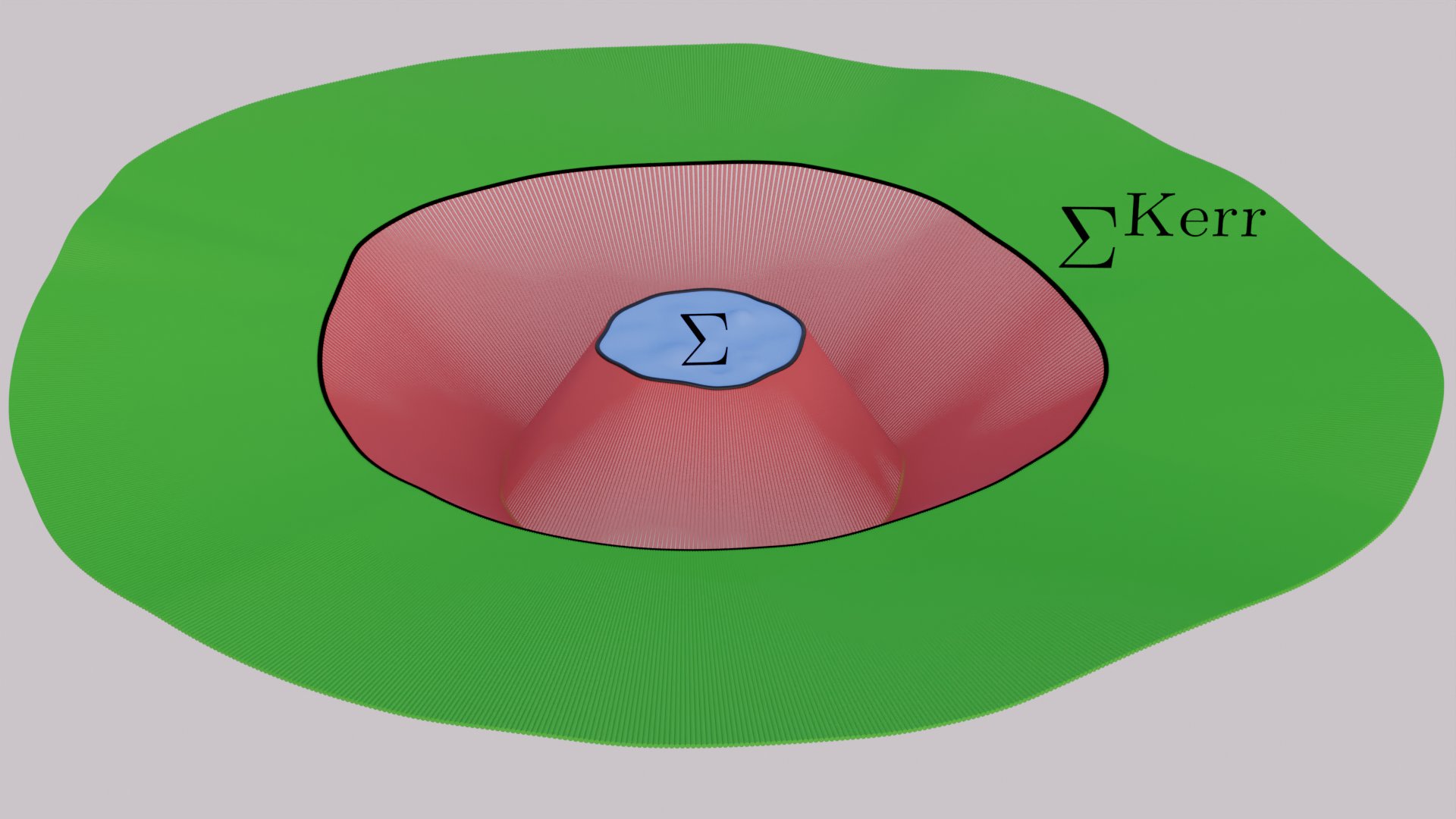} \,\,\, \includegraphics[width=7.5cm]{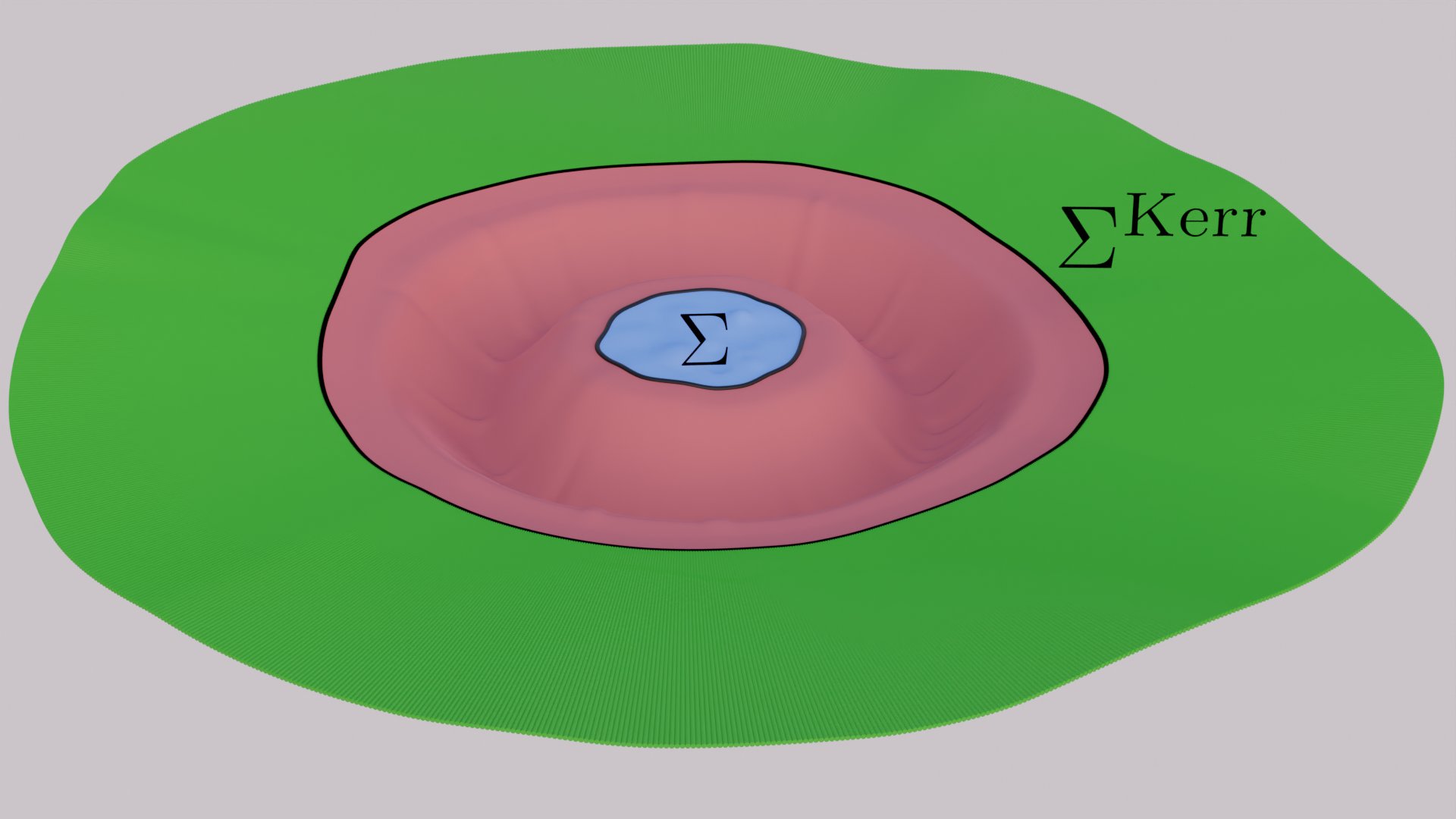}
\vspace{0.4cm} 
\caption{The bifurcated characteristic gluing of asymptotically flat data to a suitably chosen Kerr slice is shown on the left. The right picture illustrates the spacelike gluing of these data.}\label{FIG4}
\end{center}
\end{figure}
\ni In contrast to the previous Riemannian methods, our characteristic gluing approach allows us to establish an \emph{extension result} as follows.

\begin{proposition}[Extension of spacelike initial data] \label{PROPextension} Let $R\geq1$ be a real number. Let $(g,k)$ be strongly asymptotically flat spacelike initial data on $B_R$ with $M>0$, see \eqref{EQstrongAFness}. For $R\geq1$ sufficiently large, there exists an extension to Kerr, i.e. spacelike initial data $(\RRR^3,g',k')$ which isometrically contains $(B_R,g,k)$ and equals Kerr spacelike initial data outside a compact set.
\end{proposition}
\ni Corollary \ref{THMspacelikeGLUINGtoKERRv2INTRO} and Proposition \ref{PROPextension} apply similarly to spacelike initial data with non-trivial topology. Proposition \ref{PROPextension} assumes $M>0$ for the matching to Kerr. For local spacelike initial data close to Minkowski, an extension result (but not to Kerr) was proved by the second author in \cite{Czimek1}.

The methods of Theorem \ref{THMcharGluingTWOfirstintroversion199901} and Corollary \ref{THMspacelikeGLUINGtoKERRv2INTRO} also yield a novel \emph{fill-in} result, see \cite{ACR4}.
\begin{proposition}[Fill-in of spacelike initial data] \label{PROPfillin} Let $R\geq1$ be a real number. Let $(g,k)$ be strongly asymptotically flat spacelike initial data on $\RRR^3 \setminus B_R$ with $M>0$, see \eqref{EQstrongAFness}. For $R\geq1$ sufficiently large, there exists a fill-in, that is, spacelike initial data $(\RRR^3,g',k')$ which isometrically contains $(\RRR^3 \setminus B_R,g,k)$.
\end{proposition}


\ni In another direction, the codimension-$10$ characteristic gluing of Theorem \ref{THMmain1} can be upgraded to full gluing by changing the sphere data on $S_2$ using a perturbation $W$ which is not coming from a sphere perturbation or diffeomorphism. The perturbation $W$ can be chosen to be supported in any fixed angular region $K$. 
\begin{proposition}[Bifurcate characteristic gluing with localized sphere data perturbation $W$] \label{THMmain12} Let $K$ be an angular region. Consider sphere data $x_1$ and $x_2$ on spheres $S_1$ and $S_2$, respectively. Then we can characteristically glue $x_1$ to $x_2 +W$ along $\HH \cup \HHb$, where $W$ is a localized sphere data perturbation on $S_2$ used to adjust $(\mathbf{E},\mathbf{P},\mathbf{L},\mathbf{G})$ and supported in the angular region $K$.  
\end{proposition}
\ni There are many possible choices of such sphere data perturbations $W$. In Section \ref{SECWperturbation} we introduce a specific type of $W$ with advantageous properties used for the localization results in this paper.

\begin{figure}[H]
\begin{center}
\includegraphics[width=7.5cm]{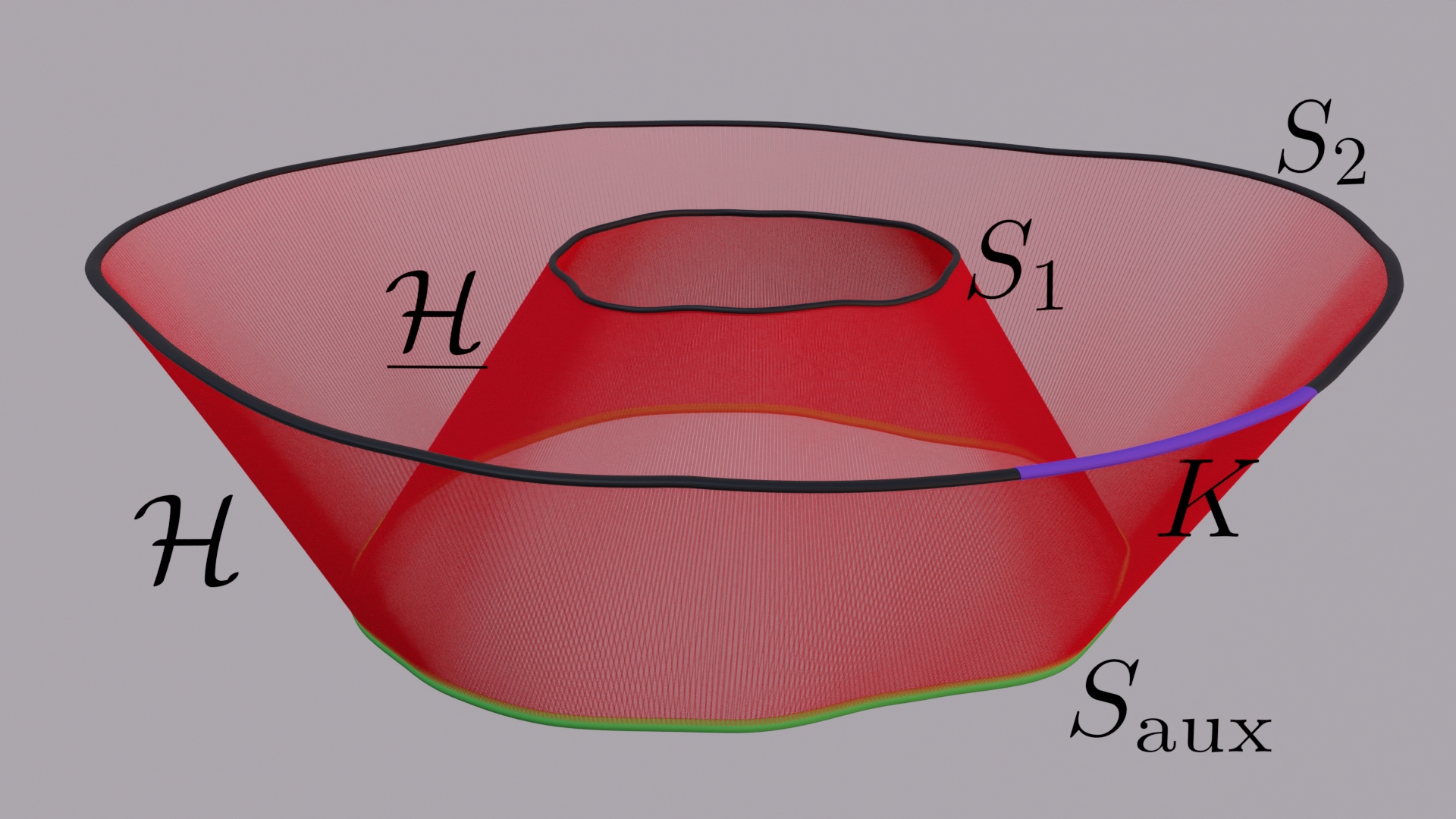} 
\vspace{0.4cm} 
\caption{The bifurcated characteristic gluing of two spheres $S_1$ and $S_2$, where the data on $S_2$ has been altered in the (arbitrarily small) region $K$.}\label{FIG5}
\end{center}
\end{figure}

\subsection{Results on the localization of characteristic and spacelike initial data} The methods of the previous sections also yield \emph{localization results}. Let $K' \subset \subset K \subset \subset \SSS^2$ be two angular regions, and let $\varphi$ be a smooth cut-off function on $\SSS^2$ such that $\varphi\equiv1$ on $K'$, and $\varphi \equiv 0$ in $K^c$. We define the corresponding angular regions along a null hypersurface in a standard way using the double null coordinates. In Section \ref{SECproofLocalization} we prove the next result concerning the localization of characteristic initial data in $K'$.
\begin{theorem}[Localized characteristic gluing] \label{THMmain4} Consider small higher-order characteristic data $x^{\mathrm{orig.}}$ given on $\HH \cup \HHb$. Let $x'_1$ be small higher-order sphere data on $S_1$ such that $x_1' \vert_{K'} = x^{\mathrm{orig.}}\vert_{S_1 \cap K'}$. Then there exists higher-order characteristic data $\tilde{x}$ on $\HH \cup \HHb$ such that $\tilde{x}$ agrees with $x^{\mathrm{orig.}}$ in the angular region $K'$ along $\HH \cup \HHb$ and 
\begin{align*} 
\begin{aligned} 
\tilde{x} \vert_{S_1} = x'_1 \text{ on } S_1, \,\, \tilde{x} \vert_{S_3} = \varphi \cdot x^{\mathrm{orig.}} \vert_{S_3} + (1-\varphi) \cdot \mathfrak{m}\vert_{S_3} + W \text{ on } S_3,
\end{aligned} 
\end{align*}
where $\mathfrak{m}$ denotes the trivial Minkowski higher-order sphere data, and $W$ is a smooth higher-order sphere data perturbation with support in the transition region $K\setminus K'$ (see Section \ref{SECWperturbation}).
\end{theorem}
\ni Theorem \ref{THMmain4} has a direct analogue for perturbative characteristic gluing (see, for example, Theorem \ref{THMmain1}), whose statement we omit here.

As a first application of Theorem \ref{THMmain4} we present an alternative proof of the Carlotto--Schoen localization of spacelike initial data \cite{CarlottoSchoen} in Section \ref{SECproofCarlottoSchoen}. Consider one-ended \emph{asymptotically flat} spacelike initial data $(\Si, g,k)$, that is, satisfying the following decay towards spacelike infinity,
\begin{align*} 
\begin{aligned} 
g_{ij}-e_{ij} = \OO(\vert x \vert^{-1}), \,\, k_{ij}(x)= \OO(\vert x \vert^{-2}),
\end{aligned} 
\end{align*}
as well as $\pr^{m'}g_{ij}=\OO(\vert x \vert^{-1-m'})$ and $\pr^{m'}k_{ij}= \OO(\vert x \vert^{-2-m'})$ for integers $0\leq m' \leq m$ and $m\geq0$ a sufficiently large integer (as needed). 

For given aperture $0<\th<\pi$ define the cone of aperture $\th$ by $$\CC_\th:= \{(x^1,x^2,x^3)\in \RRR^3: \, x^3 /\vert x \vert \geq \cos \th\}.$$

\begin{theorem}[Localized spacelike gluing] \label{PROPmain5} Consider asymptotically flat spacelike initial data $(\Si,g,k)$ together with two angles $0<\th_1<\th_2<\pi$. There is a real number $R_0>0$ such that for each $R\geq R_0$ there exists asymptotically flat spacelike initial data $(\Si,\tilde{g},\tilde{k})$, i.e.
\begin{align*} 
\begin{aligned} 
\tilde{g}_{ij}(x)-e_{ij}=\OO(\vert x \vert^{-1}), \,\, \tilde{k}_{ij}(x)=\OO(\vert x \vert^{-2}),
\end{aligned} 
\end{align*}
such that 
\begin{align*} 
\begin{aligned} 
(\tilde{g}_{ij},\tilde{k}_{ij})(x) = 
\begin{cases} 
(g_{ij},k_{ij})(x) &\text{ for } x\in \CC_{\th_1} \cup \{ \vert x \vert \leq R\}, \\
(e_{ij},0) &\text{ for } x\in \lrpar{\CC_{\th_2} \cup \{ \vert x \vert \leq 2R\} }^c.
\end{cases}
\end{aligned} 
\end{align*}
Moreover, by choosing $R\geq R_0$ sufficiently large, the energy-momentum $(\mathbf{E}_{\mathrm{ADM}},\mathbf{P}_{\mathrm{ADM}})$ of $(\tilde{g},\tilde{k})$ and $(g,k)$ can be made arbitrarily close to each other.
\end{theorem}

\ni We emphasize that the spacelike initial data constructed in Theorem \ref{PROPmain5} is asymptotically flat, and thus our construction does \emph{not} lose decay starting from asymptotically flat spacelike initial data. This is in contrast to the Carlotto--Schoen construction \cite{CarlottoSchoen} where such a loss is occurring and intertwined with their strategy of proof (coming from coercivity estimates which are crucial to solve their linearized problem), see Section 3.5 in \cite{Carlotto}, and also \cite{LeFloch}. In particular, Theorem \ref{PROPmain5} resolves \emph{Open Problem 3.18} in \cite{Carlotto}.

As in the Carlotto--Schoen construction, the localization in Theorem \ref{PROPmain5} contains an arbitrarily large subset of the original spacelike initial data. Our proof of Theorem \ref{PROPmain5} can be easily modified to localize inside the truncated \emph{outward-pointing} cone, that is, we construct asymptotically flat spacelike initial data $(\RRR^3, \tilde{g},\tilde{k})$ such that
\begin{align*} 
\begin{aligned} 
(\tilde{g}_{ij},\tilde{k}_{ij})(x) = 
\begin{cases} 
(g_{ij},k_{ij})(x) &\text{ for } x\in \CC_{\th_1} \cap \{ \vert x \vert \geq 2R\}, \\
(e_{ij},0) &\text{ for } x\in \lrpar{ \CC_{\th_2} \cap \{ \vert x \vert \geq R\} }^c,
\end{cases}
\end{aligned} 
\end{align*}
\ni in which case the mass and linear momentum of $(\tilde{g},\tilde{k})$ converge to zero as $R\to \infty$.

While Theorem \ref{PROPmain5} is stated for simplicity with cones parallel to the $x^3$-axis, any direction can be chosen as cone axis. Furthermore, the methods of Theorem \ref{PROPmain5} act exclusively on the asymptotically flat end, and thus can be applied similarly to multiple-ended spacelike initial data.

In \cite{ChruscielDelayexotic}, the Riemannian methods of Carlotto-Schoen were applied to localize asymptotically \emph{hyperbolic} spacelike initial data. We expect that our characteristic localization method of Theorem \ref{THMmain4} also applies to hyperbolic spacelike initial data. Indeed, the construction of Theorem \ref{THMmain4} can be generalized to outgoing future-complete null hypersurfaces, which -- after suitably fitting tangent (at future null infinity) to an asymptotically hyperbolic spacelike hypersurface -- should lead to localization results for asymptotically hyperbolic spacelike initial data.

As further application, Theorem \ref{THMmain4} can also be used to construct exotic spacelike initial data, see \cite{ACR4} for a proof. 
\begin{theorem}[Spacelike gluing of Minkowski ball to Kerr] \label{THMMinkowskiKerr} Let $r_0>0$ and $(\mathbf{E}_{\mathrm{ADM}})_0>0$ be a real number, and let $\mathbf{L}_{\mathrm{ADM}}\in \RRR^3$ be a vector. There is spacelike initial data $(\RRR^3,g,k)$ which isometrically contains a ball of radius $r_0>0$ of the trivial Minkowski data, $(B_{r_0},e,0)$, and agrees with Kerr spacelike initial data outside a compact set. The ADM energy $\mathbf{E}_{\mathrm{ADM}}$ and angular momentum $\mathbf{L}_{\mathrm{ADM}}$ of the Kerr spacelike initial data can be made arbitrarily close to $(\mathbf{E}_{\mathrm{ADM}})_0$ and $(\mathbf{L}_{\mathrm{ADM}})_0$.
\end{theorem}
\ni A version of Theorem \ref{THMMinkowskiKerr} is stated as Theorem 3.17 in \cite{Carlotto}, and proved therein within the class of \emph{time-symmetric} spacelike initial data (i.e. gluing Minkowski to Schwarzschild) by a delicate application of the Carlotto--Schoen spacelike localization \cite{CarlottoSchoen} and the Corvino--Schoen spacelike gluing \cite{Corvino,CorvinoSchoen}. In contrast, our proof in \cite{ACR4} relies on the localization in Theorem \ref{THMmain4}, Proposition \ref{THMmain12}, and the Kerr matching argument in Theorem \ref{PROPmain2}. Our methods of proof allow moreover to construct spacelike gluings from Minkowski to Kerr with non-trivial topology. 

\begin{figure}[H]
\begin{center}
\includegraphics[width=7.5cm]{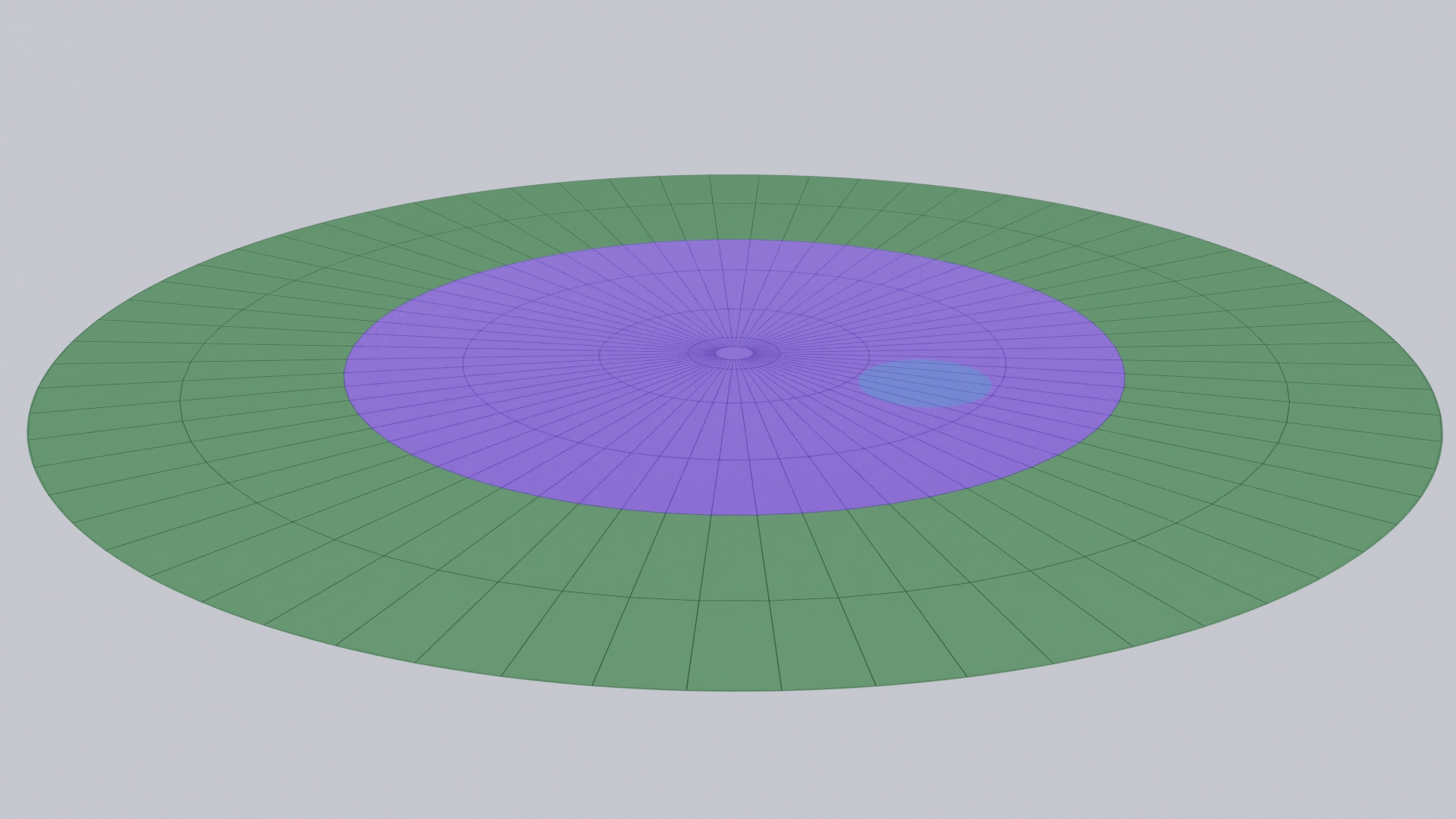} \,\,\, \includegraphics[width=7.5cm]{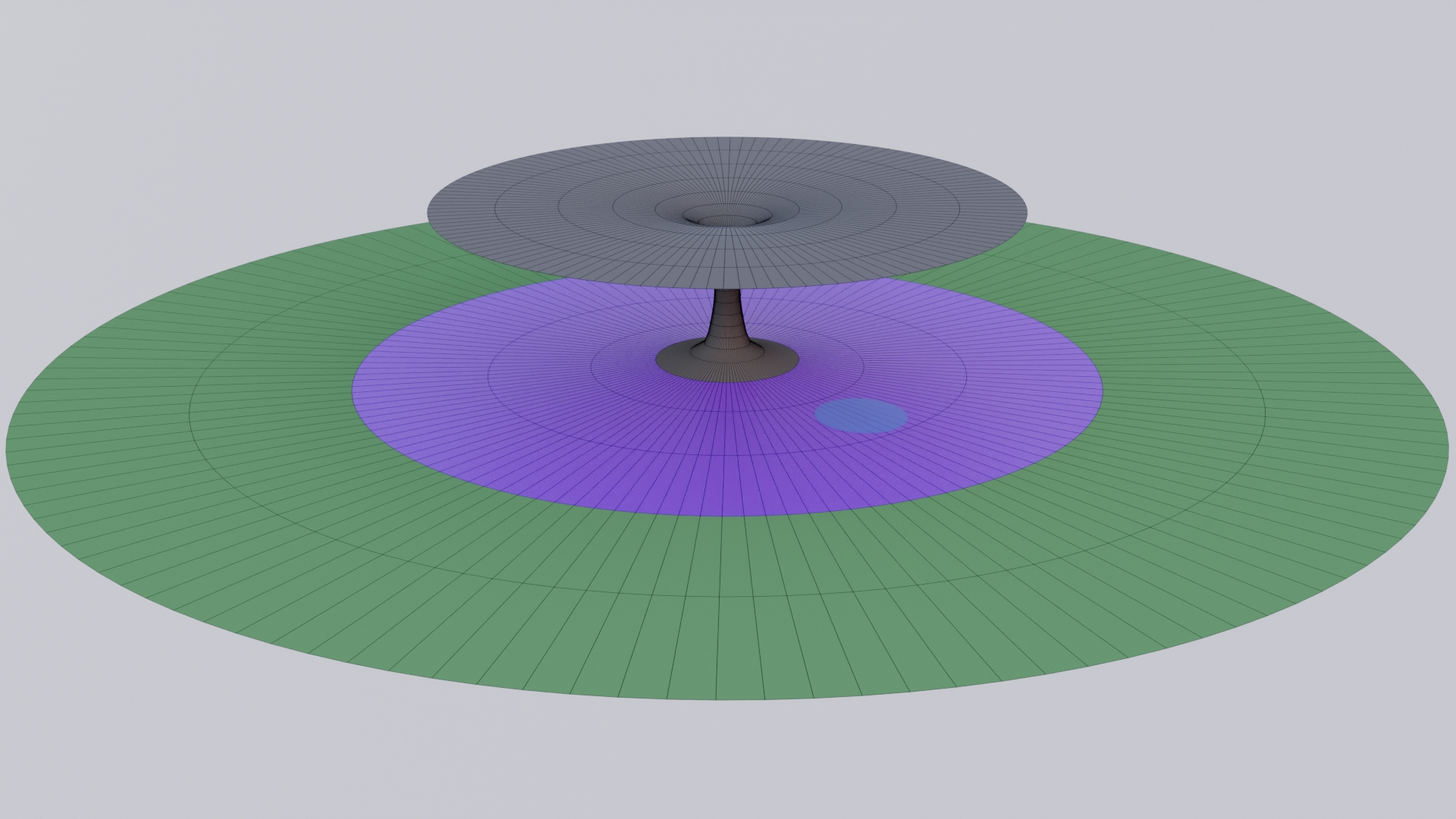}
\vspace{0.4cm} 
\caption{The left picture illustrates the spacelike gluing from a Minkowski ball to a Kerr slice. The right picture illustrates a similar gluing construction from Minkowski to Kerr that moreover contains a second Schwarzschildean asympotically flat end. The blue, green and gray regions are equipped with the Minkowski, Kerr and Schwarzschild metrics. The purple region is the transition region. }\label{FIG6}
\end{center}
\end{figure}

\subsection{Acknowledgements} S.A. acknowledges support through the NSERC grant 502581 and the Ontario Early Researcher Award. S.C. acknowledges support through the NSF grant DMS-1439786 of the Institute for Computational and Experimental Research in Mathematics (ICERM). I.R. acknowledges support through NSF grants DMS-2005464, DMS-1709270 and a Simons Investigator Award.
The authors would like to thank Mihalis Dafermos for valuable discussions.

\section{Preliminaries}

\ni For two real numbers $A$ and $B$, $A \les B$ indicates that there exists a universal constant $C>0$ such that $A \leq C \cdot B$. Greek indices range over $\a=0,1,2,3$, lowercase Latin indices over $a=1,2,3$ and uppercase Latin indices over $A=1,2$. For two real numbers $\varep>0$ and $\a \geq 0$, let $\OO(\varep^\a)$ and $\smallO(\varep^\a)$ respectively denote terms such that
$\lim\limits_{\varep\to0} \frac{\OO(\varep^\a)}{\varep^\a} <\infty$ and $\lim\limits_{\varep\to0} \frac{\smallO(\varep^\a)}{\varep^\a} =0.$
\subsection{Double null setting, Ricci coefficients and null curvature components} \label{SECdoublenull}
On a spacetime $(\MM,\g)$ let $u$ and $v$ be two \emph{optical functions} such that the intersections $S_{u,v}$ of their respective level sets $\HH_u$ and $\HHb_v$ form spacelike $2$-spheres. On $S_{u,v}$ let $\gd$ denote the induced Riemannian metric and $\Nd$ its covariant derivative, and let $r(u,v)$ denote the area radius of $(S_{u,v},\gd)$. We define the \emph{geodesic null pair} $(L',\Lb')$, the \emph{null lapse} $\Om$ and the \emph{normalized null pair} $(\widehat{L},\widehat{\Lb})$ by
\begin{align*} 
\begin{aligned} 
L' := -2 \mathbf{D}u, \,\, \Lb':= -2 \mathbf{D}v, \,\, \Om^{-2} := -\half \g(L',\Lb'), \,\, \widehat{L}:= \Om L', \,\, \widehat{\Lb} := \Om \Lb' \text{ on } \MM,
\end{aligned} 
\end{align*}
with $\D$ the covariant derivative on $(\MM,\g)$. The \emph{Ricci coefficients} are defined for $X,Y\in \mathrm{T}S_{u,v}$ by
\begin{align*} 
\begin{aligned} 
\chi(X,Y) :=&\g(\D_X\widehat{L},Y), & \chib(X,Y) :=&\g(\D_X\widehat{\Lb},Y), & \zeta(X):=& \half \g(\D_X\widehat{L},\widehat{\Lb}), \\
\eta:=& \zeta +\di \log \Om, & \om:=&D\log\Om, & \omb:=&\Du\log\Om,
\end{aligned} 
\end{align*}
with $\di$ the exterior derivative on $S_{u,v}$, and $\underline{\zeta}:=-\zeta$ and $\etab:= -\eta+2\di\log\Om$. Let further
\begin{align*} 
\begin{aligned} 
\a(X,Y) :=& \Rbf(X,\widehat{L},Y,\widehat{L}), & \beta(X):=&\half \Rbf(X,\widehat{L},\widehat{\Lb},\widehat{L}), & \rho:=& \frac{1}{4} \Rbf(\widehat{\Lb},\widehat{L},\widehat{\Lb},\widehat{L}), \\
\si \ind(X,Y) :=& \half \Rbf(X,Y,\widehat{\Lb},\widehat{L}), &\beb(X):=&\half \Rbf(X,\widehat{\Lb},\widehat{\Lb},\widehat{L}), & \ab(X,Y):=& \Rbf(X,\widehat{\Lb},Y,\widehat{\Lb}).
\end{aligned} 
\end{align*}
denote the \emph{null curvature components}, where $\Rbf$ denotes the Riemann curvature tensor of $(\MM,\g)$, and $\ind$ the area $2$-form on $(S_{u,v},\gd)$.

We define local \emph{double null coordinates} $(u,v,\th^1,\th^2)$ on $\MM$ by transporting local angular coordinates $(\th^1,\th^2)$ on a fixed sphere $S_{u_0,v_0}$ (for $v_0>u_0$) first by $L$ along $\HH_{u_0}$ and subsequently by $\Lb$ onto $\MM$. The spacetime metric $\g$ is given with respect to $(u,v,\th^1,\th^2)$ by
\begin{align} 
\begin{aligned} 
\g = - 4\Om^2 dudv + \gd_{AB} (d\th^A + b^A dv)(d\th^B+b^Bdv),
\end{aligned} \label{EQspacetimemetricDOUBLENULL}
\end{align}
where the \emph{shift vector} $b$ is an $S_{u,v}$-tangential vectorfield. We note that, by construction, $b=0 \text{ on } \HH_{u_0}$. Moreover, using $(\th^1,\th^2)$ on $S_{u,v}$, we write $\gd=\phi^2\cdot \gd_c$ where $\gd_c$ is the unique metric conformal to $\gd$ such that $\det \gd_c= \det \gac$ on $S_{u,v}$, where $\gac:= (d\th^1)^2 + (\sin\th^1)^2 (d\th^2)^2$. The standard optical functions on Minkowski are given by $u=\half(t-r)$ and $v=\half(t+r)$.

Let $D:= \Lied_L$ and $\Du :=\Lied_\Lb$ denote the projections of the Lie derivatives onto $S_{u,v}$.
\subsection{Characteristic initial data and null structure equations} \label{SECcharDATA} In the \emph{characteristic problem} for the Einstein equations (see, for example, \cite{Rendall,LukChar,LukRod1}), initial data is posed on two transversely-intersecting null hypersurfaces. The embedding of the null hypersurfaces into a Lorentzian $4$-manifold $(\MM,\g)$ and the Einstein equations stipulate constraints on the characteristic initial data, the so-called \emph{null constraint equations}. Solutions to the null constraint equations are parametrized by freely specifiable \emph{characteristic seed} defined as follows, see also Section 1 in \cite{ChrFormationBlackHoles}. Let $\HH_0$ be the outgoing null hypersurface emanating from a spacelike sphere $S_{0,1}$. Consider prescribed on $S_{0,1}$ a positive-definite symmetric $2$-tensor $\gd$, scalar functions $\trchi, \trchib, \omb, \Du\omb$, a vectorfield $\eta$, and $\gd$-tracefree symmetric $2$-tensors $\chibh$ and $\ab$. On $\HH_0$ consider prescribed the so-called \emph{characteristic seed} $(\Om,\gd_c)$, with $\gd_c$ assumed to be compatible with the prescribed $\gd$ on $S_{0,1}$.

Based on the \emph{nilpotent character} of the null constraint equations, we derive below the hierarchy of null transport equations which can be integrated using only the \emph{characteristic seed} and previously-solved-for quantities along the null hypersurface to construct the corresponding solution to the null constraint equations.

\ni \textbf{(1)} The \emph{first variation equation} and the \emph{Raychauduri equation}  imply the following \emph{linear} null transport equation for $\phi$ along $\HH_0$,
\begin{align*}
\begin{aligned}
D \gd = 2 \Om \chi, \,\, D \trchi + \frac{\Om}{2} (\trchi)^2 - \om \trchi = - \Om \vert \chih \vert^2 \,\, \Rightarrow \,\, D^2\phi -\om \Om\trchi \phi + \frac{1}{2} \Om^2\vert \chih \vert^2 \phi =0,
\end{aligned} 
\end{align*}
where $\vert \chih \vert^2_{\gd}:= \gd^{AB}\gd^{CD} \chih_{AC}\chih_{BD}$ is conformally invariant and can thus be calculated beforehand from $\gd_c$ on $\HH_0$. Integrating the equation for $\phi$ with initial data on $S_{0,1}$, the metric $\gd$, the second null fundamental form $\chi$, and the Gauss curvature $K$ of $(S_{u,v},\gd)$ are determined on $\HH_0$.

\ni \textbf{(2)} The Gauss--Codazzi equation for $\beta$ and the transport equation for $\eta$ along $\HH_0$,
\begin{align*} 
\begin{aligned} 
- \beta=\Divd \chih -\half \di \tr \chi + \chih \cdot \zeta - \half \trchi \zeta, \,\, D\eta = \Om (\chi \cdot \etab - \be),
\end{aligned} 
\end{align*}
with $(\Divd \chih)_A := \Nd^B \chih_{BA}$ and $(\chi \cdot \etab)_A:= \chih_{AB}\etab^B$, which imply that $\eta$ satisfies on $\HH_0$,
\begin{align*} 
\begin{aligned} 
D\eta + \Om \trchi \eta - \Om \lrpar{\Divd \chih - \half\di \trchi+\chih \di \log \Om + \frac{3}{2}\trchi \di \log \Om} =0.
\end{aligned} 
\end{align*}
\textbf{(3)} The \emph{Gauss equation} and the transport equation for $\Om\trchib$ along $\HH_0$,
\begin{align*} 
\begin{aligned} 
K + \frac{1}{4} \tr \chi \tr \chib - \half (\chih,\chibh) = - \rh, \,\, D (\Om \trchib) = 2 \Om^2 \Divd \etab + 2 \Om^2 \vert \etab \vert^2 - \Om^2 (\chih, \chibh) - \half \Om^2 \trchi \trchib + 2 \Om^2 \rh,
\end{aligned} 
\end{align*}
which implies that $\Om\trchib$ satisfies the following transport equation along $\HH_0$,
\begin{align*} 
\begin{aligned} 
D(\Om \trchib) + \Om\trchi (\Om\trchib) +2\Om^2\Divd (\eta-2\di \log\Om) - 2 \Om^2 \vert \eta-2\di \log\Om \vert^2 +2 \Om^2 K =0.
\end{aligned} 
\end{align*}
\textbf{(4)} The null transport equation for $\Om\chibh$ along $\HH_0$:
\begin{align*} 
\begin{aligned} 
D\lrpar{\Om \chibh} - (\Om \chih, \Om \chibh) \gd - \half \Om \trchi \Om \chibh = \Om^2 \lrpar{\Nd \widehat{\otimes}(2\di \log\Om-\eta) + (2\di \log\Om-\eta) \widehat{\otimes} (2\di \log\Om-\eta) - \half \trchib \chih},
\end{aligned} 
\end{align*}
where $(\Nd\widehat{\otimes}Y)_{AB} := \Nd_AY_B +\Nd_BY_A-(\Divd Y) \gd_{AB}$ and $(X\widehat{\otimes}Y)_{AB} := X_AY_B +X_BY_A-\gd(X,Y) \gd_{AB}$.

\ni \textbf{(5)} The null transport equation for $\omb$ along $\HHb$,
\begin{align*} 
\begin{aligned} 
D \omb =& \Om^2(2 (\eta, \etab) - \vert \eta \vert^2 -\rh),
\end{aligned} 
\end{align*}
which implies, using the Gauss equation, the following null transport equation for $\omb$ along $\HH_0$,
\begin{align*} 
\begin{aligned} 
D\omb - \Om^2 \lrpar{4(\eta, \di \log \Om)- 3 \vert \eta \vert^2 + K +\frac{1}{4} \trchi \trchib -\half (\chih, \chibh)}=0.
\end{aligned} 
\end{align*}
In Appendix B of \cite{ACR1} we moreover derive a null transport equation for $\Du\omb$ along $\HH_0$.

\ni \textbf{(6)} We have the \emph{null Bianchi equation} for $\ab$, and the \emph{curl equation} for $\eta$,
\begin{align*} 
\begin{aligned} 
\widehat{D} \aa - \half \Om \tr \chi \aa + 2 \om \aa + \Om \left( \Nd \widehat{\otimes} \beb + (4 \etab - \zeta) \widehat{\otimes} \beb + 3 \chibh \rh -3 {}^\ast \chibh \si \right) =0, \,\, \Curld \eta= - \half \chih \wedge \chibh - \si,
\end{aligned} 
\end{align*}
where $\widehat{D}\ab$ is the $\gd$-tracefree part of $D\ab$, ${}^\ast \chibh$ the left Hodge dual of $\chih$, and $\Curld \eta := \in^{AB}\Nd_A \eta_B$, which imply, together with the Gauss and Gauss--Codazzi equation, the following equation for $\ab$ on $\HH_0$,
\begin{align*} 
\begin{aligned} 
\widehat{D}\ab - \half \Om \trchi \ab + 2 \om \ab + \Om \Nd \widehat{\otimes} \lrpar{\Divd \chibh - \half \di \trchib - \chibh \cdot (\eta-\di\log\Om) + \half \trchib (\eta-\di \log\Om)}&\\
+ \Om \lrpar{9\di \log \Om -5 \eta}\widehat{\otimes} \lrpar{\Divd \chibh - \half \di \trchib - \chibh \cdot (\eta-\di\log\Om) + \half \trchib (\eta-\di \log\Om)}&\\
-3\Om \chibh \lrpar{K + \frac{1}{4} \trchi \trchib - \half (\chih,\chibh)} + 3\Om {}^*\chibh \lrpar{\Curld \eta + \half \chih \wedge \chibh}&=0,
\end{aligned} 
\end{align*}
\ni \textbf{(7)} The \emph{second variation equation} determines $\a$ along $\HH_0$ as follows,
\begin{align*} 
\begin{aligned} 
\Om \a + D\chih-\Om \vert \chih \vert^2 \gd - \om \chih=0.
\end{aligned} 
\end{align*}
\subsection{Sphere data and charges} \label{SECspheredataCharges} The sphere data (corresponding to the $C^2$-gluing problem) on a sphere $S_{u,v}$ is given by the following tuple of tensors
\begin{align} 
\begin{aligned} 
x_{u,v} = (\Om,\phi, \gd_c, \Om\trchi, \chih, \Om\trchib, \chibh, \eta, \om, D\om, \omb, \Du\omb, \a, \ab).
\end{aligned} \label{EQdefSPHEREDATA001}
\end{align}
In the $C^2$-gluing problem, the set of constraint equations the solution has to satisfy is given by the null structure equations stated in Section \ref{SECcharDATA}. By the null structure equations, sphere data describes all derivatives of the spacetime metric \eqref{EQspacetimemetricDOUBLENULL} up to order $2$. Sphere data is generally \emph{gauge-dependent}, which plays an important role later.

For the $C^{m+2}$-gluing problem (for integers $m\geq0$) the corresponding higher-order sphere data on $S_{u,v}$ is defined to be the tuple of tensors $(x_{u,v},\DD^{L,m}_{u,v}, \DD^{\Lb,m}_{u,v})$ with 
\begin{align*} 
\begin{aligned} 
\DD^{L,m}_{u,v} = \lrpar{\widehat{D}\a, \dots, \widehat{D}^m\a,D^2\om, \dots, D^{m+1}\om}, \,\, \DD^{\Lb,m}_{u,v} = \lrpar{\widehat{\Du}\ab, \dots, \widehat{\Du}^m\ab,\Du^2\omb, \dots, \Du^{m+1}\omb}.
\end{aligned} 
\end{align*}
where for $S_{u,v}$-tangential tensors $T$, $\widehat{D}T$ and $\widehat{\Du}T$ denote the $\gd$-tracefree parts of $DT$ and $\Du T$ on $S_{u,v}$, respectively. In the $C^{m+2}$-gluing problem, the solution has to satisfy \emph{higher-order} null structure equations including propagation equations for the tensors in $\DD^{L,m}_{u,v}$ and $\DD^{\Lb,m}_{u,v}$. By the higher-order null structure equations, the higher-order sphere data $(x_{u,v},\DD^{L,m}_{u,v}, \DD^{\Lb,m}_{u,v})$ determines all derivatives of the spacetime metric \eqref{EQspacetimemetricDOUBLENULL} up to order $m+2$.

The reference sphere data for Schwarzschild of mass $M\geq0$ (for $v-u>0$ large) is defined by
\begin{align*} 
\begin{aligned} 
\mathfrak{m}^M_{u,v} = \lrpar{\Om_M,r_M, \gac, \frac{2\Om_M}{r_M}, 0, \frac{2\Om_M}{r_M}, 0, 0, \frac{M}{r_M^2}, -\frac{2M\Om_M^2}{r_M^3}, -\frac{M}{r_M^2}, -\frac{2M\Om_M^2}{r_M^3}, 0, 0},
\end{aligned} 
\end{align*}
where $r_M$ is implicitly defined by $r_M(u,v)=(v-u)-2M \log(\frac{r_M(u,v)}{2M}-1)$, and $\Om_M := (1-\frac{2M}{r_M})^{1/2}$.

We associate the following ten \emph{charges} to given sphere data $x_{u,v}$. For $m=-1,0,1$, let
\begin{align*} 
\begin{aligned} 
\mathbf{E} :=& -\frac{1}{8\pi} \sqrt{4\pi} \lrpar{r^3 \lrpar{ \rho + r \Divd {\be}}}^{(0)}, &
\mathbf{P}^m :=& -\frac{1}{8\pi} \sqrt{\frac{4\pi}{3}} \lrpar{r^3 \lrpar{\rho + r \Divd {\be}}}^{(1m)},\\
\mathbf{L}^m :=& \frac{1}{16\pi} \sqrt{\frac{8\pi}{3}} \lrpar{r^3 \lrpar{ \di \trchi + \trchi (\eta-\di\log\Om) }}_H^{(1m)}, &
\mathbf{G}^m :=&  \frac{1}{16\pi}\sqrt{\frac{8\pi}{3}} \lrpar{r^3 \lrpar{ \di \trchi + \trchi (\eta-\di\log\Om) }}^{(1m)}_E,
\end{aligned} 
\end{align*}
where $r=r(x_{u,v})$ denotes the area radius of $(S_{u,v},\gd)$ and the standard (vector) spherical harmonics projections are defined with respect to the metric $\gac$ on $S_{u,v}$.

The linearizations of the above charges in $x_{u,v}$ at Minkowski spacetime are explicitly given by
\begin{align*} 
\begin{aligned} 
\dot{\mathbf{E}} :=& \frac{\sqrt{4\pi}}{16\pi} \lrpar{r^2 \omtrchibd- r^2\cdot\lrpar{\omtrchid- \frac{4}{r} \Omd} -4\phid}^{(0)}, \\
\dot{\mathbf{P}}^m :=& \frac{\sqrt{4\pi/3}}{16\pi} \lrpar{r^2 \omtrchibd- \frac{2}{r} \Divdo \lrpar{ r^2 \etad +\frac{r^3}{2} \di \lrpar{\omtrchid-\frac{4}{r}\Omd}}-r^2 \lrpar{\omtrchid-\frac{4}{r} \Omd} }^{(1m)} \\
\dot{\mathbf{L}}^m :=& \frac{\sqrt{8\pi/3}}{8\pi} \lrpar{r^2 \etad + \frac{r^3}{2} \di \lrpar{\omtrchid- \frac{4}{r} \Omd}}^{(1m)}_H, \,\, \dot{\mathbf{G}}^m := \frac{\sqrt{8\pi/3}}{8\pi}\lrpar{r^2 \etad + \frac{r^3}{2} \di \lrpar{\omtrchid- \frac{4}{r} \Omd}}^{(1m)}_E.
\end{aligned}
\end{align*}

\subsection{Scaling and norms} \label{SECnorms} In Theorem \ref{THMmain1}, we use the following norm for sphere data $x_{u,v}$,
\begin{align*} 
\begin{aligned} 
\Vert x_{u,v} \Vert_{\XX(S_{u,v})} =& \Vert \Om \Vert_{{H}^{6}(S_{u,v})}+(v-u)^{-2} \Vert \gd \Vert_{{H}^{6}(S_{u,v})} + \Vert \eta \Vert_{H^{5}(S_{u,v})}+ (v-u)\Vert \trchi \Vert_{H^{6}(S_{u,v})} \\
&+ (v-u)^{-1}\Vert \chih \Vert_{H^{6}(S_{u,v})} + (v-u)\Vert \trchib \Vert_{H^{4}(S_{u,v})} +(v-u)^{-1} \Vert \chibh \Vert_{H^{4}(S_{u,v})} \\
&+ (v-u)\Vert \om \Vert_{H^{6}(S_{u,v})}+(v-u)^2\Vert D\om \Vert_{H^{6}(S_{u,v})}+(v-u)\Vert \omb \Vert_{H^{4}(S_{u,v})} \\
&+(v-u)^2 \Vert \Du\omb \Vert_{H^{2}(S_{u,v})} + \Vert \a \Vert_{H^{6}(S_{u,v})}+ (v-u) \Vert \be \Vert_{H^{5}(S_{u,v})}+(v-u)^2\Vert \rh \Vert_{H^{4}(S_{u,v})} \\
&+(v-u)^2 \Vert \si \Vert_{H^{4}(S_{u,v})} + (v-u) \Vert \beb \Vert_{H^{3}(S_{u,v})}+\Vert \ab \Vert_{H^{2}(S_{u,v})},
\end{aligned} 
\end{align*}
where for an $k$-tensor $T$ on $S_{u,v}$, and an integer $m\geq0$, we denote
\begin{align*} 
\begin{aligned} 
\Vert T \Vert^2_{H^m(S_{u,v})} := \sum\limits_{0\leq i \leq m} (v-u)^{2(i+k-1)} \Vert \Nd^i T \Vert^2_{L^2(S_{u,v})}, 
\end{aligned} 
\end{align*}

\ni We bound the constructed solution $x$ to the null constraint equations on $\HH_{u,R\cdot[v_1,v_2]}$ in the norm
\begin{align*} 
\begin{aligned}
\Vert x_R \Vert_{\XX(\HH_R)} :=& \Vert \Om \Vert_{H^6_3(\HH_R)} +\Vert \gd \Vert_{H^6_3(\HH_R)}+ \Vert \eta \Vert_{H^5_2(\HH_R)}\\
&+ R\Vert \trchi \Vert_{H^6_3(\HH_R)}+ R^{-1}\Vert \chih \Vert_{H^6_2(\HH_R)} + R \Vert \trchib \Vert_{H^4_2(\HH_R)}+ R^{-1} \Vert \chibh \Vert_{H^4_3(\HH_R)}\\
&+R\Vert \om \Vert_{H^6_2(\HH_R)} +R^2 \Vert D\om \Vert_{H^6_1(\HH_R)}+R \Vert \omb \Vert_{H^4_3(\HH_R)}+R^2 \Vert \Du\omb \Vert_{H^2_3(\HH_R)}\\
&+  \Vert \a \Vert_{H^{6}_1(\HH_R)} + R \Vert \be \Vert_{H^{5}_2(\HH_R)}+R^2\Vert \rh \Vert_{H^{4}_2(\HH_R)} +R^2 \Vert \si \Vert_{H^{4}_2(\HH_R)} + R \Vert \beb \Vert_{H^{3}_2(\HH_R)}+\Vert \ab \Vert_{H^{2}_3(\HH_R)}, 
\end{aligned} 
\end{align*}
where for reals $u_0<v_1<v_2$, integers $m\geq0$, $l\geq0$, and $S_{u_0,v}$-tangential tensors $T$ on $\HH_{u,R\cdot [v_1,v_2]}$,
\begin{align*} 
\begin{aligned} 
\Vert T \Vert^2_{H^m_l\lrpar{\HH_{u,R\cdot[v_1,v_2]}}} :=  \int\limits_{R\cdot v_1}^{R\cdot v_2} \,\, \sum\limits_{0\leq i\leq l} (v-u_0)^{2i-1} \left\Vert D^i T \right\Vert^2_{H^m(S_{u_0,v})} dv.
\end{aligned} 
\end{align*}
Here the Lie derivative $D$ is with respect to the reference Minkowski metric on $\HH_{u,[v_1,v_2]}$. 

Higher-regularity norms of (higher-order) sphere data and (higher-order) solutions are straight-forward generalizations of the above. Specifically, the following higher-regularity norm is used to bound characteristic initial data on ingoing null hypersurfaces $\HHb:= \HHb_{R\cdot[-\de,\de],2R}$ (and similarly, $\HHb=\HHb_{R\cdot[-\de,\de],R}$), where $R\geq1$ is a real number. This norm is used in Theorem \ref{THMmain1} with $R=1$. In particular, the gauge transformations of Section \ref{SECspherePERT} are well-defined mappings with this norm.
\begin{align*} 
\begin{aligned} 
\Vert x \Vert_{\XX^+(\HHb)} :=& \Vert \Om \Vert_{H^{12}_9(\HHb)} +\Vert \gd \Vert_{H^{12}_9(\HHb)}+ \Vert \etab \Vert_{H^{11}_8(\HHb)} \\
&+R \Vert \trchib \Vert_{H^{12}_9(\HHb)}+R^{-1} \Vert \chibh \Vert_{H^{12}_8(\HHb)} + R \Vert \trchi \Vert_{H^{10}_8(\HHb)}+ R^{-1} \Vert \chih \Vert_{H^{10}_9(\HHb)}\\
&+R \Vert \omb \Vert_{H^{12}_8(\HHb)} + R^2 \Vert \Du\omb \Vert_{H^{12}_7(\HHb)}+ R\Vert \om \Vert_{H^{10}_9(\HHb)}+ R^2\Vert D\om \Vert_{H^8_9(\HHb)}\\
&+  \Vert \ab \Vert_{H^{12}_7(\HHb)} + R \Vert \beb \Vert_{H^{11}_8(\HH_R)}+R^2\Vert \si \Vert_{H^{10}_8(\HH_R)} +R^2 \Vert \rh \Vert_{H^{10}_8(\HH_R)} + R \Vert \be \Vert_{H^{9}_8(\HH_R)} +\Vert \a \Vert_{H^{8}_9(\HHb)}.
\end{aligned} 
\end{align*}

\ni Importantly, all norms are \emph{invariant} under the standard scaling of the Einstein equations, that is, $\mathbf{g}_{\mu\nu}(x) \mapsto \mathbf{g}_{\mu\nu}(R\cdot x)$. This allows us to turn asymptotic flatness conditions into small data conditions by rescaling, see Section \ref{SECcharGluingKerr}.

The charges $(\mathbf{E},\mathbf{P},\mathbf{L},\mathbf{G})$ change under the rescaling from $x_R$ on $S_R$ to ${}^{(R)}x_1$ on $S_1$ as follows,
\begin{align} 
\begin{aligned} 
(\mathbf{E},\mathbf{P})({}^{(R)}x_1) =& R^{-1}\cdot (\mathbf{E},\mathbf{P})(x_R), \,\,(\mathbf{L},\mathbf{G})({}^{(R)}x_1) = R^{-2}\cdot (\mathbf{L},\mathbf{G})(x_R).
\end{aligned} \label{EQscalingCHARGES}
\end{align}

\ni A family $(x_{-R,R})_{R\geq1}$ of sphere data is \emph{strongly asymptotically flat} if, for some real number $M\geq0$,
\begin{align} 
\begin{aligned} 
\Vert x_{-R,R} - \mathfrak{m}^M \Vert_{\XX(S_{-R,R})} = \OO(R^{-3/2}).
\end{aligned} \label{EQdefSTRONGAFfamilyspheredata}
\end{align}
Let $\de>0$ be a real number. A family $(x_{-R\cdot[-\de,\de],R})_{R\geq1}$ of solutions to the ingoing null structure equations is \emph{strongly asymptotically flat} if, for some real number $M\geq0$,
\begin{align} 
\begin{aligned} 
\Vert x_{-R\cdot[-\de,\de],R} - \mathfrak{m}^M \Vert_{\XX^+(\HHb_{-R\cdot[-\de,\de],R})} = \OO(R^{-3/2}).
\end{aligned} \label{EQdefSTRONGAFfamilyingoingdata}
\end{align}
The above rates are consistent with decay towards spacelike infinity in strongly asymptotically flat spacetimes. 
\section{Codimension-$10$ characteristic gluing}\label{SECcharGluing} \ni In this section we prove Theorem \ref{THMmain1}. We outline the strategy of the proof in Section \ref{SECoverviewProofCharGluing1}, and provide background and the full proof in Sections \ref{SECspherePERT}, \ref{SECconslaws1}, \ref{SECrepformulas}, \ref{SEClinFreedata1} and \ref{SECestimates1}. In Section \ref{SECbifurcateGluing} we prove Theorem \ref{THMbifurcate}. Detailed calculations and explicit derivations of estimates can be found in \cite{ACR1}. 

\subsection{Strategy of the proof} \label{SECoverviewProofCharGluing1} By the implicit function theorem and the fact that surjectivity is an open property, the proof of Theorem \ref{THMmain1} reduces to solving the \emph{linearized characteristic gluing problem at Minkowski}. The linearized null constraint equations at Minkowski admit infinitely-many \emph{conservation laws} which act as obstructions to characteristic gluing, see Section \ref{SECconslaws1}. By applying sphere perturbations and sphere diffeomorphisms to the sphere data on $S_2$, introduced in Section \ref{SECspherePERT}, we can manually match all but $10$ \emph{gauge-invariant charges}, denoted by $(\dot{\mathbf{E}},\dot{\mathbf{P}},\dot{\mathbf{L}},\dot{\mathbf{G}})$. The remaining quantities which do not satisfy conservation laws can be matched by an appropriate choice of the linearized free data along the gluing null hypersurface, see Section \ref{SEClinFreedata1}. This solves the linearized characteristic gluing problem up to the $10$-dimensional space of gauge-invariant charges. In Section \ref{SECestimates1} we prove perturbation and transport estimates for the gauge-invariant charges which are used in the characteristic gluing to Kerr in Section \ref{SECcharGluingKerr}.

\subsection{Sphere perturbations and diffeomorphisms} \label{SECspherePERT} Sphere data is by definition gauge-dependent. In this paper we apply two types of gauge perturbations to sphere data. 

First, we define \emph{sphere perturbations} along the ingoing null direction as follows. Let $(u,v,\th^1,\th^2)$ be a local double null coordinate system around a sphere $S_{u_0,v_0}$. For a small scalar function $f$ on $\HHb_{v_0}$, consider on $\HHb_{v_0}$ the new coordinates $(u',\th'^1,\th'^2):= (u+f(u,\th^1,\th^2),\th^1,\th^2)$. Let $(u',v',\th'^1,\th')$ be the local coordinate system constructed from $u'$ on $\HHb_2$, $v$ on $\HH_{u_0}$, $(\th'^1,\th'^2)$ on $S_{u_0,v_0}$. We denote the perturbed sphere of $S_{u_0,v_0}$ in $\HHb_{v_0}$ by $S'_{u_0,v_0}:= \{ u'=u_0, v'=v_0\}$. 

The sphere data $x'_{u_0,v_0}$ on $S'_{u_0,v_0}$ can be explicitly expressed in terms of $f$ and the characteristic initial data along $\HHb_{v_0}$. For example, we can express the null lapse $\Om'$ and the metric components $\gd'_{AB}$ of $x'_{u_0,v_0}$ as follows,
\begin{align*} 
\begin{aligned} 
\Om'^2(\th^1,\th^2) =& \lrpar{1+\pr_u f(u_0,\th^1,\th^2)} \cdot {\Om}^2(u_0+f(u_0,\th^1,\th^2),\th^1,\th^2),\\
\gd'_{AB}(\th^1,\th^2)=& {\gd}_{AB}\lrpar{u_0+f(u_0,\th^1,\th^2),\th^1,\th^2}.
\end{aligned} 
\end{align*} 
From the point of view of $f$, it holds that $x'_{u_0,v_0}$ depends only on the functions $f, \pr_u f, \pr^2_u f$ and $\pr^3_u f$ restricted to $S_{u_0,v_0}$. For the perturbation function $f$, we introduce the norm
\begin{align*} 
\begin{aligned} 
\Vert f \Vert_{\YY_f} := \Vert f \Vert_{H^8(S_{u_0,v_0})} + \Vert \pr_u f \Vert_{H^6(S_{u_0,v_0})}+\Vert \pr_u^2 f \Vert_{H^4(S_{u_0,v_0})}+\Vert \pr_u^3 f \Vert_{H^2(S_{u_0,v_0})}.
\end{aligned} 
\end{align*}
The explicit formulas for $x'_{u_0,v_0}$ yield that sphere perturbations smoothly map characteristic initial data in $\XX(\HHb_{u_0+[-\de,\de],v_0})$ and perturbation functions $f$ on $S_{u_0,v_0}$ in $\YY_f$ into sphere data $x'_{u_0,v_0}$ on $S'_{u_0,v_0}$ in $\XX(S'_{u_0,v_0})$. While sphere perturbations are ``physical" changes of the sphere and thus not intrinsic gauge-transformations of the sphere, they correspond to gauge-transformations of the surrounding spacetime, see the \emph{linearized pure gauge solutions} in \cite{DHR}.

Second, we define \emph{sphere diffeomorphisms} as follows. Given sphere data $x_{u,v}$ and coordinates $(\th^1,\th^2)$ on a sphere $S_{u,v}$, we define for a pair of scalar functions $(j^1,j^2)$ on $S_{u,v}$ the new coordinates $(\th'^1,\th'^2):= (\th^1+j^1(\th^1,\th^2),\th^2+j^2(\th^1,\th^2)$. The sphere data $x'_{u,v}$ on $S_{u,v}$ can be explicitly expressed in terms of $(j^1,j^2)$ and the sphere data $x_{u,v}$ by using the tensor transformation law. For example, the components of the induced metric $\gd$ change according to
\begin{align*} 
\begin{aligned} 
\begin{pmatrix}
\gd_{11} & \gd_{12}  \\
\gd_{12} & \gd_{22} 
\end{pmatrix}= 
\begin{pmatrix}
1+ \pr_1 j^1 & \pr_1 j^2  \\
\pr_2 j^1 & 1+ \pr_2 j^2 
\end{pmatrix}\begin{pmatrix}
\tilde{\gd}_{11} & \tilde{\gd}_{12}  \\
\tilde{\gd}_{12} & \tilde{\gd}_{22} 
\end{pmatrix}\begin{pmatrix}
1+ \pr_1 j^1 & \pr_2 j^1  \\
\pr_1 j^2 & 1+ \pr_2 j^2 
\end{pmatrix}.
\end{aligned} 
\end{align*}
For the perturbation functions $(j^1,j^2)$ we define the norm
\begin{align*} 
\begin{aligned} 
\Vert (j^1,j^2) \Vert_{\mathcal{Y}_{(j^1,j^2)}} :=& \Vert j_1 \Vert_{H^{7}(S_{u,v})}+ \Vert j_2 \Vert_{H^{7}(S_{u,v})}.
\end{aligned} 
\end{align*}
It is straight-forward to verify that the sphere diffeomorphisms smoothly map sphere data $x_{u,v}$ in $\XX(S_{u,v})$ and perturbation functions $(j^1,j^2)$ bounded in $\mathcal{Y}_{(j^1,j^2)}$ to sphere data $x'_{u,v}$ in $\XX(S_{u,v})$. Sphere diffeomorphisms change sphere data but leave the sphere invariant, and are thus gauge-transformations intrinsic to the sphere.

At the linear level, sphere perturbations and sphere diffeomorphisms are special cases of the \emph{linearized pure gauge solutions} in Section 6 of \cite{DHR}.
\subsection{Conserved charges for the linearized equations at Minkowski} \label{SECconslaws1} The following charges $\QQ_i$, $0\leq i \leq 7$ are conserved under the linearized null constraint equations at Minkowski, and thus cannot be glued by using the linearized characteristic seed.
\begin{align} 
\begin{aligned} 
\QQ_0 :=& r^2 \etad^{[1]} + \frac{r^3}{2}\di \lrpar{\omtrchid^{[1]}-\frac{4}{r}\Omd^{[1]}}, \\
\QQ_1 :=&\frac{r}{2} \lrpar{\omtrchid-\frac{4}{r}\Omd} + \frac{\phid}{r}, \\
\QQ_2 :=& r^2 \omtrchibd -\frac{2}{r}\Divdo \lrpar{r^2\etad+\frac{r^3}{2}\di\lrpar{\omtrchid-\frac{4}{r}\Omd}} -r^2 \lrpar{\omtrchid-\frac{4}{r}\Omd} +2r^3 \Kd, \\
\QQ_3:=& \frac{\chibhd}{r} -\half \lrpar{ \DDd_2^\ast \Divdo +1} \gdcd + \DDd_2^\ast \lrpar{ \etad + \frac{r}{2}\di \lrpar{\omtrchid-\frac{4}{r}\Omd}} - r \DDd_2^\ast \di \lrpar{\omtrchid-\frac{4}{r}\Omd},
\end{aligned} \label{EQdefQcharges}
\end{align}
\begin{align*} 
\begin{aligned}
\QQ_4 :=& \frac{\abd_\psi}{r} + 2 \DDd_2^\ast \lrpar{\frac{1}{r^2} \Divdo \chibhd - \frac{1}{r} \etad- \half \di \omtrchid + \DDd_1^\ast \lrpar{\ombd,0}}_{\psi},\\
\QQ_5 :=& \ombd^{[\leq1]} +\frac{1}{4r^2}\QQ_2^{[\leq1]} +\frac{1}{3r^3} \Divdo\QQ_0, \\
\QQ_6 :=& \Du\ombd^{[\leq1]} - \frac{1}{6r^3} (\Ldo-3) \QQ_2^{[\leq1]} +\frac{1}{r^4} \Divdo \QQ_0, \\
\QQ_7 :=& \Du\ombd^{[2]} +\frac{3}{2r^3} \QQ_2^{[2]}+ \frac{1}{2r^2} \Divdo \Divdo \QQ_3^{[2]} -\frac{12}{r^2} \QQ_1^{[2]} +{ \frac{3}{2r^2} \Divdo \lrpar{\eta+ \frac{r}{2}\di\lrpar{\omtrchid-\frac{4}{r}\Omd}} }^{[2]} - {\frac{3}{4r^2} \Divdo \Divdo \gdcd }^{[2]},
\end{aligned} 
\end{align*}
where $\Divdo$ and $\Ldo$ are the divergence operator and Laplace-Beltrami operator with respect to the round unit metric $\gac$ on $S_{u,v}$. Moreover, for scalar functions $f_1$ and $f_2$, we denote $\DD_1^{\ast}(f_1,f_2) := -\di f_1 + {}^{\ast}\di f_2$, and for a vectorfield $X$, $\DD_2^{\ast}(X)_{AB} := -\half \lrpar{\Nd_A X_B+\Nd_BX_A-(\Divd X) \gd_{AB}}$.

Here the (tensor) spherical harmonics are defined with respect to the round \emph{unit} metric $\gac$ on $S_{u,v}$.

It is crucial for our linearized characteristic gluing that -- except for a $10$-dimensional subspace of so-called \emph{gauge-invariant charges} -- all charges can be matched by applying linearized sphere perturbations and sphere diffeomorphisms to the linearized sphere data $\dot{x}_2$ on $S_2$. We correspondingly call these charges \emph{gauge-dependent}.

The gauge-invariant charges are given by $\QQ_0$ and $\QQ_2^{[\leq1]}$. This $10$-dimensional subspace is precisely spanned by the linearizations $(\dot{\mathbf{E}},\dot{\mathbf{P}},\dot{\mathbf{L}},\dot{\mathbf{G}})$ of the charges $({\mathbf{E}},{\mathbf{P}},{\mathbf{L}},{\mathbf{G}})$ defined in Section \ref{SECspheredataCharges}. Indeed, from Section \ref{SECspheredataCharges} and \eqref{EQdefQcharges} we have that, for $m=-1,0,1$,
\begin{align*} 
\begin{aligned} 
-\frac{8\pi}{\sqrt{4\pi}} \dot{\mathbf{E}} = -\frac{1}{2} \QQ_2^{(0)}, \,\, - \frac{8\pi}{\sqrt{\frac{4\pi}{3}}}\dot{\mathbf{P}}^m=& -\frac{1}{2} \QQ_2^{(1m)}, \,\,
\frac{16\pi}{\sqrt{\frac{8\pi}{3}}}\dot{\mathbf{L}}^m = 2 (\QQ_0)_H^{(1m)}, \,\, \frac{16\pi}{\sqrt{\frac{8\pi}{3}}}\dot{\mathbf{G}}^m = 2 (\QQ_0)_E^{(1m)}.
\end{aligned} 
\end{align*}

\subsection{Representation formulas for linearized quantities at Minkowski} \label{SECrepformulas} By integrating the linearized null structure equations at Minkowski, we get representation formulas for linearized quantities along $\HH$. In the following we state these representation formulas for simplicity in the setting of the \emph{homogeneous} linearized null constraint equations; see \cite{ACR1} for the setting of the inhomogeneous linearized equations (necessary for the implicit function theorem).

\ni \textbf{Formulas for $\phid$ and $\gdcd$.} It holds that
\begin{align} 
\begin{aligned} 
\phid(r) = r \phid(1) + 2 \int\limits_1^r \Omd dr' + \frac{r-1}{2} \lrpar{\omtrchid(1)-4\Omd(1)}, \,\, \gdcd(r) = \gdcd(1)+ 2\int\limits_1^r \frac{1}{r'^2} \chihd dr'.
\end{aligned} \label{EQrep1}
\end{align}

\ni \textbf{Formula for $\etad$.} It holds that
\begin{align} 
\begin{aligned} 
\left[ r'^2 \etad + \frac{r'^3}{2} \di\lrpar{\omtrchid - \frac{4}{r'} \Omd}\right]_1^r = \Divdo\lrpar{ \int\limits_1^r \chihd dr' }.
\end{aligned} \label{EQrep2}
\end{align}

\ni \textbf{Formula for $\ombd$.} It holds that
\begin{align} 
\begin{aligned} 
\left[ \ombd +\frac{1}{4r'^2}\QQ_2 +\frac{1}{3r'} \Divdo\lrpar{\etad+ \frac{r'}{2}\di\lrpar{\omtrchid-\frac{4}{r'}\Omd}} \right]_1^r=  \frac{1}{3} \Divdo \Divdo \lrpar{\int\limits_1^r \frac{1}{r'^3} \chihd dr'}.
\end{aligned} \label{EQrep3}
\end{align}

\ni \textbf{Formula for $\abd$.} It holds that
\begin{align} \begin{aligned}
&\left[ \frac{\abd}{r'} +\frac{2}{r'}\DDd_2^\ast \Divdo \QQ_3 - \frac{1}{2r'^2} \DDd_2^\ast \di \QQ_2 - \frac{2}{r'} \DDd_2^\ast \di \lrpar{\Ldo+2}\QQ_1 \right]^r_1 \\
&-\left[ \frac{2}{3r'} \DDd_2^\ast \lrpar{\Divdo \DDd_2^\ast + 1 + \di\Divdo} \lrpar{\etad+ \frac{r'}{2}\di\lrpar{\omtrchid-\frac{4}{r'}\Omd}} \right]^r_1\\
&+\left[ \frac{1}{r'} \DDd_2^\ast \lrpar{\Divdo \DDd_2^\ast + 1 + \di\Divdo} \Divdo\gdcd \right]^r_1\\
=& \frac{4}{3} \DDd_2^\ast \lrpar{\Divdo \DDd_2^\ast + 1 + \di\Divdo} \Divdo \lrpar{\int\limits_1^r \frac{1}{r'^3}\chihd dr'}.
\end{aligned} \label{EQrep4} 
\end{align}

\ni \textbf{Formula for $\Du\ombd$.} It holds that
\begin{align} 
\begin{aligned} 
&\left[ \Du\ombd -\frac{1}{6r'^3} \lrpar{\Ldo-3}\QQ_2+ \frac{1}{2r'^2} \Divdo \Divdo \QQ_3 +\frac{1}{r'^2} \Divdo \Divdo \DDd_2^\ast \di \QQ_1\right]_1^r\\
&-\left[ \frac{1}{4r'^2} \Divdo \lrpar{\di \Divdo -2 + \Divdo\DD_2^\ast} \lrpar{\eta+ \frac{r'}{2}\di\lrpar{\omtrchid-\frac{4}{r'}\Omd}} +\frac{1}{8r'^2} \Divdo \di \Divdo \Divdo \gdcd \right]_1^r\\
=& \frac{1}{4} \Divdo \lrpar{2-\Divdo \DDd_2^\ast} \Divdo \lrpar{\int\limits_1^r \frac{1}{r'^4}\chihd dr'}.
\end{aligned} \label{EQrep5}
\end{align}

\subsection{Solution of the linearized characteristic gluing problem} \label{SEClinFreedata1} By adding linearized sphere perturbations and sphere diffeomorphisms to the linearized sphere data on $S_2$, we can match the gauge-dependent charges, see Section \ref{SECconslaws1}. It remains to use the linearized characteristic seed $(\Omd,\chihd)$ on $\HH$ for gluing the remaining sphere data quantities (up to the $10$ gauge-invariant charges), see \eqref{EQdefSPHEREDATA001}. By the representation formulas of Section \ref{SECrepformulas}, these matching conditions turn into integral conditions on $(\Omd,\chihd)$ over $\HH$. Crucially, due to a \emph{hierarchy of radial weights} in the representation formulas (see below) the integral conditions are \emph{linearly independent}. 

\ni \textbf{(1) Matching of $(\Omd,\chihd,\omd,D\omd,\ad)$.} The gluing of $\Omd$ and $\chihd$ (and their higher angular and $L$-derivatives, such as $\omd$ and $\ad=-D\chihd$) is by setup trivial.

\ni \textbf{(2) Matching of $\phid$ and $\gdcd$.} By representation formula \eqref{EQrep1} the gluing of $\phid$ and $\gdcd$ turns into integral conditions on $\Omd$ and $\frac{1}{r'^2} \chihd$ along $\HH$. 

\ni \textbf{(3) Matching of $\omtrchid$.} The gluing of $\omtrchid$ follows subsequently from the previous matching of the gauge-dependent charge $\QQ_1 := \frac{r}{2} \lrpar{\omtrchid-\frac{4}{r} \Omd} + \frac{\phid}{r}$ and the above gluing of $\Omd$ and $\phid$.

\ni \textbf{(4) Matching of $\etad^{[\geq2]}$.} For the gluing of $\etad$, we note that the mode $l=1$ of the right-hand side vanishes, and thus the projection of \eqref{EQrep2} onto the mode $l=1$ corresponds precisely to the conservation law for $\QQ_0$. Projecting onto the modes $l\geq2$, the gluing of $\etad^{[\geq2]}$ turns into integral conditions for $\chihd$, where we use that $\Divdo$ is an elliptic operator between modes $l\geq2$. By the previous gluings, the matching of $\etad^{[1]}$ is equivalent to the matching of $\QQ_0$.

\ni \textbf{(5) Matching of $\omtrchibd^{[\geq2]}$.} By Section \ref{SECconslaws1} the projection of $\QQ_2$ (see \eqref{EQdefQcharges}) onto modes $l\geq2$ is a gauge-dependent charge, and thus matched. Using moreover \eqref{EQdefQcharges} and the previous matching of $\Omd,\phid,\gdcd, \omtrchid$ and $\etad^{[\geq2]}$, it follows that $\omtrchibd^{[\geq2]}$ is matched. By the previous gluings, the matching of $\omtrchibd^{[\leq1]}$ depends only on the matching of $\etad^{[1]}$ and $\QQ_0^{[\leq2]}$.

\ni \textbf{(6) Matching of $\chibhd$.} By the previous matching of $\QQ_3$ (see \eqref{EQdefQcharges}) and $\Omd$, $\gdcd$, $\omtrchid$ and $\etad^{[\geq2]}$, the gluing of $\chibhd$ on $S_2$ follows.

\ni \textbf{(7) Matching of $\abd$.} Using the representation formula \eqref{EQrep4} for $\abd$, and that on the right-hand side of \eqref{EQrep4} the operator $\DDd_2^\ast \lrpar{\Divdo \DDd_2^\ast + 1 + \di\Divdo} \Divdo$ is elliptic, the gluing of $\abd$ turns into an integral condition for $\frac{1}{r^3}\chihd$ over $\HH$.

\ni \textbf{(8) Matching of $\ombd$.} On the one hand, $\ombd^{[\leq1]}$ is glued by the matching of $\QQ_5$ and the previous gluings. On the other hand, the matching of $\abd$ and $\QQ_4$ (see Section \ref{SECconslaws1}), the ellipticity of the operator $\DDd_2^\ast \DDd_1^\ast(\cdot,0)$ (see the formula for $\QQ_4$ in Section \ref{SECconslaws1}) on modes $l\geq2$, and the previous gluings, imply that $\ombd^{[\geq2]}$ is glued.

\ni \textbf{(9) Matching of $\Du\ombd$.} By the matching of $\QQ_6$ and $\QQ_7$, and the previous gluings, the gluings of $\Du\ombd^{[0]}$ and $\Du\ombd^{[2]}$ follow. The gluing of $\Du\ombd^{[1]}$ depends only on the gluing of $\QQ_0$. Projecting the representation formula \eqref{EQrep5} for $\Du\ombd$ onto modes $l\geq3$, and using that the right-hand side operator $\Divdo \lrpar{2-\Divdo \DDd_2^\ast} \Divdo$ is elliptic on modes $l\geq3$, we get with the previous that $\Du\ombd^{[\geq3]}$ is glued.

To summarize the above, by adding linearized sphere perturbations and sphere diffeomorphisms on $S_2$, and appropriately choosing the linearized characteristic seed $(\Omd,\chihd)$ on $\HH$, we glued all linearized sphere data quantities but $\etad^{[1]}$ and $\omtrchibd^{[\leq1]}$, and showed that, by construction, gluing of the latter is equivalent to gluing of $\QQ_0$ and $\QQ_2$, and thus, by Section \ref{SECconslaws1}, to $(\dot{\mathbf{E}},\dot{\mathbf{P}},\dot{\mathbf{L}},\dot{\mathbf{G}})$.

From the above representation formulas we can derive explicit regularity estimates for the constructed solution to the linearized null constraint equations which are consistent with the regularity hierarchy within the norm $\XX$ on null hypersurfaces defined in Section \ref{SECnorms}.

We thus solved the linearized characteristic gluing problem up to the $10$-dimensional space of gauge-invariant charges $(\dot{\mathbf{E}},\dot{\mathbf{P}},\dot{\mathbf{L}},\dot{\mathbf{G}})$.

\subsection{Perturbation and transport estimates for $(\mathbf{E},\mathbf{P},\mathbf{L},\mathbf{G})$} \label{SECestimates1} The linearizations $(\dot{\mathbf{E}},\dot{\mathbf{P}},\dot{\mathbf{L}},\dot{\mathbf{G}})$ at Minkowski of $(\mathbf{E},\mathbf{P},\mathbf{L},\mathbf{G})$ are conserved by the linearized null constraint equations and invariant under linearized sphere perturbations and diffeomorphisms. However, linearizing at Schwarzschild of small mass $M\geq0$, the linearizations are not conserved and invariant anymore, but can be shown to change at the order of $\OO(M)$.

Subsequently, we get that transported along $\HH$ under the null constraint equations,
\begin{align} 
\begin{aligned} 
(\mathbf{E},\mathbf{P},\mathbf{L},\mathbf{G})(x_{2}) = (\mathbf{E},\mathbf{P},\mathbf{L},\mathbf{G})(x_{1}) + \OO(M\cdot \varep) + \OO(\varep^2).
\end{aligned} \label{EQchargeTRANSPORT1}
\end{align}
and that under sphere perturbations and diffeomorphisms, with $\tilde{x}_2$ denoting the sphere data on the perturbed sphere $\tilde{S}_2$,
\begin{align} 
\begin{aligned} 
(\mathbf{E},\mathbf{P},\mathbf{L},\mathbf{G})(\tilde{x}_{2}) = (\mathbf{E},\mathbf{P},\mathbf{L},\mathbf{G})(x_{2}) + \OO(M\cdot \varep) + \OO(\varep^2).
\end{aligned} \label{EQchargePERTURB1}
\end{align}
These estimates are of crucial importance for our proof of characteristic gluing to Kerr, see \eqref{EQdeff}.

\subsection{Codimension-$10$ bifurcate characteristic gluing} \label{SECbifurcateGluing} In this section we prove Theorem  \ref{THMbifurcate}. By the linear analysis of Section \ref{SECcharGluing} we obtain two classes of conserved charges: the first class of charges $\underline{\QQ}_i$, $0\leq i \leq 7$, is conserved along the ingoing null hypersurface $\HHb$ and the second class of charges ${\QQ}_i$, $0\leq i \leq 7$, is conserved along the outgoing null hypersurface $\HH$. Each of these classes splits up into the gauge-dependent and the $10$-dimensional gauge-invariant charges as in Section \ref{SECconslaws1}. 

The gauge-invariant charges on $\HH$ are related to the gauge-invariant charges on $\HHb$ in the following sense: on a sphere $S_{u,v}$ (with $r=v-u$) we have
\begin{align*}
\begin{aligned} 
\lrpar{\QQ_0}_H = -\lrpar{\underline{\QQ}_0}_H, \,\, \lrpar{\QQ_0}_E = \lrpar{\underline{\QQ}_0}_E+\frac{r}{2} \di \underline{\QQ}_2^{[1]}, \,\, \QQ_2^{[0]} = - \underline{\QQ}_2^{[0]}, \,\, \QQ_2^{[1]} = \underline{\QQ}_2^{[1]}.
\end{aligned} 
\end{align*}
Hence the two classes of gauge-invariant charges on $S_{u,v}$ are dependent in the sense that the one class of gauge-invariant charges automatically determines the other class. 

On the other hand,  the following relations hold for the gauge-dependent charges $\QQ_i$ and $\underline{\QQ}_i$, $i=0,1,2$, on $S_{u,v}$:
\begin{align*} 
\begin{aligned} 
\QQ_1 -\frac{1}{2r} \underline{\QQ}_2 - \underline{\QQ}_1 =  -2 \Omd + \half \Divdo \Divdo \gdcd- \frac{1}{r^2} \Divdo \lrpar{r^2 \etabd - \frac{r^3}{2} \di \lrpar{\omtrchibd + \frac{4}{r}\Omd }}    - \frac{1}{r} \Ldo \phid, \\
-\frac{1}{4r} \lrpar{\QQ_2 + \underline{\QQ}_2} - \half \Ldo\lrpar{\QQ_1 - \underline{\QQ}_1} = \Ldo\Omd -\frac{1}{r^2} \Divdo \lrpar{r^2 \etabd - \frac{r^3}{2} \di \lrpar{\omtrchibd +\frac{4}{r} \Omd}} -\frac{1}{r} \Ldo \phid.
\end{aligned} 
\end{align*}
The crucial observation is that the right-hand side of the above identities are such that the gauge-dependent charges $\QQ_1$ and $\QQ_2$ can be glued along $\HHb$ without any obstructions. Moreover, the remaining gauge-dependent charges  $\QQ_i$, $i=3,...,7$,  contain $\Lb$-derivatives and hence can also be glued along $\HHb$. Analogous principles hold for the gauge-dependent charges $\underline{\QQ}_i$ that manifest their gluing along $\HH$.  By the null structure equations, the above principles extend to the higher-order charges conserved under the linearized higher-order null constraint equations. We conclude that the two classes of gauge-dependent charges on $S_{u,v}$ are mutually independent in the sense that for a given prescription of the two classes, there exists regular sphere data which realizes both of them.  

We can now finish the proof of Theorem  \ref{THMbifurcate} as follows: Given higher-order sphere data on spheres $S_1$ and $S_2$, we calculate on $S_1$ the higher-order gauge-dependent charges of $\HHb$, and $S_2$ the higher-order gauge-dependent charges of $\HH$. We equip the auxiliary sphere $S_{\mathrm{aux}}$ with sphere data which realizes both the gauge-dependent charges along $\HH$ and $\HHb$. By the analysis in the previous sections and the matching of gauge-dependent charges on $S_{\mathrm{aux}}$, we can glue the higher-order sphere data on $S_1$ to the sphere data on $S_{\mathrm{aux}}$ along $\HHb$ (up to the $10$ gauge-invariant charges), and subsequently glue the resulting sphere data on $S_{\mathrm{aux}}$ to the sphere data on $S_2$ (up to the $10$ gauge-invariant charges). The regularity of the constructed solution follows by the previous analysis along each of the null hypersurfaces $\HH$ and $\HHb$. 

\section{Characteristic gluing to Kerr}\label{SECcharGluingKerr} \noindent In this section we prove Theorem \ref{PROPmain2}. The proof of Theorem \ref{THMcharGluingTWOfirstintroversion199901} is similar and omitted. We provide a geometric interpretation of $(\mathbf{E},\mathbf{P},\mathbf{L},\mathbf{G})$ in Section \ref{SECgeomINTCH}, discuss Kerr spacelike initial data in Section \ref{SECKerrspacelike}, and conclude the proof in Section \ref{SECchargluingtoKERR}. Detailed calculations and explicit derivations of estimates can be found in \cite{ACR2}. 
\subsection{Geometric interpretation of the charges $(\mathbf{E},\mathbf{P},\mathbf{L},\mathbf{G})$} \label{SECgeomINTCH} Consider strongly asymptotically flat spacelike initial data $(\Si,g,k)$, that is, for a real number $M\geq0$, in Cartesian coordinates $(x^1,x^2,x^3)$ near spacelike infinity and with $e_{ij}$ being the Euclidean metric,
\begin{align} 
\begin{aligned} 
g_{ij}(x) = \lrpar{1+\frac{2M}{\vert x \vert}} \cdot e_{ij} + \OO\lrpar{\vert x \vert^{-3/2}}, \,\, k_{ij}(x) = \OO\lrpar{\vert x \vert^{-5/2}}.
\end{aligned} \label{EQstrongAFness}
\end{align}

\ni The spacelike initial data $(g,k)$ -- together with appropriate gauge choices -- determines on the coordinate spheres $S_{-R,R} \subset \Si$ (for large $R\geq1$) sphere data $x_{-R,R}$. The constructed family of sphere data $x_{-R,R}$ is strongly asymptotically flat (see \eqref{EQdefSTRONGAFfamilyspheredata}), and crucially, the charges $(\mathbf{E},\mathbf{P},\mathbf{L},\mathbf{G})$ of the sphere data $x_{-R,R}$ can be related to the ADM invariants of $(\Si,g,k)$ by
\begin{align} 
\begin{aligned} 
(\mathbf{E},\mathbf{P},\mathbf{L},\mathbf{G})\lrpar{x_{-R,R}} = \lrpar{\mathbf{E}_{\mathrm{ADM}},\mathbf{P}_{\mathrm{ADM}},\mathbf{L}_{\mathrm{ADM}},\mathbf{C}_{\mathrm{ADM}}} + \lrpar{\OO(R^{-1}), \OO(R^{-3/2}), \smallO\lrpar{1}, \smallO\lrpar{1}}.
\end{aligned} \label{EQconvergencerates1}
\end{align}

\begin{remark}[Interpretation of the charge $\mathbf{G}$] \label{REMG1} For $R\geq1$ large, the charge $\mathbf{G}\lrpar{x_{-R,R}}$ can be expanded as
\begin{align} 
\begin{aligned} 
\mathbf{G}\lrpar{x_{-R,R}} = \mathbf{C}_{\mathrm{ADM}}^{\mathrm{loc}}(S_{-R,R}) - \, \mathrm{area}(x_{-R,R}) \cdot \mathbf{P}_{\mathrm{ADM}}^{\mathrm{loc}}(S_{-R,R}) + \smallO\lrpar{R^{-1/2}},
\end{aligned} \label{EQGexpression1}
\end{align}
where $\mathbf{C}_{\mathrm{ADM}}^{\mathrm{loc}}$ and $\mathbf{P}_{\mathrm{ADM}}^{\mathrm{loc}}$ denote the local ADM integrals of \emph{center} and \emph{linear momentum}, and $\mathrm{area}(x_{-R,R})$ denotes the area radius of $x_{-R,R}$. Applying Stokes' theorem over the end of $\Si$ to the local integral $\mathbf{P}_{\mathrm{ADM}}^{\mathrm{loc}}$, it is standard to derive the (better than expected) decay rate $\mathbf{P}_{\mathrm{ADM}}^{\mathrm{loc}}(S_{-R,R})= \OO(R^{-3/2})$, so that the second term on the right-hand side of \eqref{EQGexpression1} is of order $\OO(R^{-1/2})$, yielding the expansion for $\mathbf{G}\lrpar{x_{-R,R}}$ in \eqref{EQconvergencerates1}. In \cite{ACR2} we more generally solve the problem of characteristic gluing of families of sphere data to Kerr under the weaker (expected) decay rate $\mathbf{P}_{\mathrm{ADM}}^{\mathrm{loc}}(S_{-R,R})= \smallO(R^{-1/2})$. \end{remark}

\subsection{Kerr spacelike initial data and sphere data} \label{SECKerrspacelike} In Appendix F in  \cite{ChruscielDelay} it is shown that for any vector $\lrpar{\mathbf{E}_{0}, \mathbf{P}_0,\mathbf{L}_0,\mathbf{C}_0} \in \RRR^{10}$ with $\mathbf{E}_0 > \vert \mathbf{P}_0 \vert$, there exists a spacelike hypersurface $\Si$ in the exterior region of a Kerr spacetime such that the induced spacelike initial data $(\Si,g,k)$ is \emph{asymptotically flat with Regge-Teitelbaum conditions}, that is, in coordinates $(x^1,x^2,x^3)$ near spacelike infinity,
\begin{align} 
\begin{aligned} 
g_{ij}(x) -e_{ij} =& \OO(\vert x \vert^{-1}), & g_{ij}(x) -g_{ij}(-x) =& \OO(\vert x \vert^{-2}), \\
k_{ij}(x) =& \OO(\vert x \vert^{-2}), & k_{ij}(x) +k_{ij}(-x) =& \OO(\vert x \vert^{-3}),
\end{aligned} \label{EQdecayRatesKerr}
\end{align}
and has ADM invariants
\begin{align*} 
\begin{aligned} 
(\mathbf{E}_{\mathrm{ADM}},\mathbf{P}_{\mathrm{ADM}},\mathbf{L}_{\mathrm{ADM}},\mathbf{C}_{\mathrm{ADM}}) = \lrpar{\mathbf{E}_{0}, \mathbf{P}_0,\mathbf{L}_0,\mathbf{C}_0}.
\end{aligned} 
\end{align*}
Moreover, for vectors $(\mathbf{E}_0,\mathbf{P}_0,\mathbf{L}_0,\mathbf{C}_0)\in \RRR^{10}$ close to a fixed $(\tilde{\mathbf{E}}_{0}, \tilde{\mathbf{P}}_0,\tilde{\mathbf{L}}_0,\tilde{\mathbf{C}}_0)$ with $\tilde{\mathbf{E}}_0>\vert \tilde{\mathbf{P}}_0\vert$, the constants in \eqref{EQdecayRatesKerr} can be \emph{uniformly bounded}; see, for example, \cite{ChruscielDelay,CorvinoSchoen}.

In the above Kerr spacelike initial data we can construct, as in Section \ref{SECgeomINTCH}, Kerr sphere data $x_{-R,2R}^{\mathrm{Kerr}}$. Let $M>0$ be a real number. For vectors $(\mathbf{E}_0,\mathbf{P}_0,\mathbf{L}_0,\mathbf{C}_0)$ satisfying, for some real numbers $\varep>0$ small and $R\geq1$ large,
\begin{align*} 
\begin{aligned} 
\lrpar{R^{1/2} \vert \mathbf{E}_0 - M \vert}^2 + \lrpar{R^{1/2} \vert \mathbf{P}_0 \vert}^2 + \vert \mathbf{L}_0 - \mathbf{L}_{\mathrm{ADM}} \vert^2 + \vert \mathbf{C}_0 - \mathbf{C}_{\mathrm{ADM}} \vert^2 \leq \varep,
\end{aligned} 
\end{align*}
one can show that the sphere data $x_{-R,2R}^{\mathrm{Kerr}}$ on $S_{-R,2R}$ satisfies
\begin{align} 
\begin{aligned} 
\Vert x_{-R,2R}^{\mathrm{Kerr}} - \mathfrak{m}^M \Vert_{\XX(S_{-R,2R})} = \OO(R^{-3/2}),
\end{aligned} \label{EQstrongCONVkerrsphere1}
\end{align}
and moreover, the analysis of Section \ref{SECgeomINTCH} also applies as follows
\begin{align} 
\begin{aligned} 
(\mathbf{E},\mathbf{P},\mathbf{L},\mathbf{G})\lrpar{x_{-R,2R}^{\mathrm{Kerr}}} =& \lrpar{\mathbf{E}_{0},\mathbf{P}_{0},\mathbf{L}_{0},\mathbf{C}_{0}-3R \cdot \mathbf{P}_{0}} \\
&+ \lrpar{\OO(R^{-1}), \OO(R^{-3/2}), \OO(R^{-1/2}), \OO(R^{-1/4})}.
\end{aligned} \label{EQconvChargesKERR10}
\end{align}

\begin{remark}[Completeness of outgoing null congruences in Kerr] In \cite{ACR2} we explicitly prove that for large $R\geq1$, the constructed sphere $S_{-R,2R}$ has complete outgoing null congruence in Kerr. \end{remark}

\subsection{Proof of Theorem \ref{PROPmain2}} \label{SECchargluingtoKERR} In the following we assume for ease of presentation that the sphere data satisfies the better decay $\mathbf{P}(x_{-R,R}) = \OO(R^{-3/2})$. As mentioned in Section \ref{SECgeomINTCH}, this is the case for sphere data lying in strongly asymptotically flat spacelike initial data. The general case of sphere data with decay $\mathbf{P}(x_{-R,R}) = \smallO(R^{-1/2})$ is based on the same ideas and treated in \cite{ACR2}.

Let $(\mathbf{E}_{\mathrm{ADM}},\mathbf{P}_{\mathrm{ADM}}=0,\mathbf{L}_{\mathrm{ADM}},\mathbf{C}_{\mathrm{ADM}})$ denote the ADM invariants of the given strongly asymptotically flat spacelike initial data. In the following, for $\varep>0$ small and $R\geq1$ large, we consider vectors $(\mathbf{E}_0,\mathbf{P}_0,\mathbf{L}_0,\mathbf{C}_0) \in \RRR^{10}$ such that
\begin{align*} 
\begin{aligned} 
\lrpar{R^{1/2} \vert \mathbf{E}_0 - \mathbf{E}_{\mathrm{ADM}} \vert}^2 + \lrpar{R^{1/2} \vert \mathbf{P}_0 \vert}^2 + \vert \mathbf{L}_0 - \mathbf{L}_{\mathrm{ADM}} \vert^2 + \vert \mathbf{C}_0 - \mathbf{C}_{\mathrm{ADM}} \vert^2 \leq \varep.
\end{aligned} 
\end{align*}

\noindent For sufficiently large $R\geq1$, by \eqref{EQstrongCONVkerrsphere1} we can apply Theorem \ref{THMmain1} with $\varep = R^{-3/2}$ to characteristically glue -- up to $(\mathbf{E},\mathbf{P},\mathbf{L},\mathbf{G})$ -- the sphere data $x_{-R,R}$ to Kerr sphere data $x_{-R,2R}^{\mathrm{Kerr}}$. We denote by $(\mathbf{E},\mathbf{P},\mathbf{L},\mathbf{G})(x_{-R,2R})$ the charges transported from $S_{-R,R}$ along $\HH_{-R,[R,2R]}$ to $S_{-R,2R}$ by the null constraint equations. By Theorem \ref{THMmain1} and \eqref{EQconvChargesKERR10}, we have that the \emph{matching error of charges} can be expressed as
\begin{align} 
\begin{aligned} 
f_R\lrpar{\mathbf{E}_0,\mathbf{P}_0,\mathbf{L}_0,\mathbf{C}_0} :=& (\mathbf{E},\mathbf{P},\mathbf{L},\mathbf{G})(x_{-R,2R}) - (\mathbf{E},\mathbf{P},\mathbf{L},\mathbf{G})(x_{-R,2R}^{\mathrm{Kerr}}) \\
=& \lrpar{\mathbf{E}_{\mathrm{ADM}},\mathbf{P}_{\mathrm{ADM}},\mathbf{L}_{\mathrm{ADM}},\mathbf{C}_{\mathrm{ADM}}} - \lrpar{\mathbf{E}_{0},\mathbf{P}_{0},\mathbf{L}_{0},\mathbf{C}_{0}-3R \cdot \mathbf{P}_{0}}\\
&+ \lrpar{\OO(R^{-1}), \OO(R^{-3/2}), \smallO\lrpar{1}, \smallO\lrpar{1}}.
\end{aligned} \label{EQdeff}
\end{align}
The linear mapping 
$$\lrpar{\mathbf{E}_{0},\mathbf{P}_{0},\mathbf{L}_{0},\mathbf{C}_{0}} \mapsto \lrpar{\mathbf{E}_{0},\mathbf{P}_{0},\mathbf{L}_{0},\mathbf{C}_{0}-3R \cdot \mathbf{P}_{0}}$$ 
is a bijection, and maps $(\mathbf{E}_{\mathrm{ADM}},0,\mathbf{L}_{\mathrm{ADM}},\mathbf{G}_{\mathrm{ADM}})$ to itself. Hence it follows from a standard degree argument that for $R\geq1$ large, there exists $(\mathbf{E}_{0},\mathbf{P}_{0},\mathbf{L}_{0},\mathbf{C}_{0})$ such that the matching error vanishes, that is,
\begin{align} 
\begin{aligned} 
f_R\lrpar{\mathbf{E}_0,\mathbf{P}_0,\mathbf{L}_0,\mathbf{C}_0} = (0,0,0,0).
\end{aligned} \label{EQzerof}
\end{align}
In other words, we have characteristic gluing of sphere data from $x_{-R,R}$ on $S_{-R,R}$ to the sphere data $x_{-R,2R}^{\mathrm{Kerr}}$ corresponding to the vector $(\mathbf{E}_0,\mathbf{P}_0,\mathbf{L}_0,\mathbf{C}_0)$. By \eqref{EQdeff} and \eqref{EQzerof}, we have that
\begin{align*} 
\begin{aligned} 
(\mathbf{E}_0,\mathbf{P}_0,\mathbf{L}_0,\mathbf{C}_0) = (\mathbf{E}_{\mathrm{ADM}},0,\mathbf{L}_{\mathrm{ADM}},\mathbf{G}_{\mathrm{ADM}}) + \lrpar{\OO(R^{-1}), \OO(R^{-3/2}), \smallO\lrpar{1}, \smallO\lrpar{1}}.
\end{aligned} 
\end{align*}

\section{Spacelike gluing to Kerr} \ni In this section we prove Corollary \ref{CORmain3}. Consider strongly asymptotically flat spacelike initial data $(\Si,g,k)$, and define spheres $S_{-R,R}$ with higher-order sphere data $x_{-R,R}$ on it, see Section \ref{SECgeomINTCH}.

For $R\geq1$ sufficiently large, we can apply our bifurcate characteristic gluing to Kerr, Proposition \ref{PROPmain2}, to construct higher-order characteristic initial data on the hypersurface $\HHb_{-R,[R,2R]}\cup \HH_{[-2R,-R],2R}$ such that the higher-order sphere data on $S_{-R,R}$ agrees with $x_{-R,R}$ and with higher-order Kerr sphere data $x^{\mathrm{Kerr}}_{-2R,2R}$ on $S_{-2R,2R}$. 

Applying the local existence result \cite{LukChar,LukRod1} for the characteristic initial value problem, we can construct a spacelike hypersurface in the resulting spacetime which leads from the given spacelike hypersurface $\Si$ to a (reference) spacelike hypersurface in the Kerr spacetime we glued to. The induced spacelike initial data solves the spacelike gluing problem to Kerr for strongly asymptotically flat spacelike initial data.

\begin{figure}[H]
\begin{center}
\includegraphics[width=12cm]{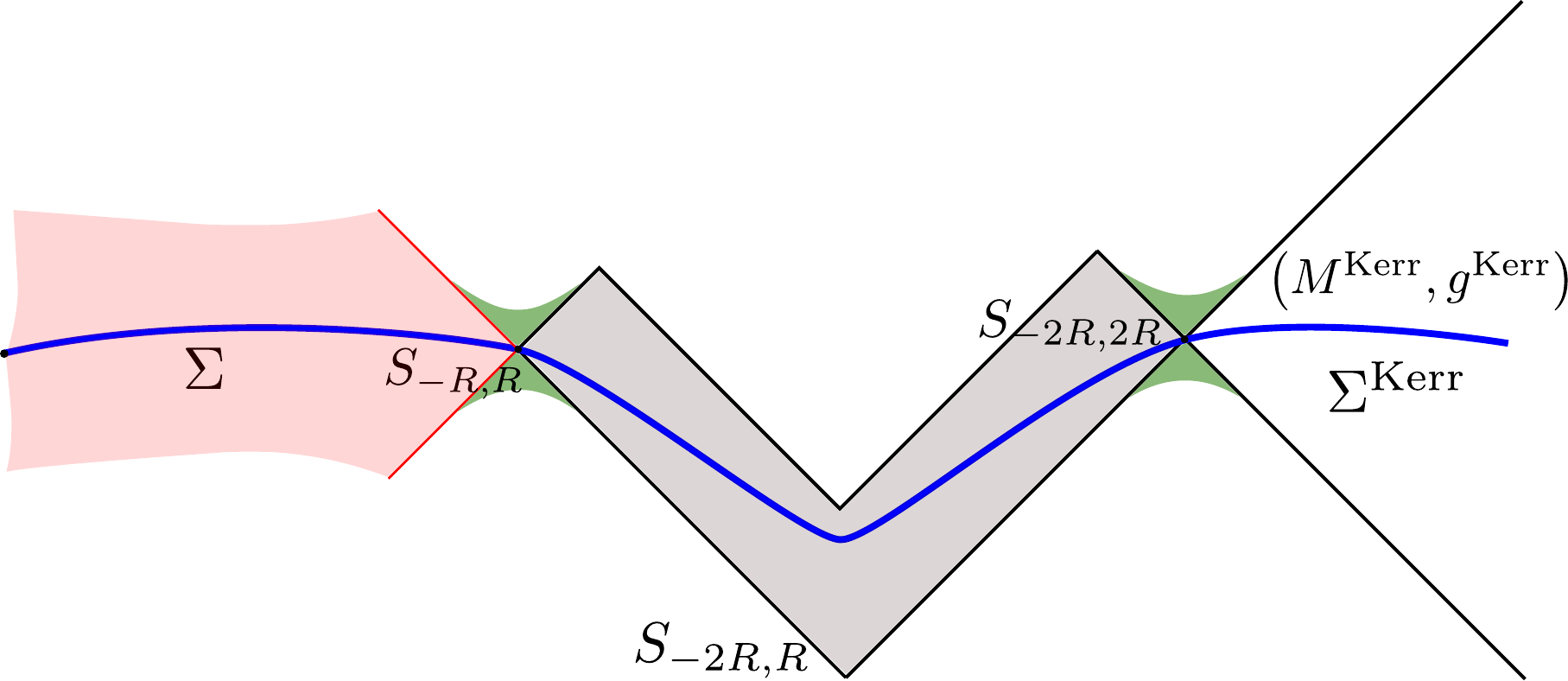}
\vspace{0.4cm} 
\caption{The proof of the spacelike gluing to Kerr.}\label{FIGKerr}
\end{center}
\end{figure}

\section{Localized characteristic gluing}\label{SECproofLocalization} \ni In this section we first prove the so-called \emph{fundamental localization lemma} in Section \ref{SECfundamentalLEMMA}, and introduce the localized sphere data perturbation $W$ in Section \ref{SECWperturbation}, before turning to the proof of Theorem \ref{THMmain4} in Section \ref{SECproofTHMmain4}.

\subsection{The fundamental localization lemma} \label{SECfundamentalLEMMA} In this section we work in the linearized setting. Denote by $\mathcal{K}$ the space of solutions $\dot{x}$ to the linearized null structure equations at Minkowski along $\HH\cup\HHb$ such that $\dot{x}\vert_{S_1}=0$ and $\dot{x}\vert_{S_3}=0$. The following is the main result of this section.
\begin{lemma}[Fundamental localization lemma] \label{LEMlemma1} Consider sphere data $\dot{x}_1\equiv0$ on $S_1$, and sphere data $\dot{x}_{1+2c}\neq 0$ on $S_{1+2c}$ such that, for some angular region $\RR$,
\begin{align*} 
\begin{aligned} 
\dot{x}_{1+2c} \vert_{\RR^c} \equiv 0, \,\, (\mathbf{E},\mathbf{P},\mathbf{L},\mathbf{G})(\dot{x}_{1+2c}) = (0,0,0,0).
\end{aligned} 
\end{align*}
The unique solution $\dot{x} \in \KK^{\perp}$ to the linearized null constraint equations on $\HH \cup \HHb$ satisfying $\dot{x} \vert_{S_1} = 0$ and $\dot{x} \vert_{S_{1+2c}} = \dot{x}_{1+2c}$ has the property that $\dot{x} \equiv 0$ in $\RR^c$ along $\HH \cup \HHb$.

\end{lemma}

\ni Indeed, the solution $\dot{x} \in \KK^{\perp}$ itself is constructed by the methods of Section \ref{SECcharGluing}. Assume for contradiction that $\dot{x}$ is non-trivial in an open angular region $\SS \subset \RR^c$. Let $\varphi: \SSS^2 \to [0,1]$ be a smooth angular cut-off function such that $\varphi \vert_\SS \equiv 1$ and $\varphi \vert_{\RR} \equiv 0$. On the one hand, cutting off the linearized free data corresponding to $\dot{x}$ on $\HH\cup \HHb$ by $\varphi$ yields linearized free data which corresponds to a non-trivial solution $\tilde{\dot{x}}$ to the linearized null constraint equations with $\tilde{\dot{x}}\vert_{S_1}\equiv0$ and $\tilde{\dot{x}}\vert_{S_3}\equiv0$, which, by definition, implies that $\tilde{\dot{x}}\in \KK$. On the other hand, for a generic choice of the cut-off function $\varphi$, it holds that $\langle \dot{x}, \tilde{\dot{x}} \rangle \neq 0$, where the scalar product is the one associated to the Hilbert space $\XX(\HH\cup\HHb)$. We conclude that $\dot{x} \notin \KK^\perp$. This contradiction finishes the proof of Lemma \ref{LEMlemma1}.

\subsection{The localized sphere data perturbation $W$} \label{SECWperturbation} In this section we specify the localized sphere data perturbation $W$ and prove estimates. More precisely, we prove the following lemma.
\begin{lemma} \label{LEMdefW} There exists a linear mapping which maps vectors $(\dot{\mathbf{E}}_0, \dot{\mathbf{P}}_0, \dot{\mathbf{L}}_0, \dot{\mathbf{G}}_0) \in \RRR^{10}$ to smooth sphere data perturbations $W$ supported in $K\setminus K'$ with only non-trivial components $\eta$ and $\Om\trchib$, and satisfying $(\dot{\mathbf{E}},\dot{\mathbf{P}},\dot{\mathbf{L}},\dot{\mathbf{G}})\lrpar{W} = (\dot{\mathbf{E}}_0, \dot{\mathbf{P}}_0, \dot{\mathbf{L}}_0, \dot{\mathbf{G}}_0)$, and bounded by
\begin{align} 
\begin{aligned} 
\Vert W \Vert_{\XX(S_{1+2c})} \les \frac{1}{1+2c} \vert \dot{\mathbf{E}}_0 \vert + \frac{1}{1+2c} \vert \dot{\mathbf{P}}_0 \vert + \frac{1}{(1+2c)^2} \vert \dot{\mathbf{L}}_0 \vert + \frac{1}{(1+2c)^2} \vert \dot{\mathbf{G}}_0 \vert.
\end{aligned} \label{EQWestimatefull}
\end{align}
\end{lemma}

\ni The following two observations are the basic ingredients of the proof of Lemma \ref{LEMdefW}. Their proofs are straight-forward and omitted.
\begin{enumerate}
\item For real numbers $(f^{(0)})_0$, $(f^{(1m)})_0$, $m=-1,0,1$, there is a smooth scalar function $f$ on $S_{1+2c}$, supported in $K \setminus K'$, such that $f^{(0)} = (f^{(0)})_0$ and $f^{(1m)}= (f^{(1m)})_0$ for $m=-1,0,1$, and bounded by
\begin{align} 
\begin{aligned} 
\sum\limits_{0\leq i \leq 4} \,\, \int\limits_{S_{1+2c}} \vert \Nd^i f \vert_{\gac}^2 d\mu_{\gac} \les  \left\vert \lrpar{f^{(0)}}_0 \right\vert^2 + \sum\limits_{m=-1,0,1} \left\vert \lrpar{f^{(1m)}}_0 \right\vert^2.
\end{aligned} \label{EQfestimate109}
\end{align}

\item For real numbers $(X_E^{(1m)})_0$ and $(X_H^{(1m)})_0$ for $m=-1,0,1$, there is a smooth vectorfield $X$ on $S_{1+2c}$, supported in $K \setminus K'$, such that $X_E^{(1m)} = (X_E^{(1m)})_0$ and $X_H^{(1m)}= (X_H^{(1m)})_0$ for $m=-1,0,1$, and bounded by
\begin{align} 
\begin{aligned} 
\sum\limits_{0\leq i \leq 5} \,\, \int\limits_{S_{1+2c}} \vert \Nd^i X \vert_{\gac}^2 d\mu_{\gac} \les \sum\limits_{m=-1,0,1}  \left\vert \lrpar{X_E^{(1m)}}_0 \right\vert^2 + \left\vert \lrpar{X_H^{(1m)}}_0 \right\vert^2.
\end{aligned} \label{EQXestimate109}
\end{align}
\end{enumerate}

\ni We turn now to the construction of $W$. By the formulas for $(\dot{\mathbf{E}},\dot{\mathbf{P}},\dot{\mathbf{L}},\dot{\mathbf{G}})$ in Section \ref{SECspheredataCharges}, and using that $\eta$ and $\Om\trchib$ are to be the only non-trivial components of $W$, it holds for $m=-1,0,1$ that
\begin{align*} 
\begin{aligned} 
\dot{\mathbf{E}}(W) =& \frac{\sqrt{4\pi}}{16\pi} \lrpar{(1+2c)^2 (\Om\trchib)^{(0)}}, &
\dot{\mathbf{P}}^m(W) =& \frac{\sqrt{4\pi/3}}{16\pi} \lrpar{(1+2c)^2 (\Om\trchib)^{(1m)}- 2(1+2c) (\Divdo \eta)^{(1m)}},\\
\dot{\mathbf{L}}^m(W) =& \frac{\sqrt{8\pi/3}}{8\pi} (1+2c)^2 \eta^{(1m)}_H, &
\dot{\mathbf{G}}^m(W) =& \frac{\sqrt{8\pi/3}}{8\pi}(1+2c)^2 \eta^{(1m)}_E,
\end{aligned} 
\end{align*}
which yields the following conditions on $\eta$ and $\Om\trchib$,
\begin{align} 
\begin{aligned} 
(\Om\trchib)^{(0)} =& \frac{1}{(1+2c)^2} \frac{16\pi}{\sqrt{4\pi}} \dot{\mathbf{E}}_0, &
(\Om\trchib)^{(1m)} =& \frac{1}{(1+2c)^2} \frac{16\pi}{\sqrt{4\pi}} \lrpar{\dot{\mathbf{P}}_0^m+ \frac{1}{1+2c} \dot{\mathbf{G}}_0^m}, \\
\eta^{(1m)}_H =& \frac{1}{(1+2c)^2} \frac{8\pi}{\sqrt{8\pi/3}} \dot{\mathbf{L}}^m_0, &
\eta^{(1m)}_E =&\frac{1}{(1+2c)^2} \frac{8\pi}{\sqrt{8\pi/3}} \dot{\mathbf{G}}^m_0.
\end{aligned} \label{EQrelations9211}
\end{align}
The above two observations (with $X=\eta$ and $f=\Om\trchib$) yield the smooth sphere data perturbation $W$. It remains to estimate $W$ in the scale-invariant sphere data norm on $S_{1+2c}$. The square of the norm for the $\Om\trchib$-component is bounded by (using \eqref{EQfestimate109} and \eqref{EQrelations9211}),
\begin{align*} 
\begin{aligned} 
\sum\limits_{0\leq i \leq 4} (1+2c)^{2i} \int\limits_{S_{1+2c}} \vert \Nd^i (\Om\trchib) \vert^2_{\ga_{1+2c}} d\mu_{\ga_{1+2c}}
=& (1+2c)^2 \sum\limits_{0\leq i \leq 4} \, \int\limits_{S_{1+2c}} \vert \Nd^i (\Om\trchib) \vert^2_{\gac} \cdot d\mu_{\gac} \\
\les& (1+2c)^2 \lrpar{\vert (\Om\trchib)^{(0)} \vert^2 + \sum\limits_{m=-1,0,1}\vert (\Om\trchib)^{(1m)} \vert^2} \\
\les& \frac{1}{(1+2c)^2} \vert \dot{\mathbf{E}}_0 \vert^2 + \frac{1}{(1+2c)^2} \vert \dot{\mathbf{P}}_0 \vert^2 + \frac{1}{(1+2c)^4} \vert \dot{\mathbf{G}}_0 \vert^2.
\end{aligned} 
\end{align*}
The square of the norm for $\eta$ on $S_{1+2c}$ is bounded by (using \eqref{EQXestimate109} and \eqref{EQrelations9211}),
\begin{align*} 
\begin{aligned} 
\sum\limits_{0\leq i \leq 5} (1+2c)^{2i} \int\limits_{S_{1+2c}} \vert \Nd^i \eta \vert^2_{\ga_{1+2c}} d\mu_{\ga_{1+2c}}
=&\sum\limits_{0\leq i \leq 5}\,\,  \int\limits_{S_{1+2c}} \vert \Nd^i \eta \vert^2_{\gac} d\mu_{\gac} \\
\les& \sum\limits_{m=-1,0,1} \lrpar{ \left\vert \eta_H^{(1m)} \right\vert^2 + \left\vert \eta_E^{(1m)} \right\vert^2} \\
\les& \frac{1}{(1+2c)^4} \cdot \vert \dot{\mathbf{L}}_0 \vert^2 + \frac{1}{(1+2c)^4} \cdot \vert \dot{\mathbf{G}}_0 \vert^2.
\end{aligned} 
\end{align*}

\ni The above two imply \eqref{EQWestimatefull} and thus finish the proof of Lemma \ref{LEMdefW}.

\subsection{Proof of Theorem \ref{THMmain4}} \label{SECproofTHMmain4} By the implicit function theorem, the proof of Theorem \ref{THMmain4} reduces to the linearized problem at Minkowski. We omit the (standard) functional setup for the implicit function theorem, and outline instead only the \emph{surjectivity of the linearization} by means of solving the following linear problem: Given a solution $\dot{x}^{\mathrm{orig.}}$ on $\HH\cup\HHb$ to the linearized null constraint equations, construct a solution $\dot{x}$ to the linearized null constraint equations on $\HH\cup\HHb$ such that $\dot{x}\vert_{S_1} =\dot{x}^{\mathrm{orig.}}_1$ on $S_1$ and
\begin{align*} 
\begin{aligned} 
\dot{x}\equiv\dot{x}^{\mathrm{orig.}} \text{ on }K' \text{ along }\HH\cup\HHb, \,\,\,\, \dot{x}\vert_{S_3} = \varphi \cdot \dot{x}^{\mathrm{orig.}}\vert_{S_3} + \dot{W} + \dot{\mathfrak{X}}_3 \text{ on } S_3.
\end{aligned} 
\end{align*}
\ni where the linearized sphere data $\dot{\mathfrak{X}}_3$ vanishes\footnote{For the implicit function theorem setup, it is straight-forward to introduce a weighted $L^2$-based Hilbert space over $\SSS^2 \setminus K'$ which incorporates vanishing towards the boundary; see, for example, \cite{Corvino,CorvinoSchoen}.} on $K'$, and $\dot{W}$ is the sphere data perturbation introduced in Section \ref{SECWperturbation} (which is supported in $K\setminus K'$).

First, by the methods of Theorem \ref{THMmain1} (see also Lemma \ref{THMmain12}), we construct a \emph{background solution} $\dot{x}'$ to the linearized null constraint equations and a sphere data perturbation $\dot{W}$ such that 
\begin{align*} 
\begin{aligned} 
\dot{x}' \vert_{S_1} =\dot{x}^{\mathrm{orig.}}_1 \text{ on } S_1, \,\,\,\, \dot{x}' \vert_{S_3} = \varphi \cdot \dot{x}^{\mathrm{orig.}}\vert_{S_3} + \dot{W} + \dot{\mathfrak{X}}_3 \text{ on } S_3.
\end{aligned} 
\end{align*}
The constructed $\dot{x}'$ does not necessarily agree with $\dot{x}^{\mathrm{orig.}}$ on $K'$ along $\HH\cup \HHb$. Let $(\Omd', \dot{\gd'_c})$ denote the linearized free data associated to $\dot{x}'$.

Second, we glue the linearized free data $(\Omd^{\mathrm{orig.}},\gdcd^{\mathrm{orig.}})$ on $\HH\cup\HHb$ associated to $\dot{x}^{\mathrm{orig.}}$ from $K'$ across the transition region $K\setminus K'$ to the constructed $(\Omd',\dot{\gd'_c})$ on $K^c$ along $\HH\cup\HHb$. We denote the glued linearized free data on $\HH\cup\HHb$ by $(\Omd'',\dot{\gd''_c})$. The corresponding solution $\dot{x}''$ to the linearized null structure equations satisfies $\dot{x}'' \vert_{S_1} =\dot{x}^{\mathrm{orig.}}_1$ on $S_1$, and $\dot{x}\equiv\dot{x}^{\mathrm{orig.}}$ on $K'$ along $\HH\cup\HHb$, and, by construction, the following difference is supported in the transition region $K\setminus K'$,
\begin{align*} 
\begin{aligned} 
\Delta := \dot{x}'' \vert_{S_3} - \lrpar{\varphi \cdot \dot{x}^{\mathrm{orig.}}\vert_{S_3} + \dot{W} + \dot{\mathfrak{X}}_3} 
\end{aligned} 
\end{align*}

Third, we apply Lemma \ref{LEMlemma1} to get a \emph{correcting solution} $\dot{x}^{\mathrm{corr.}}$ to the linearized null structure equations such that $\dot{x}^{\mathrm{corr.}}\vert_{S_1}=0$ and $\dot{x}^{\mathrm{corr.}}\vert_{S_3} = -\Delta$, and $\dot{x}^{\mathrm{corr.}}$ is supported inside the transition region $K\setminus K'$ along $\HH\cup\HHb$. Let $(\Omd^{\mathrm{corr.}}, \gdcd^{\mathrm{corr.}})$ denote the associated linearized free data.

Combining the above, the solution $\dot{x}'''$ to the linearized null constraint equations constructed from sphere data $\dot{x}_1^{\mathrm{orig.}}$ on $S_1$ and the linearized free data $(\Omd''',\dot{\gd'''_c}):= (\Omd'',\dot{\gd''_c}) + (\Omd^{\mathrm{corr.}}, \gdcd^{\mathrm{corr.}})$ on $\HH\cup\HHb$ satisfies by construction $\dot{x}''' \vert_{S_1} = \dot{x}_1^{\mathrm{orig.}}$ and 
\begin{align*} 
\begin{aligned} 
\dot{x}'''\equiv\dot{x}^{\mathrm{orig.}} \text{ on }K' \text{ along }\HH\cup\HHb, \,\,\,\, \dot{x}'''\vert_{S_3} = \varphi \cdot \dot{x}^{\mathrm{orig.}}\vert_{S_3} + \dot{W} + \dot{\mathfrak{X}}_3 \text{ on } S_3.
\end{aligned} 
\end{align*}
Thus $\dot{x}'''$ solves the linear problem. This finishes the proof of Theorem \ref{THMmain4}.
\section{Localization of spacelike initial data}\label{SECproofCarlottoSchoen} \ni In this section we prove Theorem \ref{PROPmain5} by applying Theorem \ref{THMmain4}.

\subsection{Strategy of the proof} \label{SECproofSpacelikeLoc1} Let $(\Si,g,k)$ be asymptotically flat spacelike initial data. For real numbers $R\geq1$ large, consider spheres $S_{R}$ with associated sphere data $x^{\mathrm{orig.}}_{R}$ in $\Si$. By the asymptotic flatness, that sphere data is of size $\OO(R^{-1})$ with respect to the scale-invariant norm. Let $R\geq1$ large and $c>0$ be two real numbers. The proof of Theorem \ref{PROPmain5} is based on subsequent applications of Theorem \ref{THMmain4} from $S_{(1+2c)^n\cdot R}$ to $S_{(1+2c)^{n+1}\cdot R}$ for integers $n\geq0$, outlined below.

\textbf{In the $1^{\mathrm{st}}$ step}, we apply Theorem \ref{THMmain4} to characteristically glue the sphere data $x^{\mathrm{orig.}}_R$ on $S_R$ along $\HHb_{R\cdot[1,1+c]} \cup \HH_{R\cdot [1+c,1+2c]}$ to the sphere data
\begin{align} 
\begin{aligned} 
{x}_{(1+2c)\cdot R}^{\mathbf{(1)}} := \varphi \cdot x^{\mathrm{orig.}}_{(1+2c)\cdot R} + (1-\varphi) \cdot \mathfrak{m} + W^{\mathbf{(1)}} \text{ on } S_{(1+2c)\cdot R},
\end{aligned} \label{EQfirststep1and2cdata}
\end{align}
where $\varphi$ is an angular cut-off function, and $W^{\mathbf{(1)}}$ is a localized sphere data perturbation.

\textbf{In the $n^{\mathrm{th}}$ step}, for integers $n\geq2$, we apply Theorem \ref{THMmain4} to characteristically glue along $\HHb_{(1+2c)^{n-1} \cdot R\cdot  [1,1+c]} \cup \HH_{(1+2c)^{n-1} \cdot R\cdot[1+c,1+2c]}$ the previously constructed sphere data $\tilde{x}_{(1+2c)^{n-1} \cdot R}$ on $S_{(1+2c)^{n-1} \cdot R}$ to 
\begin{align} 
\begin{aligned} 
{x}^{\mathbf{(n)}}_{(1+2c)^n\cdot R} := \varphi \cdot x^{\mathrm{orig.}}_{(1+2c)^{n} \cdot R} + (1-\varphi) \cdot \mathfrak{m} + W^{\mathbf{(n)}} \text{ on } S_{(1+2c)^n\cdot R}.
\end{aligned} \label{EQnthstepdata}
\end{align}

\ni We prove precise estimates for the constructed characteristic initial data in the next section. In particular, the above iteration is well-defined.

\begin{figure}[H]
\begin{center}
\includegraphics[width=7.5cm]{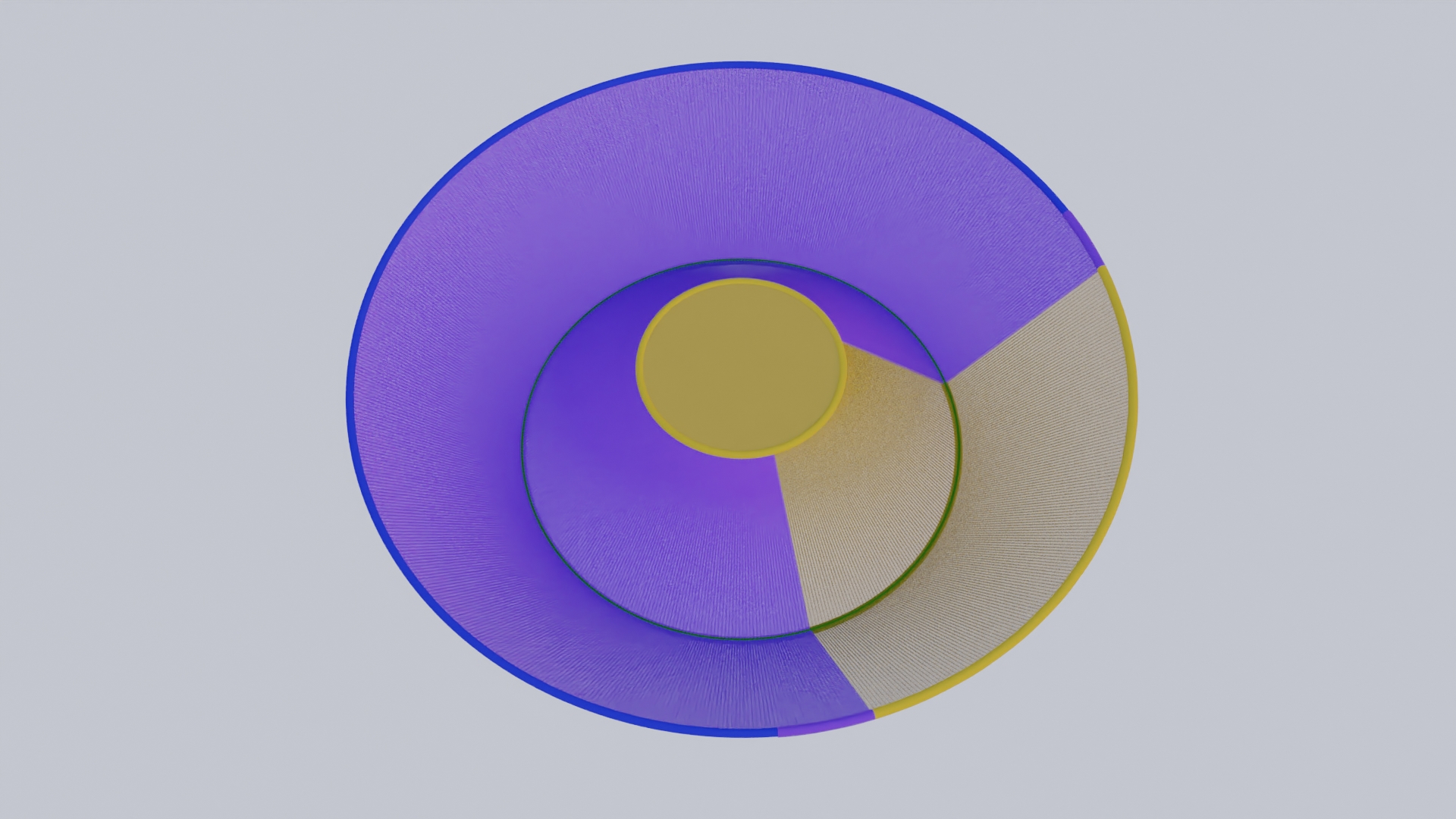} \hspace{5mm} \includegraphics[width=7.5cm]{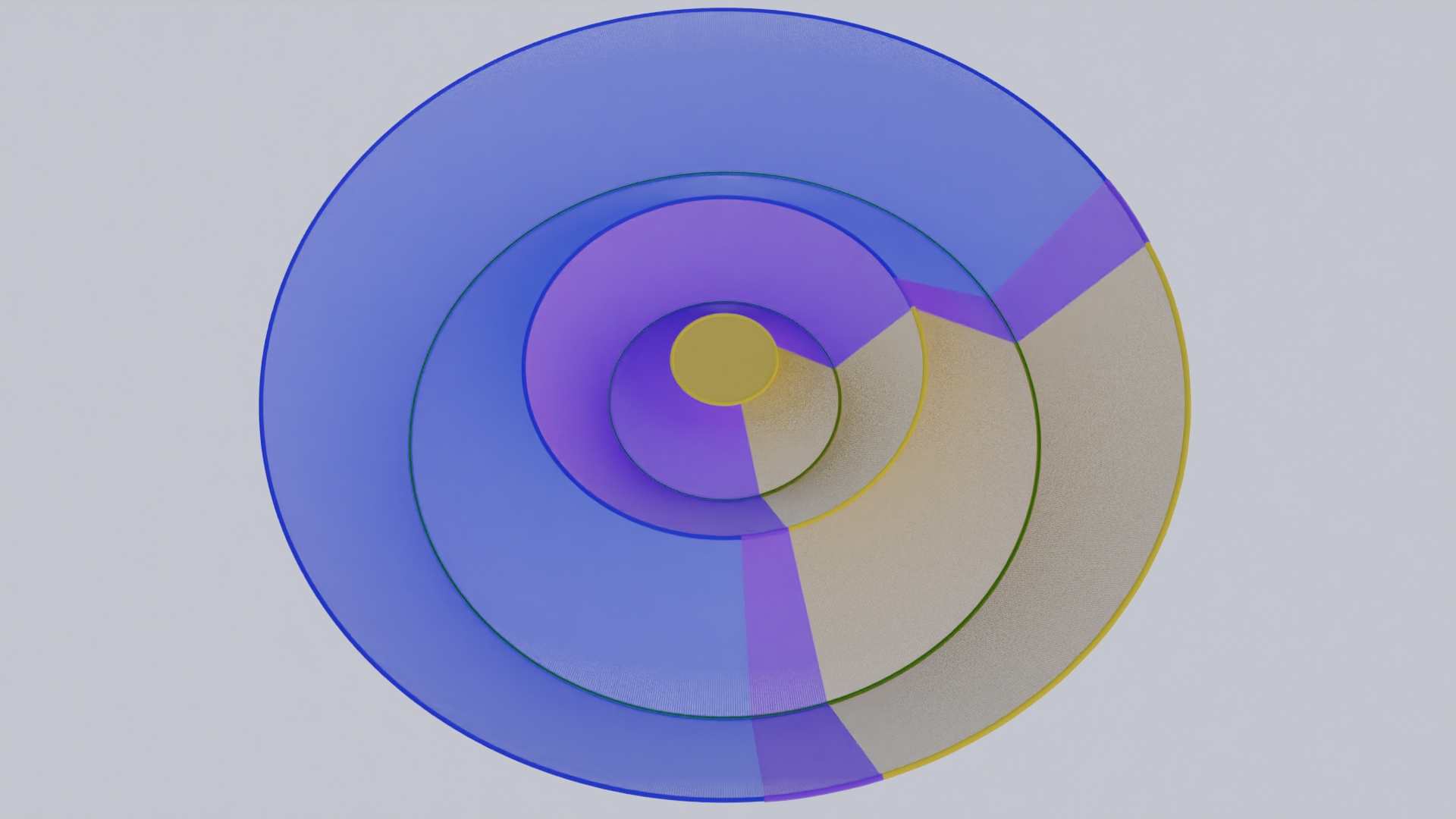} \\ \vspace{5mm}
\includegraphics[width=7.5cm]{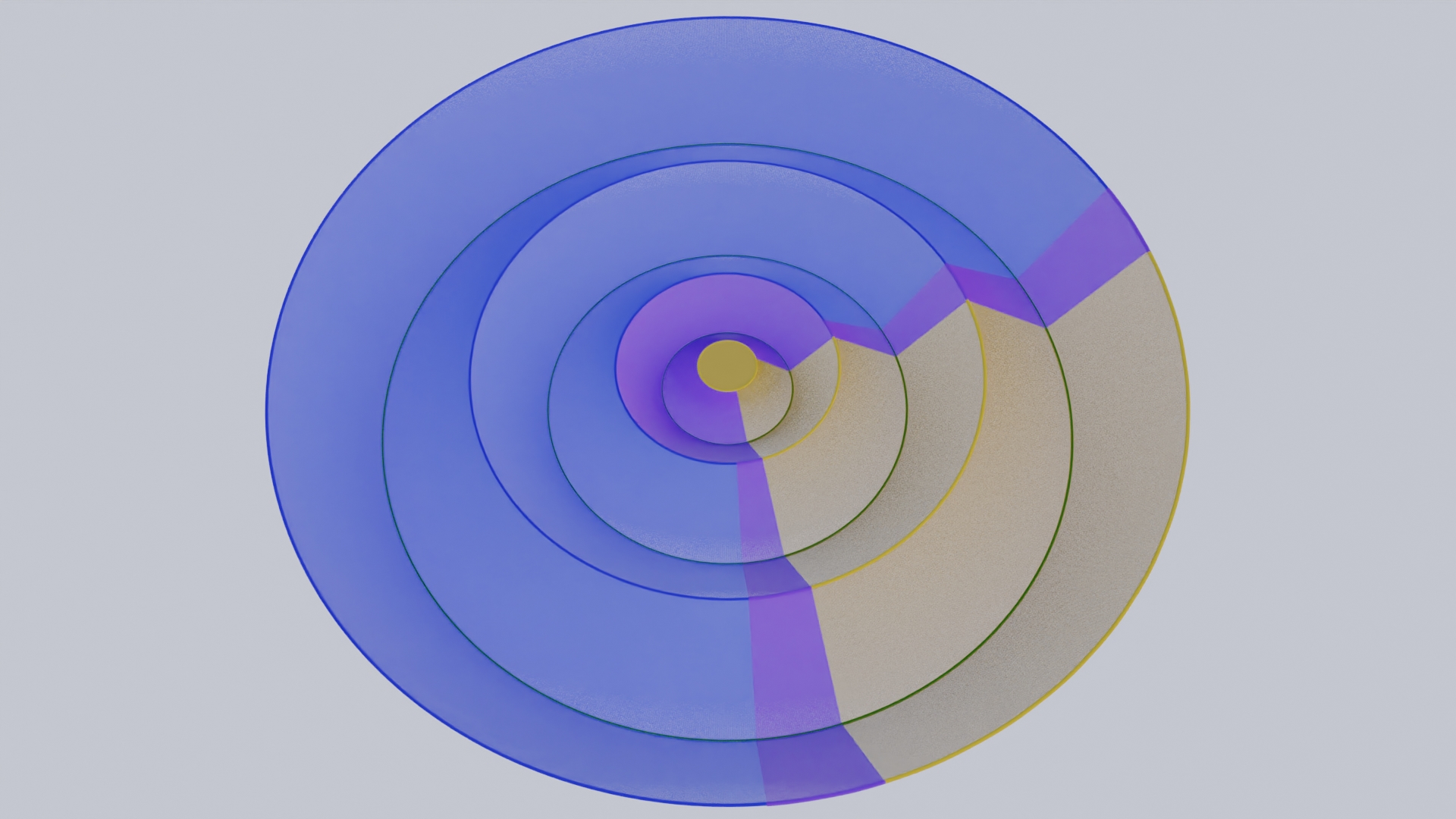} \hspace{5mm} \includegraphics[width=7.5cm]{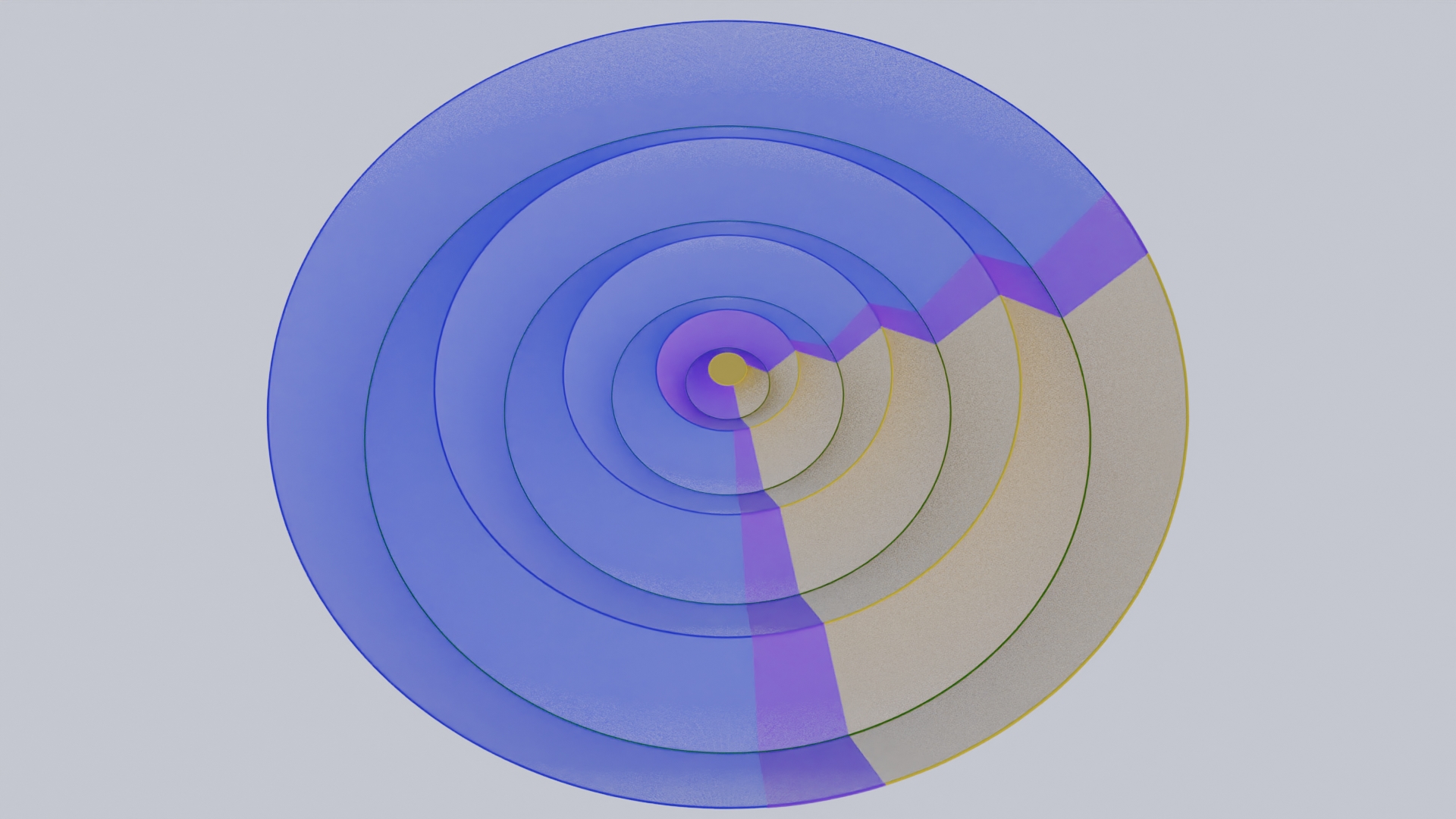} \\
\vspace{0.4cm} 
\caption{The first four steps of the localized characteristic gluing construction.  The yellow region is equipped with the given data whereas the blue region is equipped with the Minkowski data.  The purple region is the transition region.}\label{FIG7}
\end{center}
\end{figure}

\ni In Minkowski spacetime, the intersection of the future domain of influence of an angular region of aperture $\th>0$ of the null hypersurfaces $\HHb_{(1+2c)^{n-1} \cdot R\cdot  [1,1+c]} \cup \HH_{(1+2c)^{n-1} \cdot R\cdot[1+c,1+2c]}$ and the spacelike hypersurface $\{t=0\}$ contains an angular region of aperture $\th':=\th-\arcsin(c/1+c)$. This can be proved by explicitly calculating the intersections of past cones from points on $\{t=0\}$ with the null hypersurfaces $\HHb_{(1+2c)^{n-1} \cdot R\cdot  [1,1+c]} \cup \HH_{(1+2c)^{n-1} \cdot R\cdot[1+c,1+2c]}$, and deriving an upper bound on the angular width of the intersections. Alternatively, one can show that the intersections of the future cones from points on the auxiliary spheres $S_{(1+2c)^{n-1}\cdot R\cdot (1+c)}$ with the hypersurface $\{t=0\}$ are spheres centered on $\{r=(1+2c)^{n-1}\cdot R\cdot (1+c) \}$ with radius $(1+2c)^{n-1}\cdot R\cdot c$.  

\begin{figure}[H]
\begin{center}
\includegraphics[width=10cm]{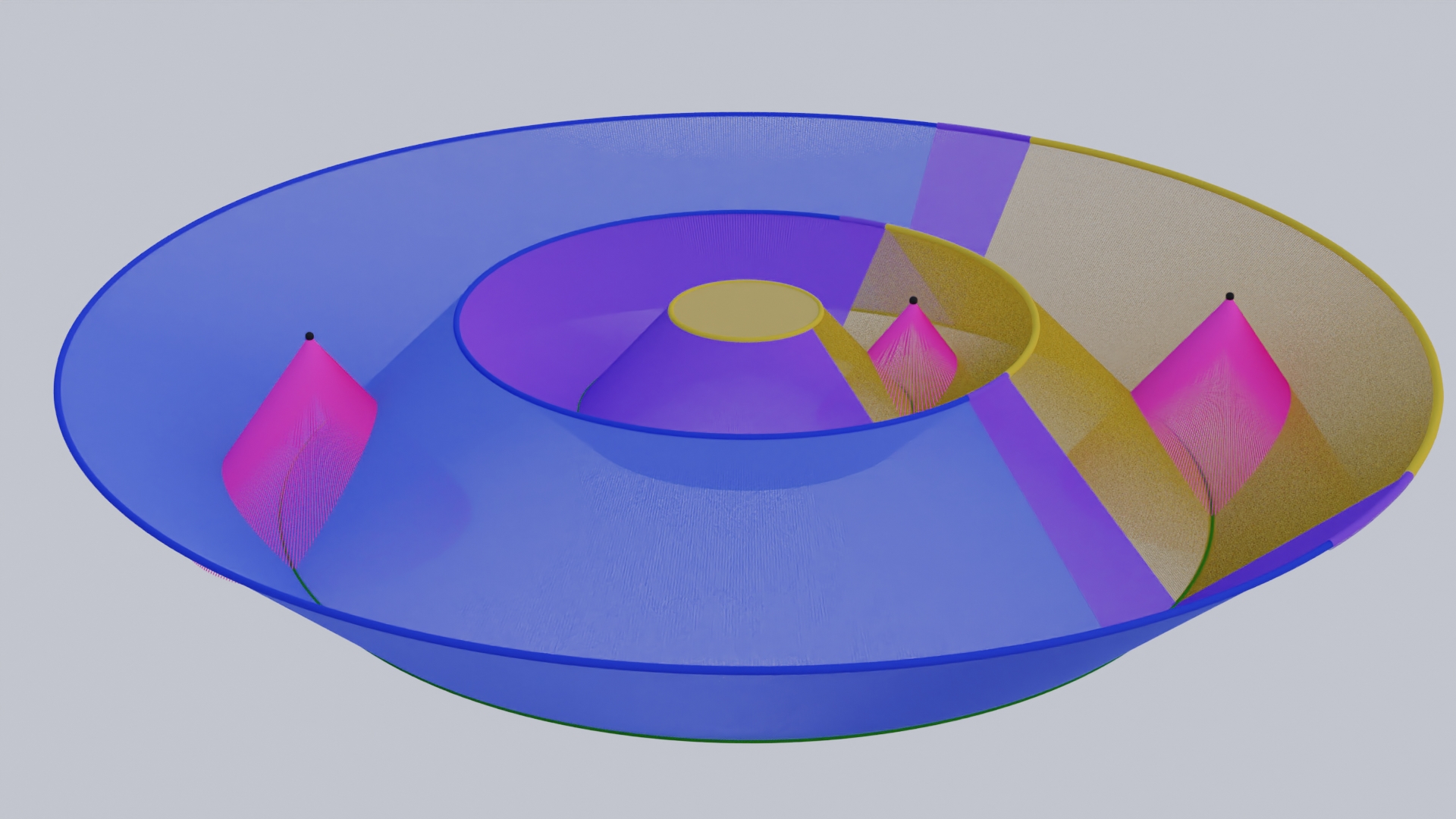} 
\vspace{0.4cm} 
\caption{The past cones from points on $t=0$ intersecting portions of the null hypersurfaces along which the characteristic gluing takes place.}\label{FIG8}
\end{center}
\end{figure}

\ni For each constructed characteristic initial data on $\HHb_{(1+2c)^{n-1} \cdot R\cdot  [1,1+c]} \cup \HH_{(1+2c)^{n-1} \cdot R\cdot[1+c,1+2c]}$ we can solve the Einstein equations to the future. By the bounds on the characteristic initial data proved in Sections \ref{SECspacelikeLOCdecayEstim} and \ref{SECconclusionCarlSproof2}, and the Cauchy stability of the characteristic initial value problem for the Einstein equations, it follows that the resulting spacetime is close to Minkowski. Moreover, by construction, the spacetime is isometric to the original in one region, and isometric to Minkowski in another (for $n\geq2$). 
\begin{figure}[H]
\begin{center}
\includegraphics[width=10cm]{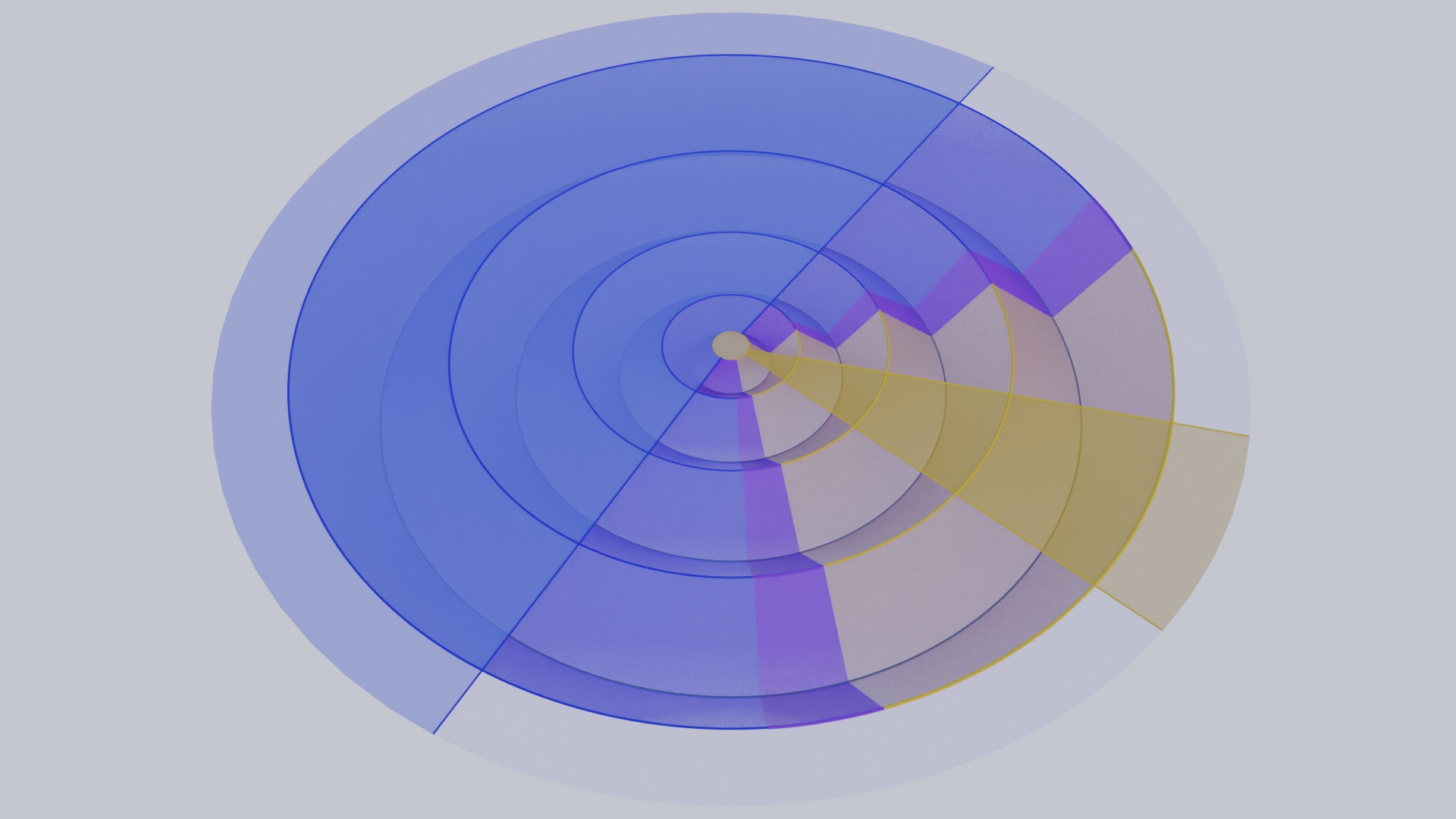}
\vspace{0.4cm} 
\caption{By the characteristic initial value problem and the domain of dependence property, the localized characteristic gluing yields localized spacelike gluing with slightly enlarged transition region.}\label{FIG9}
\end{center}
\end{figure}
\ni Hence, applying the above domain-of-influence property in Minkowski, we conclude that for given $0<\th_1<\th_2<\pi$, and for $R\geq1$ sufficiently large, there are angular regions $K$ and $K'$, a real number $c>0$, and a spacelike hypersurface $\{t=0\}$ such that the induced spacelike initial data $(\tilde{g},\tilde{k})$ on $\{t=0\}$ agree with $(g,k)$ in a cone of aperture $\th_1$ and  with the Minkowski spacelike initial data $(e,0)$ in the complement of a slightly larger cone of aperture $\th_2$.
\begin{figure}[H]
\begin{center}
\includegraphics[width=10cm]{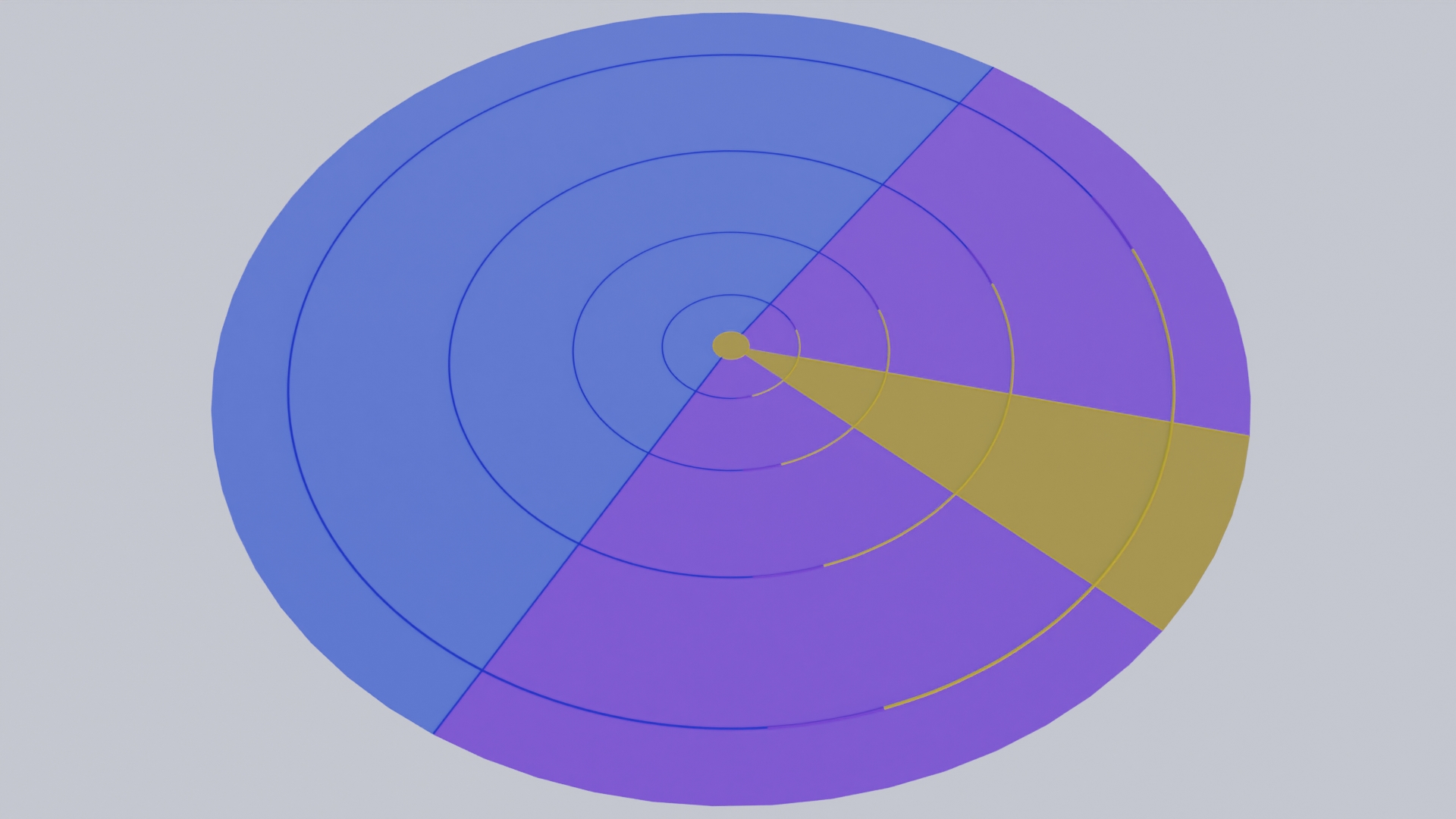} 
\vspace{0.4cm} 
\caption{The localized spacelike gluing construction from the localized characteristic gluing construction.}\label{FIG10}
\end{center}
\end{figure}

\subsection{Proof of decay estimates} \label{SECspacelikeLOCdecayEstim} In this section we work mainly in the rescaled picture, i.e. gluing from $S_1$ to $S_{1+2c}$ in each step, tacitly omitting the rescaling notation, and denote $\HHb:=\HHb_{[1,1+c]}$ and $\HH:=\HH_{[1+c,1+2c]}$. In the following we prove by induction that for $R\geq1$ large, for integers $n\geq1$, 
\begin{align} 
\begin{aligned} 
\Vert W^{\textbf{(n)}} \Vert_{\XX(S_{1+2c})} =&\OO(  (1+2c)^{-n} \cdot R^{-1}).
\end{aligned} \label{EQinductionHYP01}
\end{align}
The proof is based on the following observations. First, we recall from \eqref{EQWestimatefull} the following bound,
\begin{align} 
\begin{aligned} 
\Vert W \Vert_{\XX(S_{1+2c})} \les \frac{1}{1+2c} \cdot (\vert \dot{\mathbf{E}}(W) \vert + \vert \dot{\mathbf{P}}(W) \vert )+ \frac{1}{(1+2c)^2} \cdot (\vert \dot{\mathbf{L}}(W) \vert + \vert \dot{\mathbf{G}}(W) \vert).
\end{aligned} \label{EQWestimatefull2}
\end{align}
\ni Second, for $x^{\mathbf{(n)}}_{1+2c}$ close to $\mathfrak{m}$ on $S_{1+2c}$, it holds that
\begin{align} 
\begin{aligned} 
(\mathbf{E},\mathbf{P},\mathbf{L}, \mathbf{G})(x^{\mathbf{(n)}}_{1+2c}) = (\dot{\mathbf{E}}_0,\dot{\mathbf{P}}_0,\dot{\mathbf{L}}_0,\dot{\mathbf{G}}_0)(x^{\mathbf{(n)}}_{1+2c}) + \OO(\Vert x^{\mathbf{(n)}}-\mathfrak{m} \Vert^2_{\XX(S_{1+2c})}).
\end{aligned} \label{EQlinearexpCHARGES}
\end{align}
\ni Third, for $x^{\mathbf{(n)}}$ close to $\mathfrak{m}$ along $\HH\cup\HHb$, by the linear conservation of charges, 
\begin{align} 
\begin{aligned} 
(\mathbf{E},\mathbf{P},\mathbf{L}, \mathbf{G})(x^{\textbf{(n)}}_{{1+2c}})= (\mathbf{E},\mathbf{P},\mathbf{L}, \mathbf{G}) (x^{\textbf{(n)}}_{{1}}) + \OO \lrpar{\Vert x^{\mathbf{(n)}}-\mathfrak{m} \Vert^2_{\XX(\HH\cup\HHb)}}.
\end{aligned} \label{EQlinearconservationEPLG112}
\end{align}
Fourth, for $x^{\mathbf{(n)}}_1$ and $x^{\mathrm{orig.}}_{1+2c}$ close to $\mathfrak{m}$, by the implicit function theorem it holds that
\begin{align} 
\begin{aligned} 
\Vert x^{\mathbf{(n)}}-\mathfrak{m} \Vert_{\XX(\HH\cup\HHb)}+\Vert W^{\mathbf{(n)}} \Vert_{\XX(S_{1+2c})} \les \Vert x^{\textbf{(n)}}_{1} - \mathfrak{m} \Vert_{\XX(S_{1})} + \Vert x^{\mathrm{orig.}}_{1+2c} - \mathfrak{m} \Vert_{\XX(S_{1+2c})}.
\end{aligned} \label{EQIFTestimates}
\end{align}
Fifth, by definition of the sphere data $x^{\textbf{(n)}}_{1+2c}$, see \eqref{EQnthstepdata}, it holds that for $x^{\mathrm{orig.}}_{1+2c}$ close to $\mathfrak{m}$,
\begin{align} 
\begin{aligned} 
\Vert x^{\textbf{(n)}}_{1+2c} - \mathfrak{m} \Vert_{\XX(S_{1+2c})} \les \Vert W^{\textbf{(n)}} \Vert_{\XX(S_{1+2c})} + \Vert x^{\mathrm{orig.}}_{1+2c} -\mathfrak{m} \Vert_{\XX(S_{1+2c})}
\end{aligned} \label{EQdataEstimn}
\end{align}
Sixth, by asymptotic flatness, it holds that for $R\geq1$ large,
\begin{align} 
\begin{aligned} 
\Vert x^{\mathrm{orig.}}_{(1+2c)^n\cdot R} -\mathfrak{m} \Vert_{\XX(S_{(1+2c)^n\cdot R})} =\OO( (1+2c)^{-n} \cdot R^{-1}) \text{ for } n\geq1.
\end{aligned} \label{EQafSPHEREDATAEPLGcontrol}
\end{align} 
\textbf{Induction basis $n=1$.} From \eqref{EQIFTestimates} and \eqref{EQafSPHEREDATAEPLGcontrol} we get that
\begin{align} 
\begin{aligned} 
(\mathbf{E},\mathbf{P},\mathbf{L}, \mathbf{G}) (x^{\textbf{(1)}}_{{1}}) =\OO( R^{-1}), \,\,\,\, \Vert x^{\textbf{(1)}} -\mathfrak{m} \Vert_{\XX(\HH\cup\HHb)} =\OO( R^{-1}),
\end{aligned} \label{EQafnessandChargesS1}
\end{align}
which leads for $R\geq1$ sufficiently large by \eqref{EQlinearconservationEPLG112} to 
\begin{align*} 
\begin{aligned} 
(\mathbf{E},\mathbf{P},\mathbf{L}, \mathbf{G})(x^{\textbf{(1)}}_{{1+2c}})= \OO( R^{-1}) + \OO((\OO(R^{-1}))^2) = \OO(R^{-1}),
\end{aligned} 
\end{align*}
and thus, for $R\geq1$ sufficiently large, by \eqref{EQfirststep1and2cdata}, \eqref{EQWestimatefull2}, \eqref{EQlinearexpCHARGES}  and \eqref{EQafSPHEREDATAEPLGcontrol} further to
\begin{align*} 
\begin{aligned} 
\Vert W^{\textbf{(1)}} \Vert_{\XX(S_{1+2c})} \les& \frac{1}{1+2c} \cdot (\vert \dot{\mathbf{E}}(W^{\textbf{(1)}}) \vert + \vert \dot{\mathbf{P}}(W^{\textbf{(1)}}) \vert )+ \frac{1}{(1+2c)^2} \cdot(\vert \dot{\mathbf{L}}(W^{\textbf{(1)}}) \vert + \vert \dot{\mathbf{G}}(W^{\textbf{(1)}}) \vert)\\
\les& \frac{1}{1+2c} \cdot (\vert \dot{\mathbf{E}}(x^{\textbf{(1)}}_{1+2c}) \vert + \vert \dot{\mathbf{P}}(x^{\textbf{(1)}}_{1+2c}) \vert )+ \frac{1}{(1+2c)^2} \cdot(\vert \dot{\mathbf{L}}(x^{\textbf{(1)}}_{1+2c}) \vert + \vert \dot{\mathbf{G}}(x^{\textbf{(1)}}_{1+2c}) \vert) \\
&+\OO((1+2c)^{-1} \cdot R^{-1}) \\
\les& \frac{1}{1+2c} \cdot (\vert {\mathbf{E}}(x^{\textbf{(1)}}_{1+2c}) \vert + \vert {\mathbf{P}}(x^{\textbf{(1)}}_{1+2c}) \vert )+ \frac{1}{(1+2c)^2} \cdot(\vert {\mathbf{L}}(x^{\textbf{(1)}}_{1+2c}) \vert + \vert {\mathbf{G}}(x^{\textbf{(1)}}_{1+2c}) \vert) \\
&+\OO((1+2c)^{-1} \cdot R^{-1}) + \OO((\OO((1+2c)^{-1} \cdot R^{-1}))^2) \\
=& \OO((1+2c)^{-1}\cdot R^{-1}),
\end{aligned} 
\end{align*}
where we used \eqref{EQafSPHEREDATAEPLGcontrol} to bound
\begin{align*} 
\begin{aligned} 
\frac{1}{1+2c} \cdot (\vert \dot{\mathbf{E}}(x^{\mathrm{orig.}}_{1+2c}) \vert + \vert \dot{\mathbf{P}}(x^{\mathrm{orig.}}_{1+2c}) \vert )+ \frac{1}{(1+2c)^2} \cdot(\vert \dot{\mathbf{L}}(x^{\mathrm{orig.}}_{1+2c}) \vert + \vert \dot{\mathbf{G}}(x^{\mathrm{orig.}}_{1+2c}) \vert) = \OO((1+2c)^{-1} \cdot R^{-1}).
\end{aligned} 
\end{align*}

\ni \textbf{Induction step.} On the one hand, by \eqref{EQscalingCHARGES} it holds on $S_1$ that
\begin{align*} 
\begin{aligned} 
(\mathbf{E},\mathbf{P})(x^{\textbf{(n)}}_{{1}}) = (1+2c)^{-1} \cdot (\mathbf{E},\mathbf{P})(x^{\textbf{(n-1)}}_{{1+2c}}), \,\, (\mathbf{L},\mathbf{G})(x^{\textbf{(n)}}_{{1}}) = (1+2c)^{-2} \cdot (\mathbf{L},\mathbf{G})(x^{\textbf{(n-1)}}_{{1+2c}}). 
\end{aligned} 
\end{align*}
On the other hand, by the induction hypothesis, \eqref{EQIFTestimates}, \eqref{EQdataEstimn} and \eqref{EQafSPHEREDATAEPLGcontrol}, and abusing notation,
\begin{align*} 
\begin{aligned} 
\Vert x^{\mathbf{(n)}}-\mathfrak{m} \Vert_{\XX(\HH\cup\HHb)} \les& \Vert x^{\textbf{(n)}}_{1} - \mathfrak{m} \Vert_{\XX(S_{1})} + \Vert x^{\mathrm{orig.}}_{1+2c} - \mathfrak{m} \Vert_{\XX(S_{1+2c})} \\
\les& (\Vert W^{\textbf{(n-1)}} \Vert_{\XX(S_{1+2c})} + \Vert x^{\mathrm{orig.}}_{1} -\mathfrak{m} \Vert_{\XX(S_{1})})+ \Vert x^{\mathrm{orig.}}_{1+2c} - \mathfrak{m} \Vert_{\XX(S_{1+2c})} \\
=& \OO((1+2c)^{-n+1}\cdot R^{-1})+\OO((1+2c)^{-n}\cdot R^{-1}) \\
=& \OO((1+2c)^{-n+1}\cdot R^{-1}).
\end{aligned} 
\end{align*}
\ni Applying the above two and using \eqref{EQlinearconservationEPLG112}, we get the recursive relation
\begin{align*} 
\begin{aligned} 
(\mathbf{E},\mathbf{P})(x^{\textbf{(n)}}_{{1+2c}}) =& (1+2c)^{-1} \cdot (\mathbf{E},\mathbf{P})(x^{\textbf{(n-1)}}_{{1+2c}}) + \OO((\OO((1+2c)^{-n+1}\cdot R^{-1}))^2) \\
=& (1+2c)^{-1} \cdot (\mathbf{E},\mathbf{P})(x^{\textbf{(n-1)}}_{{1+2c}}) + \OO((1+2c)^{-2n+2}\cdot R^{-2}),
\end{aligned} 
\end{align*}
which leads, using the geometric sum formula, to
\begin{align} 
\begin{aligned} 
(\mathbf{E},\mathbf{P})(x^{\textbf{(n)}}_{{1+2c}}) =& (1+2c)^{-n+1} \cdot \lrpar{(\mathbf{E},\mathbf{P})(x^{\textbf{(1)}}_{{1+2c}}) +\OO(R^{-2}) }\\
=& (1+2c)^{-n+1} \cdot \lrpar{(\mathbf{E},\mathbf{P})(x^{\textbf{(1)}}_{{1}}) +\OO(R^{-2}) },
\end{aligned} \label{EQEPexpression1}
\end{align}
where we used \eqref{EQlinearconservationEPLG112} to relate $(\mathbf{E},\mathbf{P})(x^{\textbf{(1)}}_{{1+2c}})$ and $(\mathbf{E},\mathbf{P})(x^{\textbf{(1)}}_{{1}})$. Similarly, we get for $\mathbf{L}$ and $\mathbf{G}$ that
\begin{align} 
\begin{aligned} 
(\mathbf{L},\mathbf{G})(x^{\textbf{(n)}}_{{1+2c}}) = \OO\lrpar{(1+2c)^{-n+2} \cdot R^{-1}}.
\end{aligned} \label{EQLGexpression1}
\end{align}
The bounds \eqref{EQEPexpression1} and \eqref{EQLGexpression1} on $(\mathbf{E},\mathbf{P},\mathbf{L},\mathbf{G})(x^{\mathbf{(n)}}_{{1+2c}})$ allow to bound $W^{\mathbf{(n)}}$ on $S_{1+2c}$; see the proof of the induction basis. This finishes the proof of \eqref{EQinductionHYP01}.

\subsection{Conclusion of the proof of Theorem \ref{PROPmain5}} \label{SECconclusionCarlSproof2} From the decay estimate \eqref{EQinductionHYP01}, and \eqref{EQIFTestimates}, \eqref{EQdataEstimn} and \eqref{EQafSPHEREDATAEPLGcontrol}, we deduce that
\begin{align*} 
\begin{aligned} 
\Vert x^{\mathbf{(n)}}-\mathfrak{m} \Vert_{\XX(\HH\cup\HHb)} = \OO((1+2c)^{-n+1} \cdot R), \,\, \Vert x^{\textbf{(n)}}_{1+2c} - \mathfrak{m} \Vert_{\XX(S_{1+2c})} = \OO((1+2c)^{-n} \cdot R).
\end{aligned} 
\end{align*}
Importantly, while the above estimates are written out for the norm $\XX$ on $\HH\cup\HHb$ and $S_{1+2c}$, the same bounds hold for \emph{higher regularity norms} of \emph{higher-order sphere data}. Applying the well-posedness of the characteristic initial value problem, working in sufficiently high regularity, and scaling out, we can control the induced spacelike initial data $(\tilde{g},\tilde{k})$ on $\{t=0\}$ in the annulus 
\begin{align*} 
\begin{aligned} 
A_R := A_{[(1+2c)^{n} \cdot R, (1+2c)^{n+1} \cdot R]} = \{ x\in \RRR^3 : (1+2c)^{n} \cdot R \leq \vert x \vert \leq (1+2c)^{n+1} \cdot R\} 
\end{aligned} 
\end{align*}
as follows
\begin{align} 
\begin{aligned} 
\Vert (\tilde{g}-e,\tilde{k}) \Vert_{\XX(A_R)} = \OO( (1+2c)^{-n} \cdot R^{-1}),
\end{aligned} \label{EQasymflat}
\end{align}
where the scale-invariant norm $\XX(A_R)$ is defined by (with an integer $m\geq0$)
\begin{align*} 
\begin{aligned} 
\Vert (\tilde{g}-e,\tilde{k}) \Vert_{\XX(A_R)} := \sum\limits_{0\leq l \leq m+1} R^l \Vert \pr^l (\tilde{g}_{ij}-e_{ij}) \Vert_{L^\infty(A_R)} + \sum\limits_{0\leq l \leq m} R^{l+1} \Vert \pr^l \tilde{k}_{ij} \Vert_{L^\infty(A_R)},
\end{aligned} 
\end{align*}
where we denoted the standard tuple $\pr^l := \pr^{a_1}_1 \pr^{a_2}_2 \pr^{a_3}_3$ with $a_1+a_2+a_3 \leq l$. In particular, \eqref{EQasymflat} implies that the constructed spacelike initial data is asymptotically flat. 

The energy $\mathbf{E}_{\mathrm{ADM}}$ and linear momentum $\mathbf{P}_{\mathrm{ADM}}$ of the constructed spacelike initial data set are $\OO(R^{-1})$-close to the energy and linear momentum of the given spacelike initial data by \eqref{EQEPexpression1}. This finishes the proof of Theorem \ref{PROPmain5}.

\end{document}